\documentclass[twocolumn]{article}

\usepackage[a4paper,margin=0.8in]{geometry}
\usepackage[UKenglish]{babel}
\usepackage[usenames,dvipsnames]{xcolor}
\usepackage{amsmath}
\usepackage{amssymb}
\usepackage{wasysym}
\usepackage{bm}
\usepackage{fancyref}
\usepackage[font=small,labelfont=bf]{caption}
\usepackage{subcaption}
\usepackage[super]{nth}
\usepackage{wrapfig}
\usepackage{dblfloatfix}
\usepackage{tocloft}
\usepackage[round]{natbib}
\usepackage{graphicx}
\usepackage{hyperref}
\definecolor{darkblue}{rgb}{0,0,0.6}
\hypersetup{colorlinks,breaklinks,allcolors=darkblue}
\usepackage{fancyhdr}
\usepackage{grffile}
\usepackage{longtable}
\usepackage{multirow}
\usepackage{xspace}
\usepackage[modulo]{lineno}

\usepackage{titlesec}
\titlespacing*\section{0pt}{2.5ex}{2.5ex} 
\titlespacing*\subsection{0pt}{2.5ex}{1ex} 
\titlespacing*\subsubsection{0pt}{2ex}{0.5ex} 
\titleformat{\section}{\Large\bfseries}{\thesection.}{0.5em}{}
\titleformat{\subsection}{\large\bfseries}{\thesubsection.}{0.5em}{}
\titleformat{\subsubsection}{\bfseries}{\thesubsubsection.}{0.5em}{}

\newcommand{\aap}{Astron. Astrophys.}
\newcommand{\aj}{Astrophys. J.}
\newcommand{\apj}{Astrophys. J.}
\newcommand{\apjl}{Astrophys. J. Lett.}

\newcommand{\icarus}{Icarus}

\newcommand{\mnras}{Mon. Not. R. Astron. Soc.}

\newcommand{\planss}{Planet. Space Sci.}

\newcommand{\psj}{Planet. Sci. J}

\newcommand\nonumfootnote[1]{%
  \begingroup
  \renewcommand\thefootnote{}\footnote{#1}%
  \addtocounter{footnote}{-1}%
  \endgroup
}

\newcommand{\swift}{{\sc Swift}\xspace}
\newcommand{\woma}{{\sc WoMa}\xspace}
\newcommand{\seagen}{{\sc SEAGen}\xspace}
\newcommand{\rebound}{{\sc Rebound}\xspace}

\newcommand{\smars}{{\scriptsize\mars{}}}

\begin{document}

\twocolumn[
\begin{@twocolumnfalse}
\begin{center}

\phantom{.}\\
\vspace{1.5\baselineskip}
\textbf{\Large Origin of Mars's moons by disruptive partial capture of an asteroid}
\\
\vspace{1.2\baselineskip}

\href{http://orcid.org/0000-0001-5383-236X}{Jacob A. Kegerreis}$^{1\,*}$,
\href{http://orcid.org/0000-0001-6513-1659}{Jack J. Lissauer}$^{1}$,
\href{http://orcid.org/0000-0001-5416-8675}{Vincent R. Eke}$^{2}$,
\href{http://orcid.org/0000-0002-4630-1840}{Thomas D. Sandnes}$^{2}$,
\href{http://orcid.org/0009-0008-3020-801X}{Richard C. Elphic}$^{1}$.
\\
\vspace{0.9\baselineskip}

{
\footnotesize
$^{1}$NASA Ames Research Center, Moffett Field, CA 94035, USA.
\\
$^{2}$Institute for Computational Cosmology, Physics Department, Durham University, Durham, DH1 3LE, UK.
}

\end{center}

\vspace{0.8\baselineskip}

\noindent
The origin of Mars's small moons, Phobos and Deimos, remains unknown.
They are typically thought either to be captured asteroids or to have accreted from a debris disk produced by a giant impact.
Here, we present an alternative scenario wherein fragments of a tidally disrupted asteroid are captured and evolve into a collisional proto-satellite disk.
We simulate the initial disruption and the fragments' subsequent orbital evolution.
We find that tens of percent of an unbound asteroid's mass can be captured and survive beyond collisional timescales, across a broad range of periapsis distances, speeds, masses, spins, and orientations in the Sun--Mars frame.
Furthermore, more than one percent of the asteroid's mass could evolve to circularise in the moons' accretion region.
This implies a lower mass requirement for the parent body than that for a giant impact, which could increase the likelihood of this route to forming a proto-satellite disk that, unlike direct capture, could also naturally explain the moons' orbits.
These three formation scenarios each imply different properties of Mars's moons to be tested by upcoming spacecraft missions.
\vspace{0.5\baselineskip}
\vspace{2\baselineskip}

\end{@twocolumnfalse}
]

\nonumfootnote{
  $^*$Corresponding author,
  \href{mailto:jacob.kegerreis@durham.ac.uk}{jacob.kegerreis@durham.ac.uk}.
}
\vspace{-2.2\baselineskip}

\section{Introduction}
\label{sec:intro}

Most hypotheses for the uncertain formation of Mars's small moons
fall into two categories: direct capture and a giant impact.
The moons' spectral properties suggest that
they could be primitive asteroids caught by the planet
\citep{Hunten1979,Landis2009,Pajola+2012,Pajola+2013,Higuchi+Ida2017,Fornasier+2024}.
However, their near-circular and near-equatorial orbits
more naturally align with accretion from a disk around Mars,
typically assumed to have arisen from a large impact
\citep{Craddock2011,Citron+2015,Hyodo+2017a,Canup+Salmon2018}
or alternatively from a primordial disk \citep{Rosenblatt+Charnoz2012},
although the latter may not be compatible with their spectra \citep{Ronnet+2016}.
Distinguishing between these two scenarios is the primary goal of the upcoming
JAXA Martian Moons eXploration (MMX) mission \citep{Kuramoto+2022,Kuramoto2024}.

Here, we present and explore a third possibility:
the capture of tidally disrupted material from an unbound asteroid
passing within Mars's Roche limit,
which then evolves into a collisional disk.
A similar mechanism has been shown to be a plausible route for forming
the rings of giant planets \citep{Dones1991,Hyodo+2017c},
and a much smaller asteroid than that needed for an impact
might be sufficient to produce a successful proto-satellite disk.

The combined mass of Phobos and Deimos is
$1.2 \times 10^{16}$~kg $= 2 \times 10^{-8}~M_\smars$ (Table~\ref{tab:moon_stats}),
a far smaller fraction of Mars's mass
than our Moon is of Earth's mass ($>$1\%).
The moons orbit with low eccentricities and inclinations,
on either side of the corotation radius for
a synchronous orbit with Mars's spin, at $\sim$$6~R_{\smars}$.
Phobos therefore migrates inwards under tidal interactions with Mars,
and Deimos migrates outwards,
so their orbits would have been closer together in the past
\citep{Murray+Dermott1999}.
The current migration rate of Deimos is too slow to be directly observed,
while Phobos is approaching Mars at a rate of $\sim$1.8~cm~yr$^{-1}$.

\subsection{Previous work and plausible disk masses}
\label{sec:intro:hypotheses}

A variety of studies have modelled the accretion of Phobos and Deimos analogues
from debris disks, and/or the incipient formation of such disks with a giant impact.
However, a consensus has not yet been reached for the range of disk masses
that could produce the moons,
nor the details of a corresponding impact event that would create such a disk.
This is due in part to the complex potential sculpting of the thin, outer disk
by resonances with large moons that can temporarily form
in the inner disk before falling back to Mars \citep{Rosenblatt+2016}.
Phobos and Deimos are thought to have accreted from the outer disk
in order to both explain their orbits either side of synchronous,
and reconcile a giant impact origin with their odd spectral properties
by their accretion from small, solid grains
that condense directly from gas in the outer regions,
instead of from slow-cooling magma in the inner regions \citep{Ronnet+2016}.

\begin{table*}[t]
    \centering
    \begin{tabular}{c | ccccccc}
        \hline\hline
         & $M$ (kg) & $M$ ($M_\smars$) & $R$ (km) & $\rho$ (g cm$^{-3}$)
         & $a$ ($R_\smars$) & $e$ & $i$ ($^\circ$) \\
        \hline
        Phobos & $1.07 \times 10^{16}$ & $1.66 \times 10^{-8}$ & 11.1 & 1.86
         & 2.76 & 0.015 & 1.1 \\
        Deimos & $1.48 \times 10^{15}$ & $2.30 \times 10^{-9}$ & 6.3 & 1.47
         & 6.92 & $3 \times 10^{-4}$ & 0.93 \\
        \hline\hline
    \end{tabular}
    \caption{
        Properties of Phobos and Deimos.
        The mass, $M$; mean radius, $R$; density, $\rho$;
        semi-major axis, $a$; eccentricity, $e$; and inclination, $i$, to Mars's equator
        (JPL Horizons System, \href{https://ssd.jpl.nasa.gov/}{ssd.jpl.nasa.gov}; \citealt{Ernst+2023}).
        Mars's mass and radius are $M_\smars = 6.417 \times 10^{23}$~kg
        and $R_\smars = 3,389$~km.
        \label{tab:moon_stats}}
\end{table*}

At the lower end of potentially successful post-impact disk masses,
\citet{Canup+Salmon2018} found that Phobos- and Deimos-like satellites could
accrete from a $\sim$$2 \times 10^{19}$~kg ($3 \times 10^{-5}$~$M_{\smars}$) disk.
This is much smaller than the
$\sim$$5 \times 10^{20}$~kg ($8 \times 10^{-4}$~$M_{\smars}$)
or $\sim$$10^{20}$--$10^{22}$~kg ($10^{-4}$--$10^{-2}$~$M_{\smars}$)
disks favoured by previous works \citep{Rosenblatt+2016,Hesselbrock+Minton2017}.
The latter proposed that a much larger inner moon would form first,
which subsequently migrates inwards and is tidally destroyed to form a new disk
\citep{Hesselbrock+Minton2017}.
This then spawns a smaller satellite that is torqued outwards,
until the disk disperses and the moon falls in again,
eventually leading to today's Phobos after several cycles.
\citet{Hesselbrock+Minton2017} also suggested that this origin of Phobos
moving outwards from the Roche limit may be the only way
to explain Deimos's low eccentricity,
which could otherwise be excited to much higher values
had Phobos originated near synchronous orbit
and migrated inwards to its present position instead.
However, the Phobos-cycle scenario might imply
the continued existence of a debris ring in orbit today,
which is not observed \citep{Madeira+2023}.

A potential issue for a larger disk mass
could be the growth of inner moons that
preclude the formation of Phobos and Deimos analogues,
as found by \citet{Canup+Salmon2018} using extended accretion models
that allow multiple bodies to form at the Roche limit.
However, subsequent work examined the importance
of fragmentation between satellites in evolution models,
instead of the perfect mergers that were previously assumed,
which might allow Deimos to survive in more massive disks after all \citep{Pouplin+2021}.
Furthermore, Deimos's inclination ($1.8^\circ$ with respect to its Laplace plane)
could be explained by resonant interactions with a large inner moon,
aligning with the Phobos-cycle scenario \citep{Cuk+2020b}.
In summary, the mass of a post-impact disk required to form the moons remains uncertain,
but the relevant range is likely around $\sim$$10^{19}$--$10^{21}$~kg.

For the alternative scenario we consider here,
the radial structure of a disk produced via disruptive partial capture
could be less centrally dominated than that from a giant impact.
Debris could thus be distributed out to the corotation radius
more efficiently than by an impact,
which could allow for a lower mass, higher likelihood parent.
In that case, Mars-moon analogues might form from a lower total-mass disk,
which compounds the uncertainty on the total disk mass required.
However, the accretion of moons from different, non-impact disk structures
has not yet been studied in detail.
Therefore, for this exploratory study,
we test scenarios across several orders of magnitude of the parent asteroid's mass
to examine the potential formation of a wide range of disk masses.

As an aside, it was also recently suggested
that the moons could have formed from
the splitting of a larger progenitor moon \citep{Bagheri+2021},
to match the simulated past tidal evolution of the moons.
However, the splitting event has not yet been directly modelled,
and some $N$-body simulations have indicated that moons formed in this way
would recollide within a few thousand years \citep{Hyodo+2022}.

\subsection{Tidal disruption and partial capture}
\label{sec:intro:disruptive_capture}

A satellite on a circular orbit inside a planet's Roche limit
will be torn apart by tidal forces,
unless it is kept together by sufficient cohesive strength.
However, a body that temporarily passes within the Roche limit
on an eccentric or unbound orbit
may have only some, all, or none of its mass disrupted
\citep{Dones1991,Asphaug+Benz1996,Kegerreis+2022},
depending on the properties of the body and its trajectory.

Previous theoretical work showed that enough material from
a tidally disrupted large comet (or rather a centaur)
could be captured to form Saturn's rings \citep{Dones1991}.
\citet{Hyodo+2017c} then used standard-resolution SPH simulations
to compare with the predicted disruption around Saturn or Uranus,
and modelled the initial orbital evolution of resulting fragments
in the planet's oblate potential.
The break-up of comet Shoemaker--Levy 9 around Jupiter
also follows this mechanism and was similarly modelled using SPH
\citep{Asphaug+Benz1996,Movshovitz+2012}.
Compared with these giant-planet systems, the smaller Hill sphere of Mars
($R_{\rm Hill} = 289.9~R_\smars$)
could reduce the effectiveness of disruptive capture here,
given Mars's lower mass and closer proximity to the Sun.
On the other hand, Mars's higher density allows for (non-impacting) asteroid periapses
much deeper within its Roche limit
of $\sim$2.6~$R_{\smars}$ for rocky material,
offering greater capture potential.

To examine the plausibility of such a scenario producing
a proto-satellite disk around Mars,
we combine high-resolution smoothed particle hydrodynamics (SPH) simulations
of the tidal disruption of unbound asteroids
with long-term integrations of the fragments' subsequent orbits,
using the SPH and gravity code \swift \citep{Kegerreis+2019,Schaller+2024}
and the $N$-body code \rebound \citep{Rein+Liu2012},
respectively (\S\ref{sec:methods}).

Collisions between the captured fragments, including disruptive ones, will, on average,
lead debris to circularise and evolve towards damped orbits
in a more highly collisional disk \citep{Walsh+Levison2015,Teodoro+2023}.
We test how (1) the mass of initially captured fragments
and (2) the mass that survives in orbit
change with the asteroid's periapsis distance, speed, mass, composition, and spin,
as well as the orientation in the Sun--Mars frame.

\section{Methods}
\label{sec:methods}

\subsection{SPH simulations}
\label{sec:methods:sph_sims}

Smoothed particle hydrodynamics \citep[SPH;][]{Lucy1977,Gingold+Monaghan1977}
is a Lagrangian method widely used to model diverse systems
in planetary science, astrophysics, and other fields.
Materials are represented by many interpolation points or `particles'
that are evolved under hydrodynamical forces,
with pressures computed via an equation of state (EoS), and gravity.
SPH is particularly well-suited for modelling asymmetric geometries
and/or large dynamic ranges in density and distribution of material,
such as the messy disruptions and collisions considered here.
However, standard resolutions in planetary simulations,
using $10^{5}$--$10^{6}$ SPH particles,
can fail to converge on even large-scale outcomes
\citep{Genda+2015,Hosono+2017,Kegerreis+2019,Kegerreis+2022}.
Here, we run simulations with up to $\sim$$10^{7}$ SPH particles,
to resolve the detailed outcomes of tidal disruption events
and to test for numerical convergence,
using the open-source hydrodynamics and gravity code \swift%
\footnote{
  \swift is publicly available at \url{www.swiftsim.com}.
}
\citep{Kegerreis+2019,Schaller+2024}.

For this study, we use a standard `vanilla' form of SPH \citep{Kegerreis+2019}
plus the \citet{Balsara1995} switch for the artificial viscosity%
\footnote{
  The complete simulation input parameters used are provided
  at \href{https://github.com/jkeger/gihrpy}{github.com/jkeger/gihrpy},
  including the code used to create the initial conditions and analyse the results.
}.
As in previous works, we neglect material strength,
since asteroids of this (and much smaller) mass
are far above the size range for which material strength
would be comparable with gravitational cohesion
\citep{Holsapple+Michel2008,Harris+DAbramo2015}
-- although this assumption will be valuable to test in future work.

The interparticle distance in the simulations for this study is much smaller
than in, for example, a more common lunar-forming impact SPH simulation,
owing to the much smaller asteroid and thus particle masses.
However, the sound speed remains approximately the same,
so the timestep is reduced by about an order of magnitude.
For this reason, we use a primary resolution of $10^{6.5}$ particles,
with a small number of extra high resolution simulations with $10^{7}$,
compared with the $\sim$$10^{8}$ particles that are usually feasible
with planetary \swift.
Given our fiducial asteroid mass of $10^{20}$~kg,
using $10^{6.5}$ particles sets a particle mass of $\sim$$3 \times 10^{13}$~kg,
such that $\sim$400 particles would make up the combined mass of Phobos and Deimos.
Mars is represented as a point-mass potential with mass $6.417 \times 10^{23}$~kg.

\subsection{Asteroid initial conditions}
\label{sec:methods:init_cond}

Given the masses of the parent asteroids that we consider
and the low densities of Phobos and Deimos (Table~\ref{tab:moon_stats}),
we use a rocky, undifferentiated, strengthless, spherically symmetric asteroid
for the primary suite of simulations,
with some comparison tests of others that have a differentiated metallic core.
These materials are modelled with the updated ANEOS forsterite
and Fe$_{85}$Si$_{15}$ core-analogue EoS \citep{Stewart+2020}, respectively,
which encompass thermodynamically consistent models
of multiple phases and improved fits to experimental data.

We generate the asteroids' internal profiles by integrating inwards
while maintaining hydrostatic equilibrium%
\footnote{
  The \woma code \citep{RuizBonilla+2021}
  for producing spherical and spinning planetary profiles and
  SPH initial conditions is publicly available with documentation and examples
  at \href{https://github.com/srbonilla/WoMa}{github.com/srbonilla/WoMa},
  and the python module \texttt{woma} can be installed directly with
  \href{https://pypi.org/project/woma/}{pip}.
}
at an isothermal 500~K.
We then place the roughly equal-mass particles
to precisely match the resulting density profiles
using the stretched equal-area method%
\footnote{
  \seagen \citep{Kegerreis+2019} is publicly available at
  \href{https://github.com/jkeger/seagen}{github.com/jkeger/seagen},
  or as part of \woma.
}
\citep{Kegerreis+2019},
and its modified version for spinning bodies \citep{RuizBonilla+2021}.
Before simulating the Mars encounter,
a brief settling simulation is first run for each body for 1~h in isolation,
to allow any final relaxation of the particles to occur.
The specific entropies of the particles are kept fixed during settling,
enforcing that the particles relax themselves adiabatically.

\subsection{Tidal disruption encounters}
\label{sec:methods:encounter_scenarios}

\begin{figure*}[t]
\centering
\begin{minipage}[t]{0.485\textwidth}
    \centering
    \includegraphics[width=\textwidth, trim={2.32cm 6.8cm 3.55cm 6.5cm}, clip]{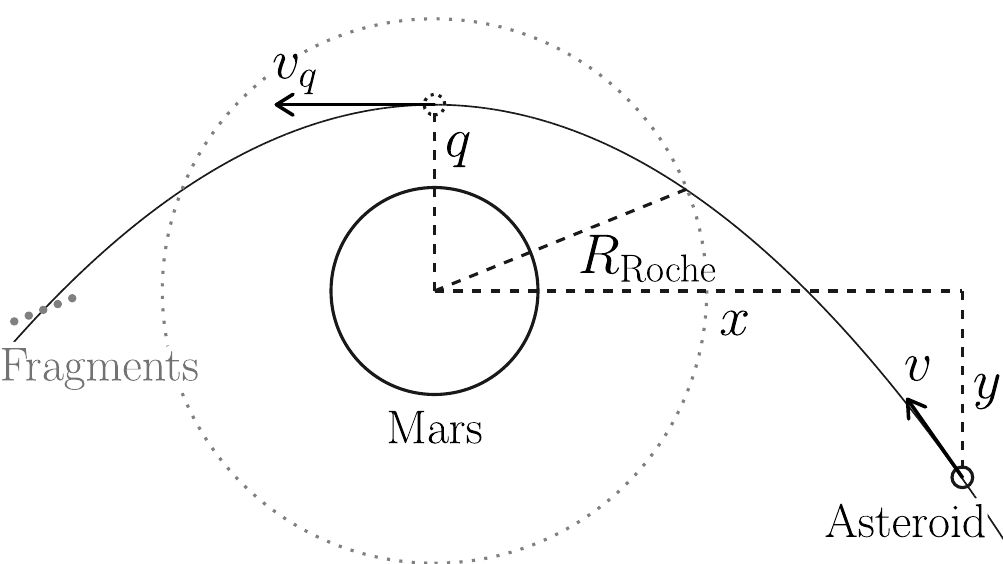}
    \caption{
    The scenario parameters for a tidal disruption simulation.
    The asteroid approaches Mars on an orbit defined by the periapsis distance, $q$,
    and the speed at infinity, $v_\infty$, with mass $M_{\rm A}$.
    The speed at periapsis is $v_q = \sqrt{v_\infty^2 + 2 \mu / q}\,$,
    where $\mu \equiv G (M_\smars + M_{\rm A})$.
    The dotted circle indicates the Roche limit, $R_{\rm Roche}$.
    An asteroid may also be given some spin angular momentum, $L$.
    The initial positions are set such that the time to periapsis is $1.5$~hours,
    as described in Appx.~\S\ref{sec:orbit_equations}.
    \label{fig:disruption_setup}}
\end{minipage}\hfill
\begin{minipage}[t]{0.485\textwidth}
    \centering
    \includegraphics[width=0.85\textwidth, trim={3cm 6.5cm 4.5cm 5cm}, clip]{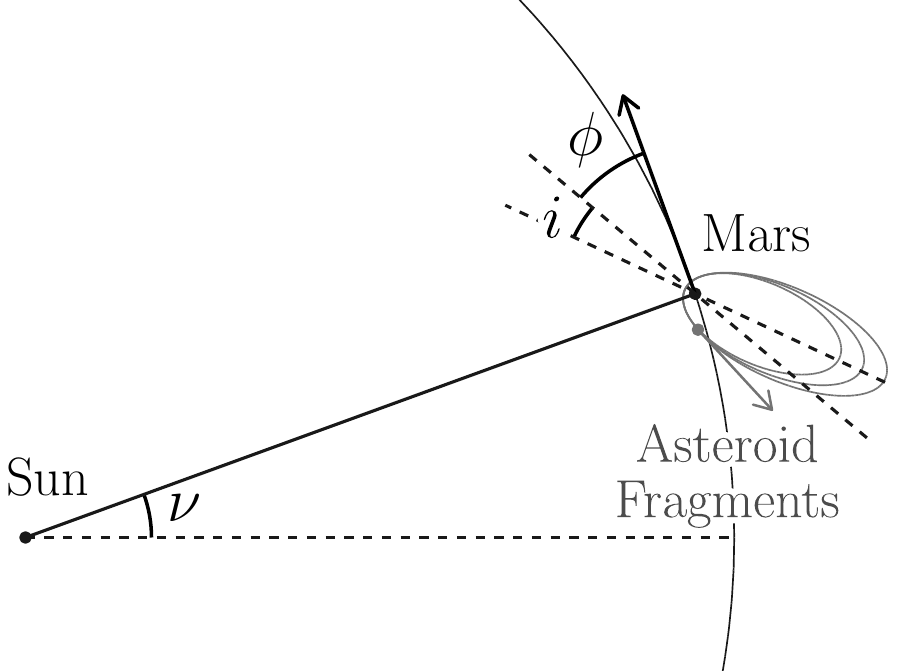}
    \caption{
    The scenario parameters for an orbital evolution simulation.
    Grey lines indicate the initial post-disruption orbits of the debris.
    The periapsis of the initial encounter (Fig.~\ref{fig:disruption_setup})
    is oriented at a longitude $\phi$ from Mars's velocity,
    with an inclination $i$ with respect to Mars's orbit around the Sun.
    Mars is oblate and is given an obliquity, $o$.
    In the case of an eccentric orbit for Mars,
    its initial true anomaly, $\nu$, is also specified.
    \label{fig:orbit_integration_setup}}
\end{minipage}
\vspace{-1em}
\end{figure*}

The general scenario is illustrated in Fig.~\ref{fig:disruption_setup}.
Our primary suite of simulations covers
a range of periapsis distances and speeds at infinity, extending from
$q = 1.1$--$2.4~R_\smars$ and $v_\infty = 0.0$--$1.6$~km~s$^{-1}$,
simulated using $10^{6.5}$ SPH particles,
with a fiducial asteroid mass of $10^{20}$~kg.
The full sets of simulation parameters and results
are listed in Table~\ref{tab:results}.
The simulations start at 1.5~h before periapsis and run for 25~h,
by which time any fragments have separated and the mass function is stable.

We then examine spinning asteroids
with spin angular momenta, $L$, of
1/8, 1/4, 1/2, 3/4, 1 times
the maximum stable spin angular momentum \citep{RuizBonilla+2021},
$L^{\rm max} = 1.2 \times 10^{27}$~kg~m$^2$~s$^{-1}$ for the fiducial mass.
These correspond to spin periods of $17.0$, $8.6$, $4.7$, $3.6$, $3.0$~h,
a not-uncommon range observed for asteroids in the Solar System today
\citep{Szabo+2022,Durech+Hanus2023}.
For reference, the orbital angular momenta of the encounters considered here
are far higher, at around $10^{30}$~kg~m$^2$~s$^{-1}$.
In previous works, a spinning body has only been given an angular momentum
parallel to that of the orbit -- in the $\pm$$z$ direction \citep{Hyodo+2017c}.
This is expected to have the greatest effect on the captured mass,
but we also test a subset of scenarios with $L$ in the $x$ and $y$ directions.

The mass of the asteroid is also varied,
from $10^{18}$--$10^{21}$~kg in steps of $0.5$~dex,
for a subset of the periapsis and speed scenarios.
We also test a differentiated fiducial-mass asteroid
with a 30\%-by-mass metallic core.
Finally, we run simulations with resolutions of $10^{5}$--$10^{7}$ SPH particles
in steps of $0.5$~dex to examine the numerical convergence of the results.
A set of repeat simulations with reoriented initial conditions was also run.
With infinite resolution these should yield identical outcomes,
but in some circumstances can produce non-negligibly different results
even at high resolutions \citep{Kegerreis+2020,Kegerreis+2022},
so they can help to constrain the uncertainty
on the detailed outcomes of individual scenarios.

\subsection{Identifying debris objects}
\label{sec:methods:fof}

The fragments produced by the disruption simulations
are identified using a standard friends-of-friends (FoF) algorithm,
where particles within a certain distance of each other
-- the `linking length', $l_{\rm link}$ -- are grouped together \citep{Davis+1985}.
In these scenarios, where the disrupted bodies have separated
relatively cleanly by the time of analysis,
the selected groups are not sensitive to the choice of linking length.
For our fiducial resolution of $10^{6.5}$ particles,
we set $l_{\rm link} = 4$~km,
which scales for other resolutions and asteroid masses
with the cube root of the particle mass.
We confirm the lack of sensitivity to this choice by repeating the analysis
for $l_{\rm link} = 3$ and $5$~km,
which produce identical masses for the vast majority of fragments
(\S\ref{sec:results:disruption}).

Note that to compute the mass on initially captured orbits,
we estimate the orbits of SPH particles in FoF groups
using their group's centre-of-mass position and velocity,
as detailed in \S\ref{sec:orbit_equations},
and then include any remaining particles not in groups.
If every particle's own orbit were computed instead,
then a single cohesive object could be artificially split
into some bound and some unbound individual-particle orbits.

\subsection{Long-term orbit integrations}
\label{sec:methods:orbit_integrations}

To investigate whether disrupted material could survive in the system
and to characterise the timescales and properties of collisions,
we take the outputs of SPH simulations
and evolve the resulting FoF-identified fragments
using the $N$-body code \rebound \citep{Rein+Liu2012}.
The general scenario is illustrated in Fig.~\ref{fig:orbit_integration_setup}.
The full sets of parameters and results are given in Table~\ref{tab:reb_results}.
We include the Sun in these evolution models,
which was neglected in previous work around more distant, giant planets,
but can significantly affect the fragments' orbits here.

\begin{figure*}[t]
  \centering
  \includegraphics[
    width=\textwidth, trim={64.3mm 25mm 51.3mm 27.3mm}, clip]{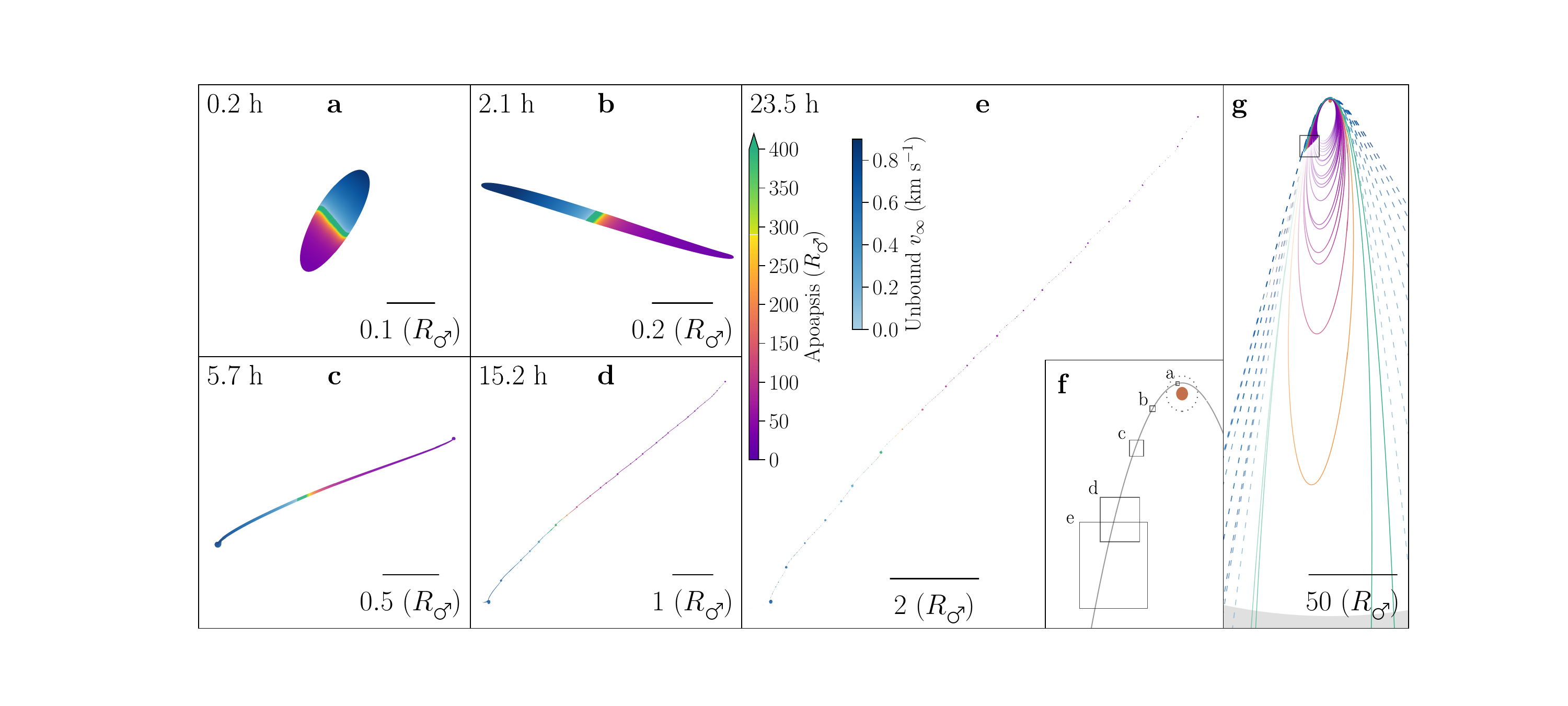}
  \\\vspace{-0.8em}
  \caption{
    Illustrative snapshots from a tidal disruption simulation
    showing the $10^{7}$ SPH particles in the example scenario
    with a parabolic encounter periapsis of $1.6$~$R_\smars$.
    Each SPH particle is coloured by its apoapsis if bound
    and its velocity at infinity if unbound.
    The white line break between yellow and green in the apoapsis colour bar
    indicates the Hill sphere radius.
    The time since periapsis is given in the top-left corner.
    Panel {f} shows a zoomed-out view
    including Mars, the Roche limit, the initial orbit,
    and the outlined locations of {a}--{e}'s axes.
    Panel {g} shows the predicted orbits of the 30 largest fragments,
    using the same colours, the outside of Mars's Hill sphere with grey shading,
    and an outline of panel {e}'s axes.
    Dashed lines indicate unbound orbits.
    \label{fig:snaps_r16_v00}}
  \vspace{-0.7em}
\end{figure*}

In each case, we include all fragments from the input SPH simulation
with a mass of at least $3 \times 10^{15}$~kg
and a bound apoapsis within $1.2$ times the radius of Mars's Hill sphere
($R_{\rm Hill} = 289.9~R_\smars$),
taking their positions and velocities relative to Mars
directly from the SPH outputs.
Most objects that exit the Hill sphere are rapidly lost from the Mars system
but, in some scenarios, fragments with a predicted initial apoapsis
slightly outside $R_{\rm Hill}$ can still survive long-term.
The mass cutoff corresponds to $\sim$100 SPH particles
for our base resolution,
which is approaching but conservatively above
the limit of what should be resolved by the SPH simulations,
where each particle has a typical $\sim$50 neighbours
within its smoothing kernel.
The included fragments also comprise the vast majority
of the total mass in most cases (Table~\ref{tab:reb_results}).
Depending on the scenario, this is typically of order 100 objects.
For example, in the scenario illustrated in Fig.~\ref{fig:snaps_r16_v00},
$99.7$\% of the initially captured mass is in fragments above the threshold mass
to be tracked in the orbit integrations.

The fragments' initial orbits have apoapses that can reach the Hill sphere.
As such, and unlike previous works, we include the Sun,
in addition to Mars's oblate $J_2$ potential.
We evolve the system for 5~kyr (order $\sim$100,000 fragment orbits),
using the IAS15 high-order, adaptive-timestep integrator \citep{Rein+Spiegel2015},
which can accurately handle the highly eccentric orbits here,
and yields fractional energy changes
by the end of a simulation of order $10^{-9}$.
We remove particles that collide with Mars
or escape to beyond twice the Hill radius.

In order to evaluate the evolution timescales of the system,
we track all collisions between particles.
These collisions can be highly disruptive (\S\ref{sec:results:collisions}),
so a simple hard-sphere model would not be appropriate,
and the modelling of all cascading debris is beyond the scope of this paper.
We discard repeat collisions between the same particles in the analysis.
Similarly, we do not model directly the subsequent tidal disruption of the initial fragments.
Many of their orbits continue or return to pass through the Roche limit,
so further disruption would be expected in a real system;
distributing smaller masses on a spread of orbits
around that of the (now bound) incoming fragment,
as discussed further in \S\ref{sec:discussion:limitations}.

Our primary sets of simulations explore a range of
longitude orientations, $\phi$ (Fig.~\ref{fig:orbit_integration_setup}),
from $0$--$345^\circ$, and inclinations (with respect to Mars's solar orbit), $i$,
from $0$--$90^\circ$, both in $15^\circ$ increments,
for a representative example disruption scenario
with $q = 1.6~R_\smars$, $v_\infty = 0$~km~s$^{-1}$.
For a subset of the longitude and inclination scenarios,
we then use the fragments from different SPH simulations as the inputs,
to investigate how the varied debris
from different tidal disruption scenarios evolves long-term.

The precise history of Mars's obliquity and eccentricity is not well known,
but their values could have spanned ranges of about
0$^\circ$--60$^\circ$ and 0--0.12, respectively \citep{Laskar+2004}.
For the scenarios above,
we set a default obliquity of $o_\smars = 30^\circ$,
and an eccentricity of $e_\smars = 0$,
for Mars's semi-major axis of $1.524$~au.
We then re-run a subset of models with $o_\smars = 0^\circ$ and $60^\circ$,
or with $e_\smars = 0.1$ and a range of now-relevant initial true anomalies
$\nu_\smars$ from $0^\circ$ to $315^\circ$, in $45^\circ$ increments,
to test how sensitive to these uncertainties our results could be.

\subsection{Subsequent collision simulations}
\label{sec:methods:collision_sims}

As a brief additional investigation,
we run a small number of \swift SPH simulations of collisions
taken from the \rebound integrations,
as described in \S\ref{sec:results:ang_mom_evol}.
Once we select the collision to model,
we extract the positions and velocities of the two fragments at the time of contact
and compute their positions 200~s earlier as the starting point for the SPH simulation,
over which time the effects of the Sun's gravity are negligible.
For simplicity and to retain a consistent mass resolution,
we use the corresponding FoF groups of particles for the two fragments
from the end of the tidal disruption simulation
rather than creating new objects from scratch with \woma.
As for the primary simulations,
a settling simulation is first run for each body in isolation, here for 10~ks,
to allow any final settling to occur before the collision.
The centre-of-mass positions and velocities of the fragments are then set around Mars
to match the collision in the long-term integration.
We run the simulation for 3~ks to then examine the orbital elements of the debris,
using the same code configuration as for the primary simulations.
For these disruptive collisions of low-mass colliding bodies,
the results at 3~ks are the same as after 2~ks,
with a $<$$0.1$\% change to the resulting mass
with $a_{\rm eq} > 6~R_\smars$ (\S\ref{sec:results:ang_mom_evol}).

\section{Results}

We first present the outcomes of the SPH simulations of tidal disruptions
and examine the trends across different encounter orbits and asteroid properties
in \S\ref{sec:results:disruption}.
Then, we discuss the results of the long-term integrations
of the initial fragments orbiting around Mars
in \S\ref{sec:results:orbit_evol}.
Further discussion of the limitations and implications
then follows in \S\ref{sec:discussion}.
Note that, unless otherwise specified,
the inclination of a particle's orbit is considered
with respect to the plane of Mars's orbit around the Sun.

\begin{figure*}[t]
\centering
\begin{minipage}[t]{0.485\textwidth}
    \centering
    \includegraphics[
    width=\columnwidth, trim={9mm 9mm 9mm 9mm}, clip]{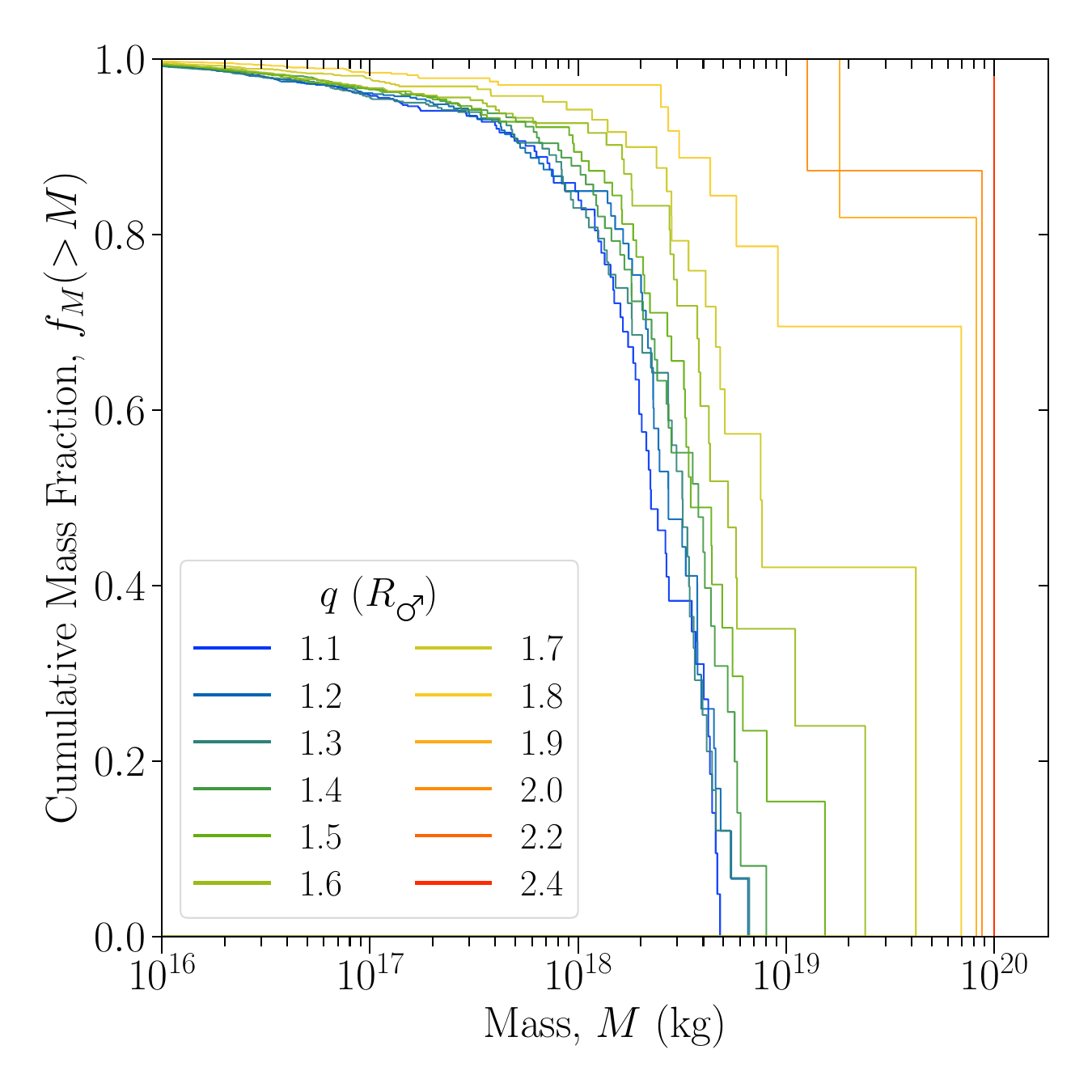}
    \caption{
    The cumulative mass function of disrupted-asteroid fragments,
    coloured by the periapsis as given in the legend,
    23.5~h after periapsis,
    from simulations of fiducial $10^{20}$~kg asteroids on parabolic orbits
    with no initial spin.
    The corresponding distributions of only the fragments on captured orbits
    are shown in Fig.~\ref{fig:mass_funcs_capt}.
  \label{fig:mass_funcs}}
\end{minipage}\hfill
\begin{minipage}[t]{0.485\textwidth}
    \centering
    \includegraphics[
    width=\columnwidth, trim={9mm 9mm 9mm 9mm}, clip]{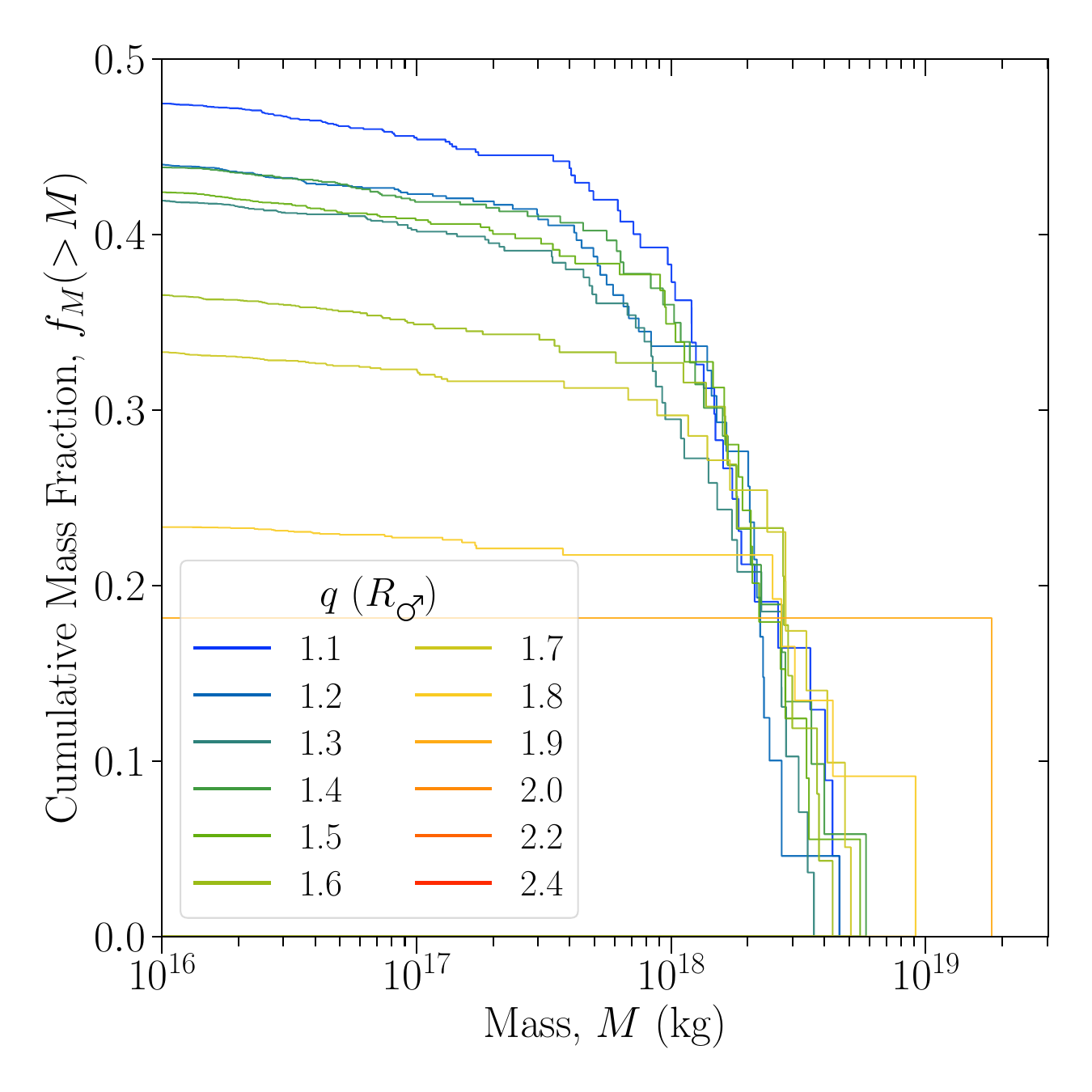}
    \caption{
    The cumulative mass function of disrupted fiducial-asteroid fragments,
    as in Fig.~\ref{fig:mass_funcs},
    but only counting the fragments on bound orbits
    with apoapses inside the Hill sphere.
    Simulations with high periapses $\geq 2.0~R_\smars$ and no initial spin
    produce no fragments on initially captured orbits.
  \label{fig:mass_funcs_capt}}
\end{minipage}
\vspace{-1em}
\end{figure*}

\subsection{Tidal disruption and initial capture}
\label{sec:results:disruption}

How much material can be initially captured onto bound orbits in different scenarios,
and with what distributions of fragments?


The typical behaviour and outcome of an encounter
are illustrated in Fig.~\ref{fig:snaps_r16_v00},
for a representative example of a non-spinning, $10^{20}$~kg asteroid
on a parabolic orbit with a periapsis $q = 1.6~R_\smars$.
The asteroid is distorted by Mars's tidal field,
stretches into a long string,
and splits into a series of self-gravitating fragments on bound and unbound orbits.
The initial orbital elements of these fragments
vary monotonically with their location along the chain,
with the division between bound and unbound orbits
close to the centre of mass for a parabolic asteroid.
In this example, $46.6$\% of the asteroid's mass is disrupted onto bound orbits,
and $36.8$\% is initially captured with apoapses interior to the Hill sphere.

The spatial distribution of fragments is relatively symmetric about the centre of mass
if an encounter is highly disruptive, i.e., with a low periapsis and speed.
As the disruption becomes less extreme,
the fragment distributions become less symmetric,
with a lower number of larger fragments on the side farther from the planet.
As the input periapsis and/or speed is increased further,
only the material closest to the planet is stripped off,
and eventually the asteroid remains entirely intact and unbound.

This strong dependence of the fragment mass distribution
on the disruption scenario is shown in Fig.~\ref{fig:mass_funcs},
for fiducial asteroids without spin.
The most disruptive scenarios produce a range of fragments
with up to $\sim$5--10\% the mass of the asteroid,
while a periapsis above $\sim$$2$~$R_\smars$
allows these non-spinning asteroids to stay mostly intact.
In all cases, fragments larger than $10^{18}$~kg ($>$1\% of the parent mass)
make up over $3/4$ of the total mass.
This is in contrast to the distributions that arise from collisional debris,
where the mass is typically dominated
by the smaller bodies instead \citep{Teodoro+2023}.

For disruptive, low-periapsis scenarios,
the fragments on initially captured orbits follow a similar distribution as the total
-- with about half the magnitude, and usually not including the largest fragment --
as shown by the lower-$q$ lines in Fig.~\ref{fig:mass_funcs_capt}.
In contrast, for higher periapses, the captured distributions
are missing the broad higher-mass range of primarily escaping fragments.
In general, the captured mass functions are insensitive to the periapsis
up to $\sim$$1.6$~$R_\smars$,
with most of the mass comprised of $\sim$$1$--$5 \times 10^{18}$~kg objects.

The majority by mass of the initially captured debris
is typically distributed across of order 100 fragments,
up to $\sim$1000 for some rapid-spin scenarios (Table~\ref{tab:results}).
For the middle fragment-mass range with sizeable counts of well-resolved objects,
around $10^{15.5}$--$10^{17}$~kg,
the cumulative number distribution is well described by a power law
$N \propto M^b$ with exponents around $-0.45$ to $-0.5$,
as shown in Fig.~\ref{fig:count_dist} (Appx.~\ref{sec:extended_results}).
As noted above, these contrast with the steeper ($b \lesssim -1$) distribution
that typically arises for impact debris \citep{Teodoro+2023}.
This reflects the less violent nature of tidal distortion and splitting,
which results in multiple large fragments of similar sizes to each other,
compared with more disruptive massive impacts
that typically produce just one or two large remnants initially,
followed by a more hierarchical distribution of small fragments and reaccreted debris.
Unlike in a highly shocked giant impact,
our simulations confirm that the asteroids are not much thermally altered
by these tidal encounters, and are at most heated by a few tens of Kelvin.

\begin{figure}[t]
  \centering
  \includegraphics[
    width=\columnwidth, trim={7.5mm 6mm 8mm 7mm}, clip]{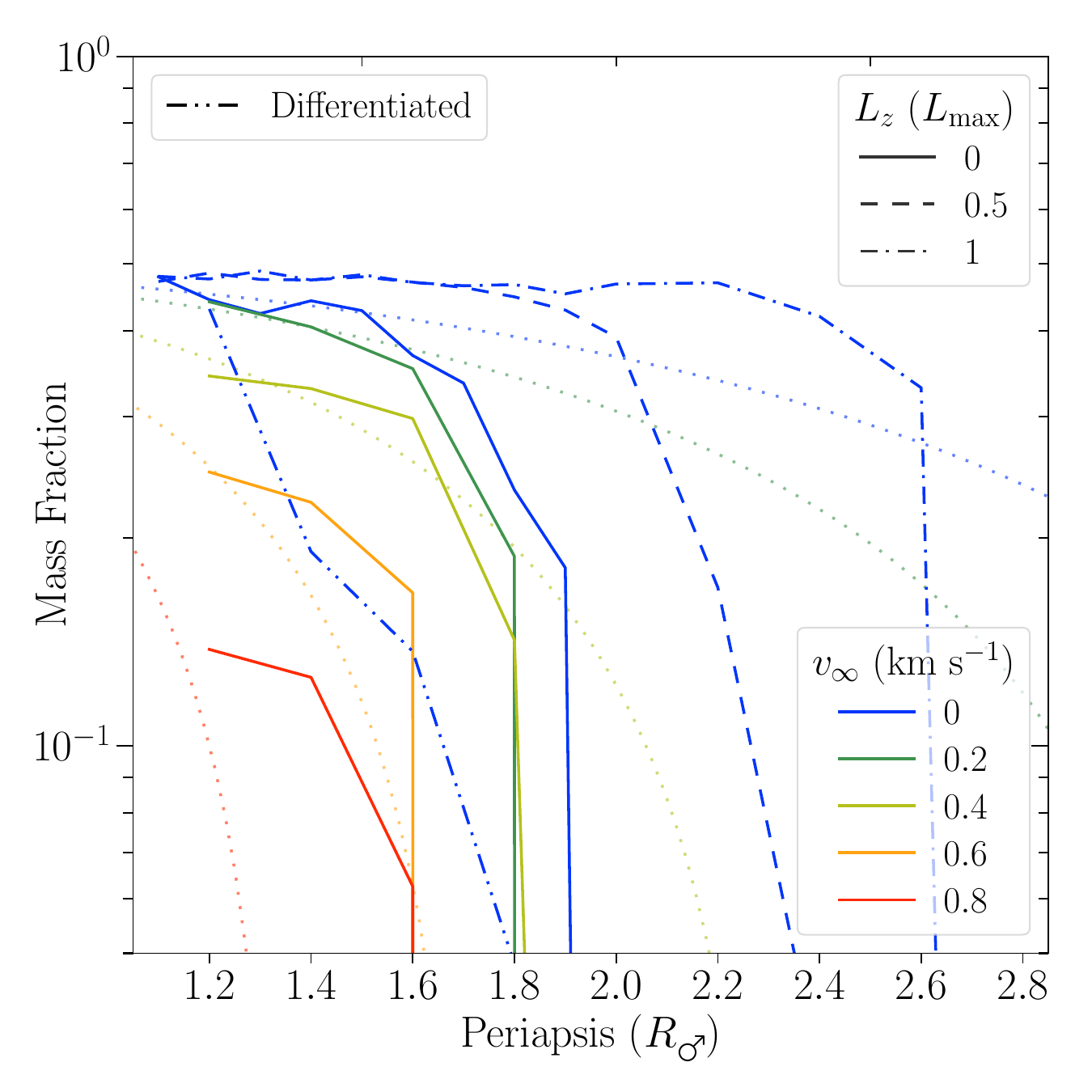}
  \\\vspace{-0.7em}
  \caption{
    The initially captured mass fraction of a $10^{20}$~kg asteroid
    that is tidally stripped onto bound orbits within Mars's Hill sphere,
    as a function of periapsis distance.
    The colours indicate the speeds at infinity,
    the dashed and dash-dot lines indicate
    a spin angular momentum in the $z$ direction
    (parallel to that of the orbit),
    and the dash-dot-dot line shows differentiated asteroids,
    as detailed in the legends.
    Dotted lines show the predictions \citep{Dones1991} for no spin.
  \label{fig:m_capt_r_p_L_z}}
  \vspace{-1em}
\end{figure}

Inspecting the total initially captured mass,
a large fraction (but $<$50\%) of the parent asteroid
can be captured across a broad range of periapses and a modest range of unbound speeds,
as shown in Fig.~\ref{fig:m_capt_r_p_L_z}.
Full results are presented in Table~\ref{tab:results}.
For low periapses and speeds,
the captured masses from non-spinning asteroids
align with previous theoretical predictions \citep{Dones1991,Hyodo+2017c},
which neglected self-gravity and assumed that the orbital energy
is distributed uniformly and symmetrically across the debris.
The gradual breakdown of this symmetry at higher periapses and speeds
breaks these assumptions and, for example, far less of the asteroid
is disrupted and captured at higher periapses than predicted.

\subsubsection{Spinning asteroids}
\label{sec:results:spin_disruption}

The spin, mass, and composition of the asteroid
can dramatically affect both the fragment size distribution
and the initially captured mass.
However, the general behaviour and results are similar
and often result in greater capture than the fiducial cases.

A spin angular momentum (AM) parallel to that of the orbit
(i.e., in the $+z$ direction; see Fig.~\ref{fig:disruption_setup})
means that the part of the asteroid closest to the planet is moving backwards
relative to the centre of mass.
This gives the near side a lower orbital energy
so it is more readily stripped away onto bound orbits,
with a steeper gradient in energy across the body
to the more readily unbound far side,
which allows efficient capture out to much larger periapses.
For $L_z$ close to the maximum stable spin,
material can be captured out to a periapsis of $\sim$$2.6$~$R_\smars$,
as shown in Fig.~\ref{fig:m_capt_r_p_L_z},
compared with the no-spin limit of $\sim$$1.9$~$R_\smars$.
Even with half the maximum spin AM,
the capture region is significantly extended to $\sim$$2.3$~$R_\smars$.
A positive $L_z$ also enables capture at higher speeds,
up to $\gtrsim$$1$~km~s$^{-1}$ even for periapses out to at least 2~$R_\smars$,
compared with $\lesssim$0.8~km~s$^{-1}$ and only lower periapses without spin.
For spin AM in the $-z$ direction,
negligible mass is captured at any periapsis
for $L_{-z} \geq \tfrac{1}{2}~L_{\rm max}$.

When the spin AM is not (anti-)parallel to that of the orbit,
there is no longer as straightforward a benefit or drawback
in terms of the relative velocity of the material to the direction of motion,
as shown in Fig.~\ref{fig:m_capt_L_i} (Appx.~\ref{sec:extended_results}).
For low and medium periapses,
the captured mass is mostly unchanged by rapid rotation (apart from in $-z$),
with just slightly higher typical values than the no-spin case.
For high periapses where no mass is captured without spin,
only spin AM in the $z$ direction is sufficient to enable capture.
However, we tested here only a single mass for the spinning asteroids.
The effects of rotation on disruption and capture will increase further
for larger bodies with the same spin AM in units of $L_{\rm max}$
-- or with the same rotation period,
since the minimum period corresponding to $L_{\rm max}$ is insensitive to the mass.
The larger the spinning object, the faster its outer material is moving
with respect to the orbital speed of the centre of mass,
and the more readily it can be stripped and captured.

The distributions of fragment sizes are also affected by initial rotation,
although the overall trends remain similar to the fiducial cases,
especially for highly disruptive lower periapses.
The mass of the largest fragment decreases with increasing $L_z$,
and we find a greater number of similar-size
$\sim$$1$--$3 \times 10^{18}$~kg objects (Fig.~\ref{fig:mass_funcs_spin}).
For higher $L_z \gtrsim \tfrac{1}{2}~L_{\rm max}$,
the distributions are largely unchanged.
The size distributions are far less sensitive to $L_x$,
with no systematic change to the single largest mass.
Increasing $L_y$ does yield a reduction of the largest-fragment's mass
but, unlike for $L_z$, significant effects only start to appear
for $L_y \gtrsim \tfrac{1}{2}~L_{\rm max}$.

In addition to the fragments' masses, their orbits are also affected.
Whereas in the $z$ or no-spin cases the initial orbits are all in one plane,
for spin AM in the $x$ or $y$ directions
the fragments are given a spread of inclinations up to $\sim$$10^\circ$.

\subsubsection{Asteroid mass and composition}
\label{sec:results:mass_disruption}

The relative mass distributions of fragments from different-size asteroids
are similar to the fiducial case,
with the same $>$$3/4$ of the mass being composed of $>$1\% parent-mass fragments
as in Fig.~\ref{fig:mass_funcs}.
However, the spread of the fragments' initial orbits in eccentricity
and semi-major axis increases for larger parents,
with eccentricities of only $\sim$0.97--0.99 for $10^{18}$~kg
extending to $\sim$0.8--0.99 for $10^{21}$~kg.
The total mass-fraction initially captured thus also increases with the asteroid's size,
from $\sim$30\% for $10^{18}$~kg to $\sim$45\% for $10^{21}$~kg
(Table~\ref{tab:results}),
in general agreement with the predicted radius dependence \citep{Dones1991}
-- although we find that more than predicted can be captured at lower masses.
We will consider the consequences for the fragments' long-term survival
in \S\ref{sec:results:evol_trends},
but leave a more thorough investigation of parent-mass effects for future work,
since for this exploratory study it is sufficient to determine that
tens of percent of an asteroid can be disrupted and captured
across a wide range of masses in the relevant regime
for a martian proto-satellite disk.

The mass initially captured from differentiated asteroids
with metallic cores decreases more rapidly with increasing periapsis
than the fiducial case, as shown in Fig.~\ref{fig:m_capt_r_p_L_z}.
For a low periapsis, the trend with speed at infinity remains similar
to the undifferentiated case and the theoretical predictions \citep{Dones1991}.
Tens of percent of the asteroid's total mass can still be captured,
but only out to periapses of $\sim$$1.6~R_\smars$,
and for that distance or higher the core remains fully intact
and only mantle is removed.
The core can be split into fragments at lower periapses,
but only at $1.2~R_\smars$ do we find significant ($>$$10$\%) capture
of disrupted core material.
The largest fragment mass from a differentiated asteroid,
even at a low periapsis, is $\gtrsim 3 \times 10^{19}$~kg,
compared with the fiducial $\sim$$5 \times 10^{18}$~kg.
In general, the distributions are dominated by a smaller number of larger fragments,
such that at a given periapsis they are more similar to the undifferentiated distribution
for a $\sim$$0.4~R_\smars$ higher periapsis (see Fig.~\ref{fig:mass_funcs}).
Note that we only tested differentiated asteroids
with the fiducial mass and no initial spin
so, like in the undifferentiated cases,
greater disruption could be possible in other scenarios.

\begin{figure}[t]
  \centering
  \includegraphics[
  width=\columnwidth, trim={7mm 7mm 7mm 7mm}, clip]{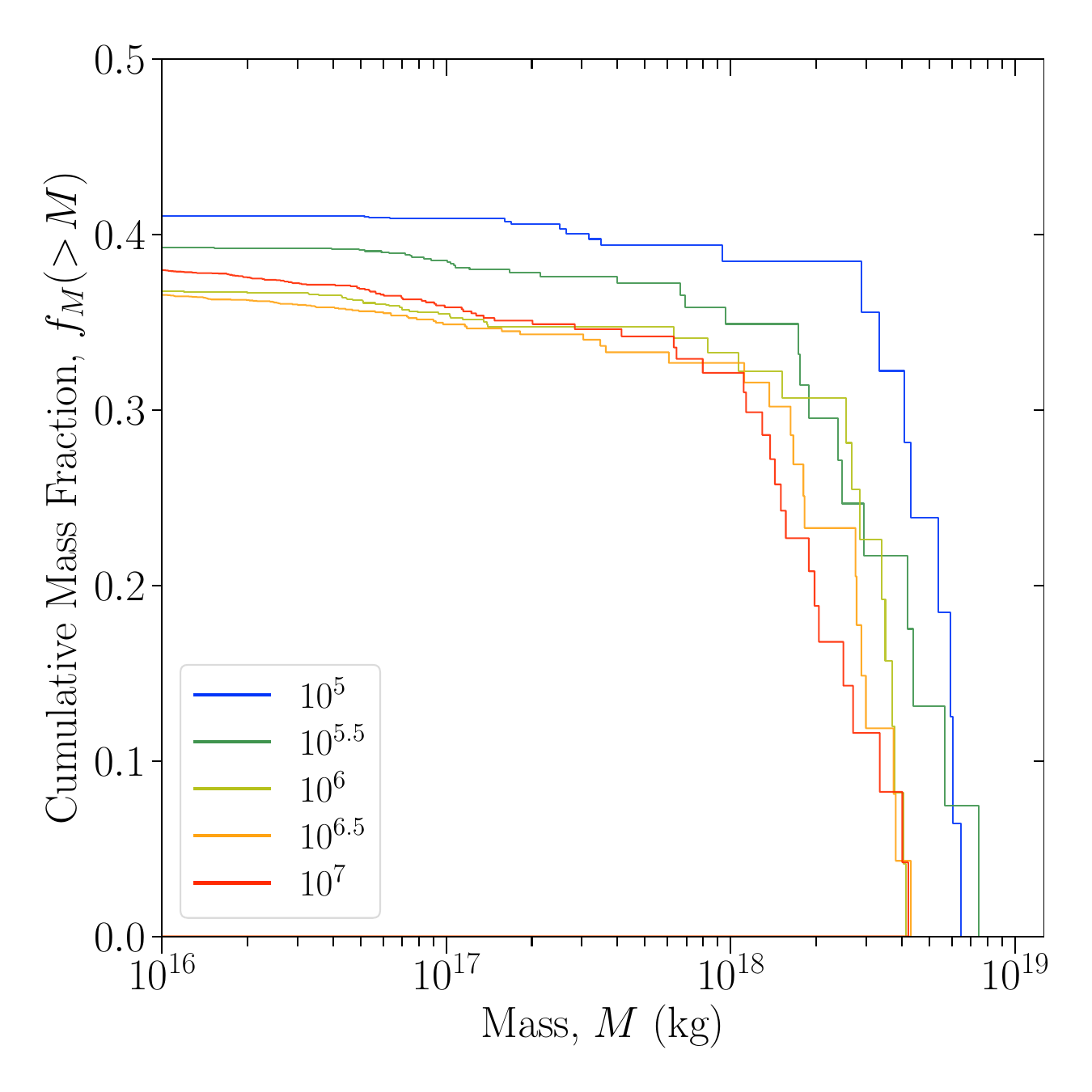}
  \\\vspace{-1em}
  \caption{
    The cumulative mass function of initially captured fragments
    for different numerical resolutions,
    as in Fig.~\ref{fig:mass_funcs_capt},
    for the reference $q = 1.6~R_\smars$ disruption scenario.
  \label{fig:mass_funcs_res}}
  \vspace{-1em}
\end{figure}

\subsubsection{Numerical convergence}
\label{sec:results:numerical_disruption}

Our primary resolution of $10^{6.5}$ particles can plausibly resolve
fragments reliably down to a mass of around $3 \times 10^{15}$~kg ($\sim$100 SPH particles),
and below this mass the numbers of fragments indeed begin to tail off
from the power-law trends (Fig.~\ref{fig:count_dist}).
In practice, the formation and evolution of SPH objects
much larger than this limit may not have numerically converged,
and supposedly-resolved features may not be reliable
\citep{Hosono+2017,Kegerreis+2019,Kegerreis+2022}.
However, here it appears that the fragment mass functions have approximately converged,
as shown in Fig.~\ref{fig:mass_funcs_res},
with similar results for $10^6$, $10^{6.5}$, and $10^{7}$ particles,
although some minor differences remain, as expected.
In most scenarios, lower resolutions produce
a sparser distribution of larger fragments,
and can yield masses for the largest remnants
around a factor of two greater than at high resolution.
The total mass of initially captured material is less sensitive to the resolution,
and appears to have converged by $\sim$$10^6$ particles to within a few percent
at low periapses and $<$1\% at higher periapses (Table~\ref{tab:results}).

The reoriented repeat simulations (\S\ref{sec:methods:encounter_scenarios})
also yield consistent mass distributions,
with a 7\% relative standard deviation for the mass of the largest fragment
-- far below the variation between asteroid scenarios.
The relative standard deviation for the total initially captured mass is 4\%.
Combined with the general agreement between resolutions of $10^6$ or more particles,
this suggests a more than sufficient reliability for our SPH results.

\begin{figure*}[t]
  \centering
  \includegraphics[
  width=\textwidth, trim={25mm 7mm 29mm 25mm}, clip]{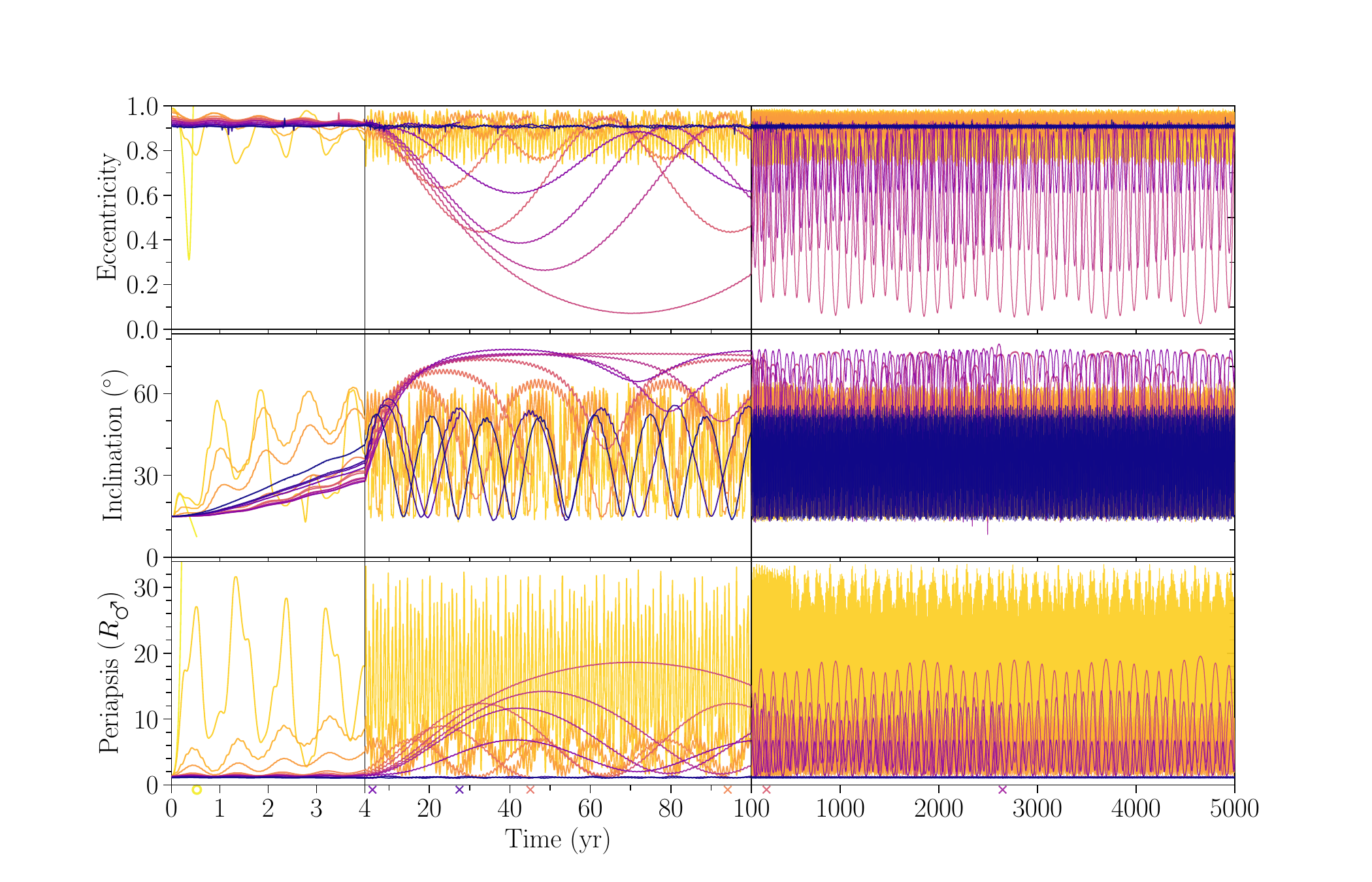}
  \\\vspace{-0.3em}
  \caption{
    The evolution of 15 representative fragments orbiting Mars
    in the example scenario with $\phi = 90^\circ$ and $i = 15^\circ$,
    coloured qualitatively in initial-apoapsis order
    from high in yellow to low in dark purple
    to demonstrate the range of typical behaviour.
    Note the piecewise-linear time axis changes scale
    to aid inspection of the range of evolution timescales.
    The $\circ$ and $\times$ marker(s) below the axes
    indicate when the same-colour fragment is removed
    by escaping beyond $2 R_{\rm Hill}$ or colliding with Mars, respectively.
  \label{fig:evol_e_i_r_p}}
  \vspace{-0.5em}
\end{figure*}

\subsection{Orbital evolution}
\label{sec:results:orbit_evol}

How much of the initially captured material can survive long-term in different scenarios,
and what are the timescales for evolution and collisions?


The fragments' orbits are driven by
solar and oblate-Mars perturbations to evolve
over a range of short and long timescales,
as shown in Fig.~\ref{fig:evol_e_i_r_p}
for a representative example with an orientation longitude of $\phi = 90^\circ$,
where $\phi$ is the angle of the input encounter periapsis away from Mars's velocity
(see Fig.~\ref{fig:orbit_integration_setup}),
and inclination $i = 15^\circ$ to Mars's orbit around the Sun,
using the Fig.~\ref{fig:snaps_r16_v00} example scenario fragments.
Their primary evolution includes:
harmonic oscillation of angular momentum with a period of half Mars's year;
von-Zeipel--Lidov--Kozai (ZLK)-like oscillations
\citep{vonZeipel1910,Lidov1962,Kozai1962,Ito+Ohtsuka2019}
in eccentricity and inclination over tens to hundreds of years;
precession that randomises orbit orientations on a similar spread of timescales;
and some even longer, kiloyear evolution timescales,
though most fragments that survive for that time
have already suffered a significant collision%
\footnote{
  As described in \S\ref{sec:methods},
  in these integrations, only the information from a collision is recorded
  and the two particles are allowed to continue in orbit,
  with subsequent re-collisions discounted.
},
as detailed in \S\ref{sec:results:evol_timescales}.

Most of the collisions between these eccentric and inclined fragments
occur at relative speeds of $\sim$$0.3$--$3$~km~s$^{-1}$
(Fig.~\ref{fig:v_rel_c_hist}),
as discussed further in \S\ref{sec:results:collisions}.
As a simple diagnostic, we define a `significant' collision
as one having enough energy to disrupt the larger object
\citep{Leinhardt+Stewart2012}.
Given these speeds and accounting for the bodies' masses,
this criterion applies to $\sim$70--90\% of collisions across scenarios.
Although the colliding fragments can thus readily break each other apart
and distribute orbital energy,
their collision speeds rarely reach the level required
for shock melting or vaporisation of silicates \citep{Ahrens+OKeefe1972},
depending on their porosity \citep{Zeldovich+Raizer1967}.

In the example scenario of Fig.~\ref{fig:evol_e_i_r_p},
68\% of the initially captured $3.7 \times 10^{19}$~kg
survives long enough to undergo significant collisions,
and regardless of collisions
20\% remains on apparently safe orbits after 5000~yr.
In this and most scenarios,
only a small fraction of particles escapes the system unbound,
usually within the first year,
while the majority of removed particles are lost by impacting Mars
throughout the first $\sim$1000~yr (Fig.~\ref{fig:t_surv_hist}).

The initial orbits of the debris are highly eccentric,
with $e \gtrsim 0.9$ and a smooth range of apoapses
from beyond the Hill sphere to as low as $20~R_\smars$,
as illustrated by Fig.~\ref{fig:snaps_r16_v00}g.
For asteroids with a rapid initial rotation,
the minimum eccentricity can be lowered to $\sim$0.8,
with initial apoapses down to $13~R_\smars$.
The most tightly bound fragments have low periapses
that increase their risk of colliding with Mars,
while fragments on higher energy orbits have high apoapses
that make them more susceptible to being lost beyond Mars's Hill sphere.
Nevertheless, we find that many fragments can survive long-term
while being perturbed onto a wider range of orbital elements.

\begin{figure*}[t]
  \centering
  \includegraphics[
  width=0.495\textwidth, trim={39mm 7mm 45mm 7mm}, clip]{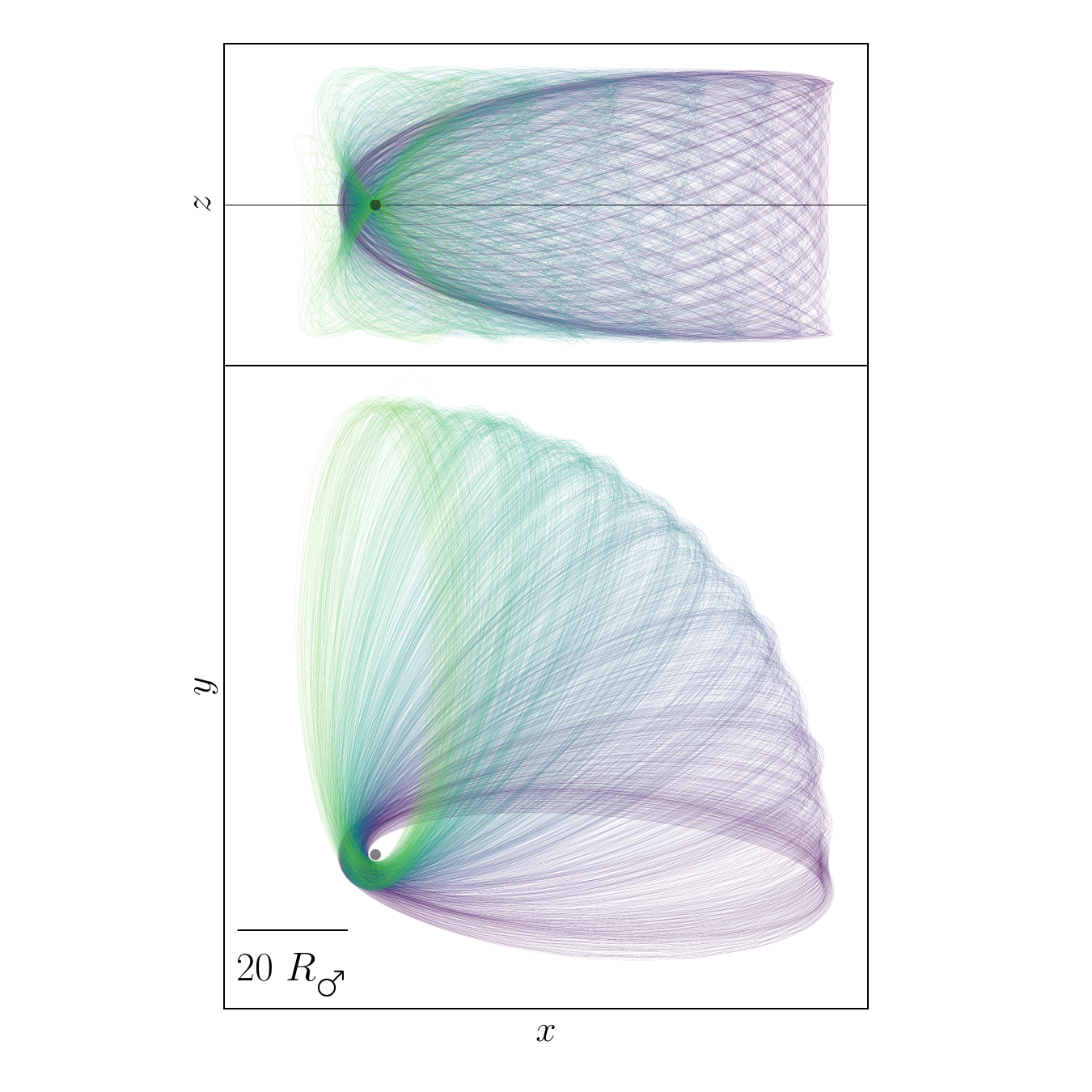}
  \includegraphics[
  width=0.495\textwidth, trim={39mm 7mm 45mm 7mm}, clip]{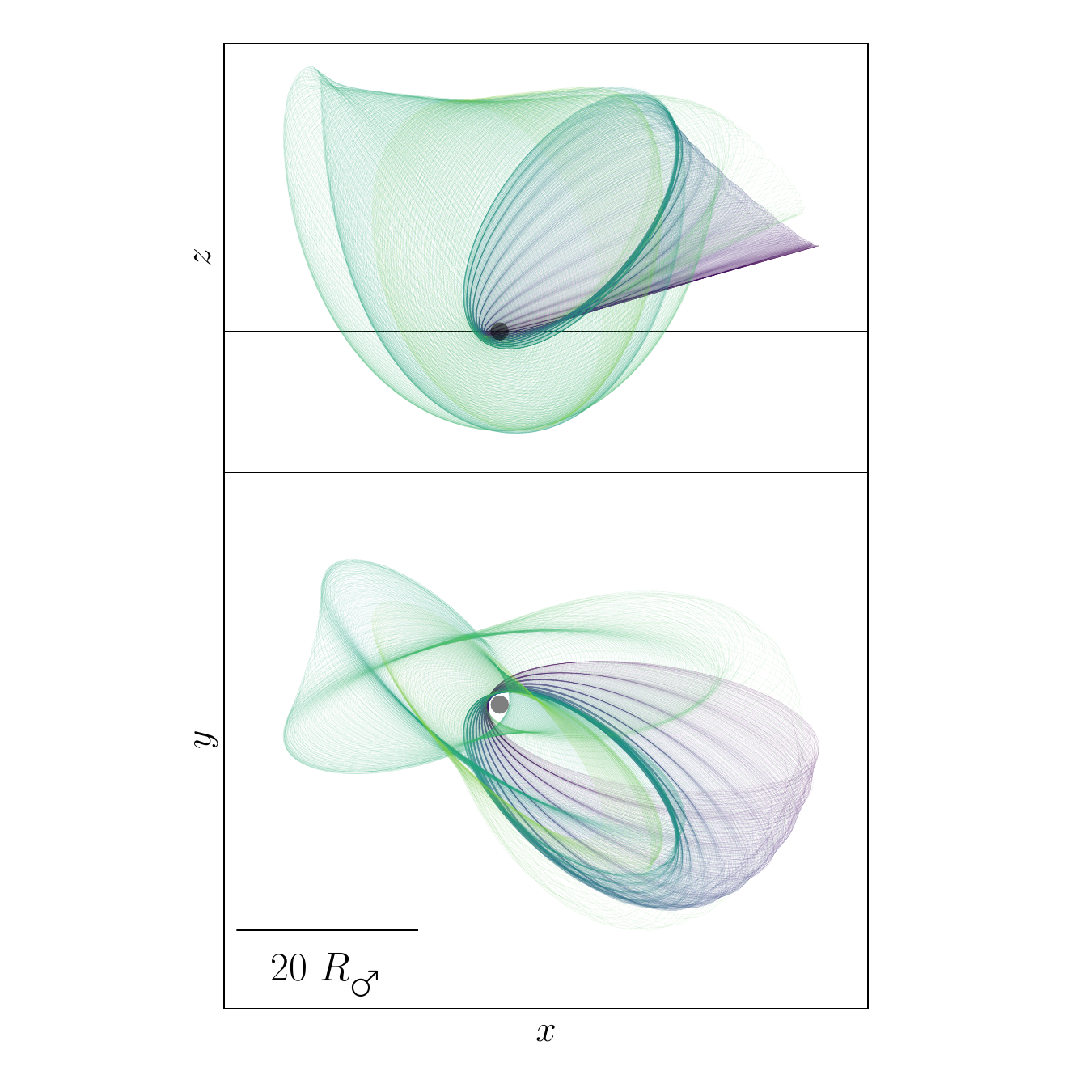}
  \\\vspace{-0.5em}
  \caption{
    The paths traced out by two representative example surviving fragments
    over the first 200~yr ($\sim$4,000 and 13,000 orbits respectively)
    of the orbital integration from the example
    $\phi = 90^\circ$, $i = 15^\circ$ scenario,
    projected onto the $x$--$y$ and $x$--$z$ planes
    in the bottom and top panels, respectively.
    The colour indicates the time evolution from purple to green.
    The grey circle shows Mars, to scale;
    note that the left panel view is more zoomed-out than the right,
    as indicated by the scale bars.
    The thin black line in each $x$--$z$ projection
    shows the $z=0$ plane of Mars's orbit around the Sun.
    \label{fig:orbit_trace}}
  \vspace{-0.5em}
\end{figure*}

\subsubsection{Evolution timescales}
\label{sec:results:evol_timescales}

The fragments' orbits evolve on a variety of year to kiloyear timescales.
The early orbits of two representative fragments
under the solar and oblate-Mars perturbations
are traced out in full in Fig.~\ref{fig:orbit_trace}.
As also illustrated by Fig.~\ref{fig:evol_e_i_r_p},
the periapsis and size of the orbit oscillate rapidly,
with a period of half of Mars's year,
while the inclination and eccentricity evolve and the longitude precesses
on timescales of tens to hundreds of years.
The first, higher-orbit fragment
($a \approx 44~R_\smars$, period $\approx 20$~days)
undergoes relatively consistent, prograde precession,
and goes on to survive the full 5,000 years,
while the second fragment, closer to Mars
($a \approx 19~R_\smars$, period $\approx 6$~days),
initially precesses retrograde then experiences more dramatic orbital changes,
and survives a further 2,400 years beyond the first 200
that are shown in Fig.~\ref{fig:orbit_trace},
ignoring the collisions with other fragments that it would suffer during that time,
before eventually colliding with Mars.

Across the different scenarios, the primary features
of the fragments' orbital evolution stay broadly consistent,
as in the Fig.~\ref{fig:evol_e_i_r_p} and \ref{fig:orbit_trace} examples.
In general, most fragments have their average eccentricities reduced
and inclinations increased from their initial values.
On short timescales, solar perturbations drive a harmonic oscillation
in each orbiting fragment's angular momentum,
with a period of half of Mars's year and an amplitude of
\begin{equation}
  \Delta h = \dfrac{15}{8} n_\smars a^2 e^2 \;,
\end{equation}
where $n_\smars = \sqrt{G (M_{\odot} + M_\smars) / a_\smars^3}$
is the planet's mean motion around the Sun \citep{Benner+McKinnon1995}.
The predicted evolving periapsis is then $q = h^2 / (\mu (1 + e))$,
where $\mu \equiv G (M_\smars + m)$ and $m$ is the orbiting fragment's mass,
which here is negligible.
This agrees well with our results,
although this expression fits progressively less closely for
satellites with higher inclinations and semi-major axes.
For unfavourable orientations $\phi$
as discussed in \S\ref{sec:results:evol_trends},
these oscillations can drive many fragments to early collisions with Mars.

On longer timescales, ZLK-like oscillations
\citep{vonZeipel1910,Lidov1962,Kozai1962,Ito+Ohtsuka2019}
affect the fragments' eccentricities and inclinations,
with an estimated timescale of
\begin{equation}
  \tau_{\rm ZLK} = \dfrac{P_\smars^2}{P}
    \,\left(1 - e_\smars^2\right)^{3/2} \;, \label{eqn:LidovKozai}
\end{equation}
where $P$ is the fragment's orbital period.
This generally aligns with our numerical results,
although we find that the timescales can differ by factors of $\sim$$1$--$10$
from Eqn.~\ref{eqn:LidovKozai}
for the very high-$e$ fragments and the high-mass perturbing `secondary' of the Sun here.
As illustrated in Fig.~\ref{fig:evol_e_i_r_p},
this corresponds to typical primary oscillation periods of around ten to a hundred years.

The solar and oblate-Mars perturbations also cause the fragments' orbits
to precess at diverse rates,
and in opposite directions depending primarily on the distance from Mars,
as illustrated by the Fig.~\ref{fig:orbit_trace} examples.
This yields a rapidly randomised spread of orbit orientations,
which advances the collisional evolution of the system.
We find that the typical precession timescale for fragments
ranges from $\sim$50--500~yr,
depending primarily on the particle's initial orbit
and the inclination of the scenario.
The analytical predictions from Mars's $J_2$ moment
for the precession rates of the argument of periapsis
and the longitude of the ascending node are \citep{Kaula1966}, respectively,
\begin{align}
  \dot{\omega} &= 3 \sqrt{\dfrac{\mu}{a^3}} \left(\dfrac{R_\smars}{a (1 - e^2)}\right)^2
    \left(1 - \dfrac{5}{4} \sin(i)^2 \right) J_2
  \\
  \dot{\Omega} &= -\dfrac{3}{2} \sqrt{\dfrac{\mu}{a^3}} \left(\dfrac{R_\smars}{a (1 - e^2)}\right)^2
      \cos(i) \,J_2  \;.
\end{align}
This yields timescales of $2 \pi / \dot{\omega} \approx 100$~yr
and $2 \pi / \dot{\Omega} \approx 200$~yr for a typical orbiting fragment,
in line with our simulation results.

Depending on the scenario, we also see some evolution
on even longer, kiloyear timescales.
However, such effects rarely result in the loss of fragments
that have survived to that point,
and furthermore are occurring on timescales longer than it typically takes
for a still-orbiting fragment to suffer a significant collision.

\subsubsection{Collisions}
\label{sec:results:collisions}

As the surviving fragments evolve and their orbits precess,
they can collide with each other.
As a simple diagnostic for the potential evolution of the system
towards a more uniform proto-satellite disk,
we define a `significant' collision as one with more than
the disruption energy ($Q'^\star_{\rm RD}$) \citep{Leinhardt+Stewart2012}
at which the largest remnant would have less than half its original mass.
The masses and $\sim$$0.3$--$3$~km~s$^{-1}$ relative speeds
of these fragments lead to $\sim$$3/4$ or more of the collisions
being labelled as significant in most scenarios,
with typical collision energies from $\sim$$0.1$--$1000~Q'^\star_{\rm RD}$,
assuming a moderate $45^\circ$ impact angle.
As noted above, these collisions are usually not fast enough to cause
widespread melting or vaporisation \citep{Ahrens+OKeefe1972},
although the likely porous nature of the debris
could allow for increased shock heating \citep{Zeldovich+Raizer1967}.

Even the smallest fragments included in our integrations
can occasionally hit the largest fragments with enough energy to disrupt them.
We did not include fragments of masses smaller than $3 \times 10^{15}$~kg
because of the reduced numerical reliability in the tidal disruption simulations
below a conservative $\sim$100 SPH particles,
in addition to their minimal contribution to the total mass.
If smaller fragments were accounted for, then the collision frequency would increase,
and one can make a simple estimate of how many collisions might be missing
and what fraction of them could still be significant.
For the larger $>$$10^{18}$~kg fragments,
we find of order ten collisions by just the smallest $10^{15.5}$--$10^{16.5}$~kg fragments
in a typical fiducial scenario, $\sim$60\% of which have significant energy.
The number of fragments increases to lower masses approximately as
$\propto$$M^{-1/2}$ (Fig.~\ref{fig:count_dist}).
So, for an order of magnitude step down in mass to $10^{14.5}$--$10^{15.5}$~kg fragments,
we might predict around $10^{1/2} \approx 3$ times as many collisions
($\sim$30 collisions with the largest fragments).
The smaller masses translate to lower-energy and less-destructive collisions
so, assuming the same range of relative speeds for these collisions,
these values could yield a proportion of $\sim$10\% with significant energy.
At these speeds, even tinier fragments would not disrupt
the large fragments that we considered for the above estimate,
but they could still disrupt smaller objects
and thus influence the system's collisional evolution.
Therefore, the inclusion of even smaller fragments than we model
could affect the detailed evolution of the debris,
but is unlikely to make a great difference.
Further limitations and their potential ramifications are discussed in
\S\ref{sec:discussion:limitations}.

\begin{figure}[t]
  \centering
  \includegraphics[
    width=\columnwidth, trim={9mm 9mm 9mm 9mm}, clip]{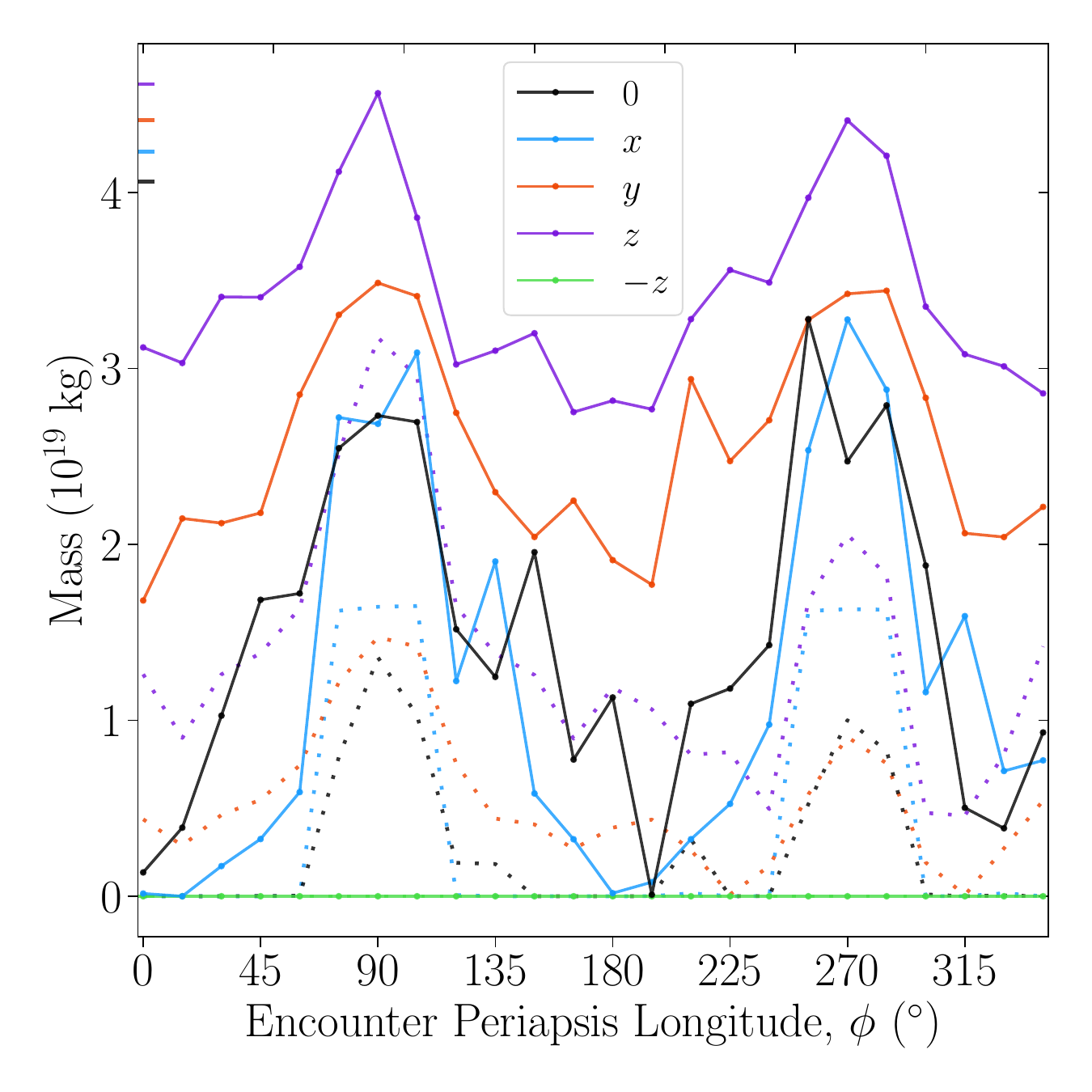}
  \\\vspace{-0.3em}
  \caption{
    The mass of fragments that survive long enough
    to undergo significant collisions (solid lines; \S\ref{sec:results:collisions}),
    as a function of the encounter longitude
    of the input periapsis away from Mars's velocity.
    The dotted lines show the mass that survives in orbit for 5~kyr
    neglecting collisions.
    The input $10^{20}$~kg asteroid disruption
    is the same as in Fig.~\ref{fig:snaps_r16_v00}
    for each integration shown by the black line,
    with the reference $q = 1.6~R_\smars$, $v_\infty = 0$,
    $i = 15^\circ$, $o_\smars = 30^\circ$ scenario.
    The coloured lines show the same for input disruptions of
    asteroids with spin angular momenta of $\tfrac{1}{2}~L_{\rm max}$
    in the direction shown in the legend.
    The same-colour ticks on the left axis show the initially captured mass.
  \label{fig:m_col_phi}}
  \vspace{-1em}
\end{figure}

\subsubsection{Survival trends}
\label{sec:results:evol_trends}

The fraction of orbiting fragments that survive long-term
varies with the orientation of the tidal encounter in the Mars--Sun system
and the details of the input disruption event.
For approximately half of the encounter longitudes, $\phi$,
the short-timescale AM oscillations (\S\ref{sec:results:evol_timescales})
have a phase such that the fragments' periapses begin by being reduced,
which causes many of them to collide with Mars within the first year.
For $\phi$ closer to $90^\circ$ and $270^\circ$,
the oscillation phase is such that their periapses are initially raised,
allowing more fragments to survive on safer orbits.
This is illustrated in Fig.~\ref{fig:m_col_phi},
which shows as a function of $\phi$
both the mass of fragments that are still in orbit after 5~kyr
and -- more importantly for the potential production of a proto-satellite disk --
those that survive long enough to undergo significant collisions.

With less and decreasing significance,
the inclination of the encounter, Mars's obliquity,
and Mars's eccentricity and starting true anomaly
also influence the fragments' survival,
as discussed later in this section.
Regardless, in all cases we find ranges of $\phi$ for which
tens of percent of the asteroid's mass can survive and collide (Table~\ref{tab:reb_results}).

\begin{figure*}[t]
  \centering
  \includegraphics[
    width=\textwidth, trim={19mm 9mm 9mm 9mm}, clip]{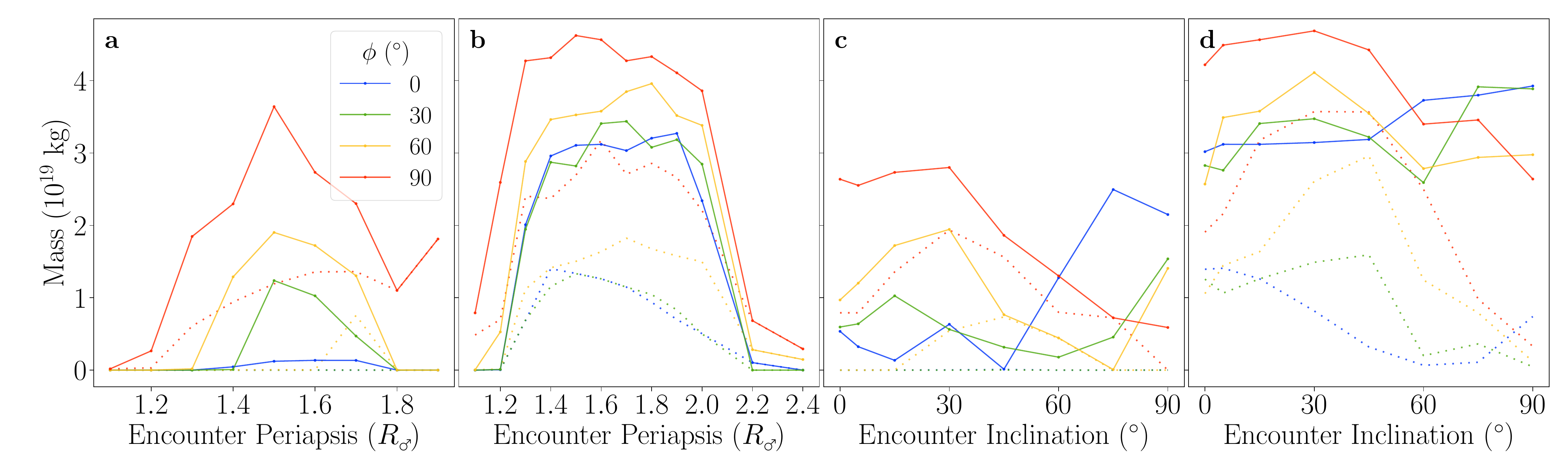}
  \caption{
    The mass of fragments that survive long enough in orbit
    to undergo significant collisions,
    as a function of the disrupted $10^{20}$~kg asteroid's
    periapsis in panels {a} and {b},
    and its inclination with respect to Mars's orbit around the Sun
    in panels {c} and {d},
    for different longitude orientations as given in the legend.
    Dotted lines show the mass still in orbit after 5~kyr,
    as in Fig.~\ref{fig:m_col_phi}.
    The input asteroids for panels {a} and {c} have no spin,
    and for panels {b} and {d} have $L_{z} = \tfrac{1}{2}~L_{\rm max}$.
    Note that the initially captured mass also changes with the input periapsis
    (see Fig.~\ref{fig:m_capt_r_p_L_z}).
  \label{fig:m_col_r_p_in_i}}
  \vspace{-0.5em}
\end{figure*}

Shifting focus from the orientation to the input disruption,
the parent asteroid's periapsis, speed, spin, composition, and mass
affect both the initially captured mass of material
and the distributions of fragment sizes and orbits.
Nevertheless, in most cases the orbital evolution of these different fragment populations
remains broadly consistent with the Fig.~\ref{fig:evol_e_i_r_p} example,
while the total fraction of the initial mass that survives can differ.

Perhaps most straightforwardly, a higher initial periapsis
makes it easier for fragments to avoid colliding with Mars,
so a greater fraction can survive long-term in orbit.
Counteracting this, a higher asteroid periapsis leads
to less tidal disruption and fewer initially captured fragments.
This trade-off means that, for no initial spin,
the greatest mass of surviving collisional material
arises for a middling periapsis around $1.4$--$1.6~R_\smars$,
as shown in Fig.~\ref{fig:m_col_r_p_in_i}a.
As the speed at infinity increases,
the potential disk-forming mass decreases.
Nevertheless, even without spin,
tens of percent of a parent can survive for $v_\infty$ up to $\sim$$0.6$~km~s$^{-1}$,
including for all asteroid masses across the tested range.

For all tested spin AM magnitudes and directions aside from $-z$,
the total mass of initially captured material remains similar
(Fig.~\ref{fig:m_capt_L_i}).
However, the mass that subsequently survives evolution towards a collisional disk
can be far more affected by the spin-altered fragments' size distributions and orbits.
A $\tfrac{1}{2}~L_{\rm max}$ spin AM in $y$ or $z$
significantly increases the long-term surviving and colliding mass over the no-spin case,
to $>$15\% and $>$25\% of the asteroid for all $\phi$,
as shown in Fig.~\ref{fig:m_col_phi}.
A spin AM of this magnitude in $x$ has a lesser effect,
and in $-z$ no mass is initially disrupted regardless.
Similarly, a $z$-direction spin AM allows far more mass to survive
across a wider range of asteroid periapses.
The sensitivity to the input periapsis is less extreme than the no-spin case,
and significant material can survive for $q$ out to $\sim$$2~R_\smars$
even for the most unfavourable $\phi$,
as shown in Fig.~\ref{fig:m_col_r_p_in_i}b.
The parameter space for high survival rates is even broader for faster spins,
but these far-from-extreme $\tfrac{1}{2}~L_{\rm max}$ rotation periods of $\sim$5~hours
\citep{Szabo+2022,Durech+Hanus2023} already have a strong effect.

As $L_z$ increases, following the greater initial-disruption effects
for the parent asteroid, more of the resulting fragments can also survive long-term
across different encounter periapses and longitude orientations,
as shown in Fig.~\ref{fig:m_col_L_z_r_p} (Appx.~\ref{sec:extended_results}).
The collisional mass is significantly increased by
just $L_z \approx \tfrac{1}{4}$ or $\tfrac{1}{2}~L_{\rm max}$,
with additional but diminishing gain from more rapid rotation.
This includes encounter longitudes
that precluded any long-term survival with no spin, but now become viable.

\begin{figure}[t]
  \centering
  \includegraphics[
    width=\columnwidth, trim={9mm 9mm 9mm 9mm}, clip]{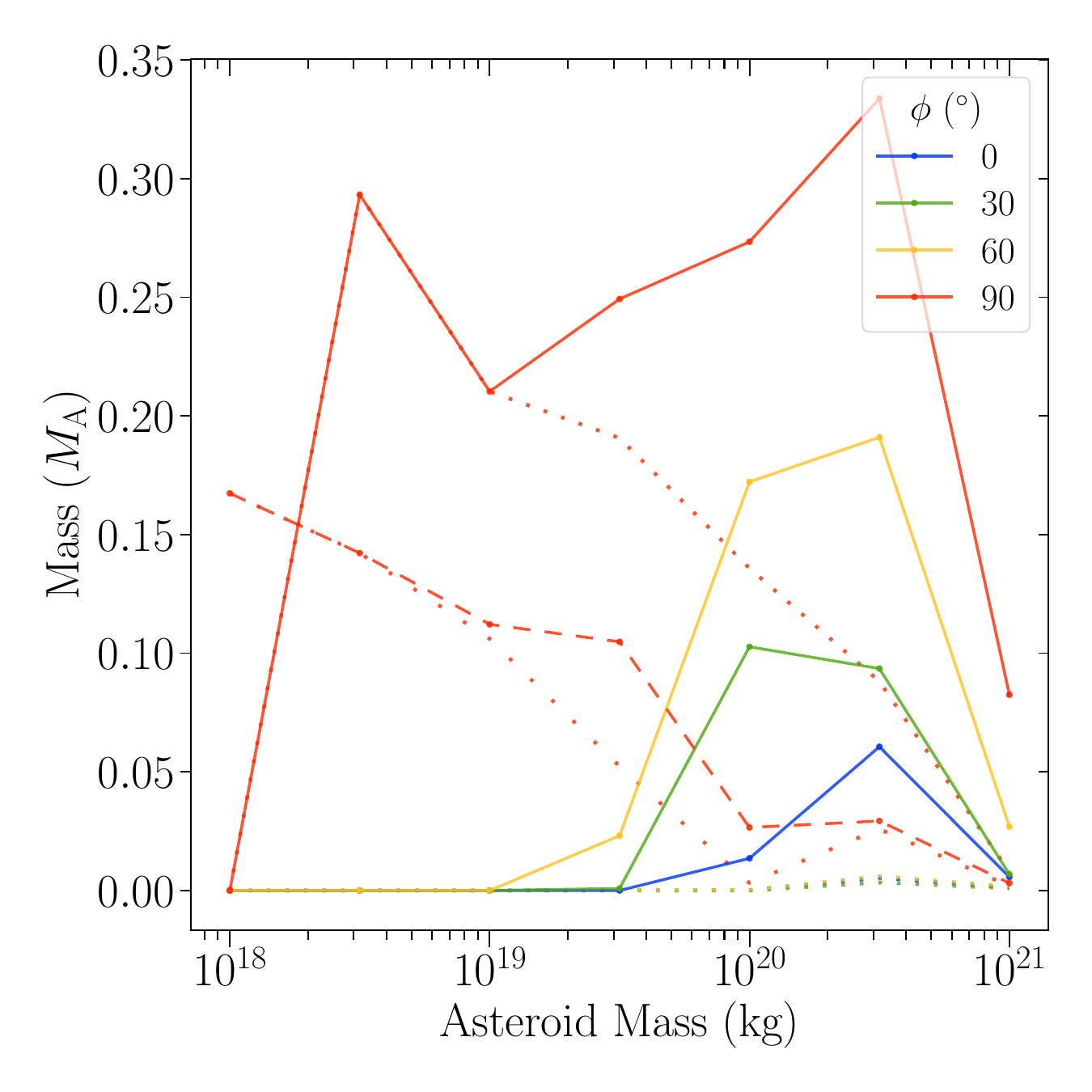}
  \\\vspace{-0.3em}
  \caption{
    The collisional and surviving mass of fragments
    as a function of and normalised by the parent asteroid mass
    in otherwise the same reference scenario
    with no initial spin and $q = 1.6~R_\smars$ for the solid lines.
    Dotted lines show the mass still in orbit after 5~kyr,
    as in Fig.~\ref{fig:m_col_phi}.
    The colours show the encounter longitude, $\phi$,
    with the fiducial $i = 15^\circ$ and $o_\smars = 30^\circ$,
    and the dashed line indicates $q = 1.2~R_\smars$ instead.
  \label{fig:m_col_mass}}
  \vspace{-1em}
\end{figure}

For parent asteroid masses from $10^{18.5}$--$10^{20.5}$~kg,
the collisional mass-fraction is broadly similar for favourable $\phi$
and a mid-range periapsis of $1.6~R_\smars$,
as shown in Fig.~\ref{fig:m_col_mass}.
For even smaller parents at this periapsis distance,
not enough material is stripped onto sufficiently low-$e$ orbits
(see \S\ref{sec:results:mass_disruption}),
so almost all of the material escapes unbound from Mars in the first few years.
For even larger parents, the much greater spread to both low and high $e$
leads more massive fragments to hit Mars or escape early on
-- although this lower fractional mass
still corresponds to a greater absolute mass.
In contrast, for $q = 1.2~R_\smars$ the surviving and collisional mass fractions
are highest for $10^{18}$~kg and decrease with the parent mass,
because with low periapses the wider dispersion of orbits from larger asteroids
leads increasing numbers of fragments to collide with Mars.
Note that these results are without any initial spin,
which, as for the fiducial mass,
could dramatically increase the viable parameter space,
and will have even greater effects for larger asteroids
(\S\ref{sec:results:spin_disruption}).

Turning to the secondary orientation parameters of the encounter:
in addition to the primary effects of the longitude and periapsis,
the input inclination also affects the fragments' evolution.
However, the large variations in inclination that are induced
by the Sun and by oblate-Mars
drive broadly similar behaviour regardless of the initial value,
as noted in \S\ref{sec:results:evol_timescales}.
For $\phi$ that are otherwise favourable at low $i$,
the mass that survives on collision timescales decreases with $i$,
but tens of percent can still survive even for $i = 90^\circ$,
as illustrated in Fig.~\ref{fig:m_col_r_p_in_i}c,d.
Conversely, for $\phi$ that are disfavoured at low $i$ (see Fig.~\ref{fig:m_col_phi}),
the surviving and colliding mass can increase with $i$.
This is because the amplitudes of the rapid periapsis oscillations
become greatly reduced for fragments orbiting well away from the solar plane,
such that they are no longer perturbed by the Sun to early collisions with Mars.
The evolution of individual large-mass fragments that happen either to or to not
survive or to have a significant collision adds stochasticity to all these trends.
Full results are given in Table~\ref{tab:reb_results}.

The obliquity of Mars also affects the survival of orbiting fragments,
as does its initial true anomaly if on an eccentric (here $e_\smars = 0.1$) orbit.
More mass tends to survive long-term for low Mars obliquities,
and slightly more mass may survive if Mars begins near periapsis
(Table~\ref{tab:reb_results}).
However, these trends are not as substantial
as those from the longitude and inclination.
As such, the uncertainties on the historical values of these parameters
across the tested range \citep{Laskar+2004} are not significant for our conclusions.

As an indication of the uncertainty on any one result,
e.g., from the precise distribution and initial orbits of fragments,
we repeat the same long-term integration scenario using each of the
eight reoriented repeat SPH simulations as the input
(with 4\% relative deviation on the initial captured mass;
\S\ref{sec:results:numerical_disruption}).
For the example scenario with $\phi$ of $60^\circ$ and $90^\circ$,
the standard deviations of the surviving mass are 18\% and 14\% of the mean, respectively.
Thus, while the precise outcomes of stochastic collisions
and orbit perturbations are unsurprisingly sensitive to the details of the initial fragments,
this empirical uncertainty is well below the level of the scenario-parameter trends.

\subsubsection{Angular momentum distribution and direct SPH collision simulations} \label{sec:results:ang_mom_evol}

These results demonstrate that disrupted fragments
can readily survive long-term around Mars
and potentially evolve into a collisional disk.
To consider what structure such a disk could have,
a simple estimate for the eventual evolution of collisional orbiting material,
in the absence of other perturbations,
is commonly made by assuming that each piece of debris
will end up on the equivalent circular orbit
with the same angular momentum, yielding $a_{\rm eq} = a (1 - e^2)$,
or, in terms of the initial periapsis, $a_{\rm eq} = q (1 + e)$.

In the fiducial scenarios here, most of the bound fragments
have initial eccentricities between $0.9$ and $0.99$,
and periapses around $0.9$--$1.5~R_\smars$
-- that is, clustered around the encounter periapsis of the parent asteroid
with a typical scatter of a few tens of percent.
This gives typical initial $a_{\rm eq}$ around $1.7$--$3~R_\smars$.
However, the Sun's influence can significantly change the orbital AM
of a fragment, as discussed in \S\ref{sec:results:evol_timescales}.
Furthermore, the disruptions and collisions that occur between surviving fragments
will redistribute the resulting debris
onto orbits with a range of energies and angular momenta
beyond that of a simple, singular $a_{\rm eq}$ estimate for each body.
In general, fragments that have their AM reduced by the solar perturbations
are more easily removed by colliding with Mars,
so those that survive will preferentially be fragments that had their AM increased.

\begin{figure}[t]
  \centering
  \includegraphics[
  width=\columnwidth, trim={9mm 9mm 9mm 9mm}, clip]{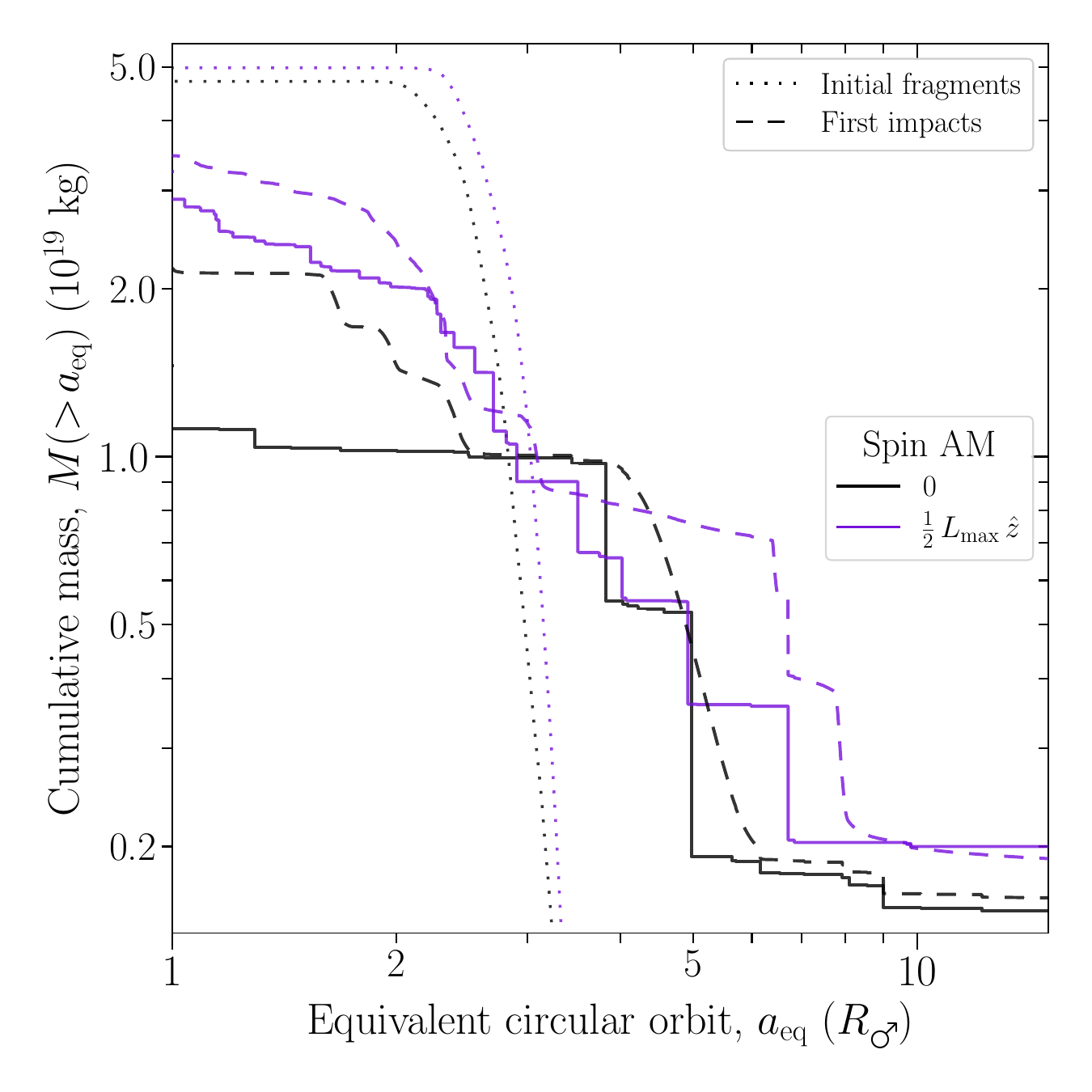}
  \\\vspace{-0.3em}
  \caption{
    The cumulative radial mass distribution of fragments still in orbit after 5~kyr
    under the assumption that they will evolve by collisional energy dissipation
    to circularise onto an equivalent orbit with the same angular momentum,
    for the reference input and evolution scenarios
    with either no or $\tfrac{1}{2}~L_{\rm max}$ spin AM,
    shown by the solid lines,
    as given in the legends.
    The dotted lines show the same for the fragments' initial orbits.
    The dashed lines replace the individual contributions
    of the large fragments that suffer significant collisions
    with those of the post-collision debris
    from direct SPH simulations (\S\ref{sec:methods:collision_sims}).
  \label{fig:a_eq_hist}}
  \vspace{-1em}
\end{figure}

This is illustrated in Fig.~\ref{fig:a_eq_hist},
which shows the $a_{\rm eq}$ predictions
for the distribution of mass in a circularised disk
from the orbiting fragments at the end of two example integrations
and, for comparison, at the start immediately after the input disruption.
While a real proto-satellite disk's evolution will be more complex,
this demonstrates the general effects that the gravitational perturbations can have
on typical distributions of fragment orbits,
with significant additions to their angular momenta.
It is particularly encouraging that material can readily survive evolution onto orbits
with $a_{\rm eq}$ around and beyond the corotation radius of $\sim$$6~R_\smars$
where Deimos currently orbits and may have accreted.

Furthermore, these simple estimates ignore the many fragments
that suffered significant collisions before 5~kyr.
To briefly explore their potential importance, we run new SPH simulations to model directly
the first significant collisions suffered by fragments (\S\ref{sec:methods:collision_sims})
where the larger has a mass of at least $10^{17}$~kg
and, to discard highly grazing collisions with minimal effects,
with an impact parameter of less than $0.8$.
This corresponds to collisions that collectively involve
between $61.4$\% and $71.2$\% of the orbiting mass in these two example cases.
As shown in Fig.~\ref{fig:a_eq_hist}, accounting for this first portion of
collisional evolution both increases and spreads out the predicted mass distribution.
For example, the mass with $a_{\rm eq} > 6~R_\smars$
increases from $1.9$ and $3.6 \times 10^{18}$~kg,
when considering only the still-orbiting mass,
to $2.1$ and $7.2 \times 10^{18}$~kg with the simulations of the first collisions.

This remains a highly simplified estimate, as we have not yet modelled all the collisions
and tidal disruptions that both the initial fragments
and their subsequent child fragments would experience,
and the Sun and Mars's oblateness would continue to perturb the orbits of the debris
throughout the process of circularisation
and damping to the Laplace plane.
That being said, the comparable behaviour and survivability
that we find across the wide range of tested initial fragment distributions
suggest that many of the later-produced bodies
could similarly be expected to evolve and survive.
These limitations are discussed further in \S\ref{sec:discussion:limitations}.
For now, starting with this standard $a_{\rm eq}$ approach,
we conclude that more than a percent of a parent asteroid's mass
could plausibly evolve into the outer regions of a proto-satellite disk.
This suggests that a far smaller parent asteroid could be sufficient
for the production of Phobos and Deimos, with a mass of down to around $10^{18}$~kg
in favourable encounter and spin scenarios,
or $\sim$$10^{19}$~kg and above for even wider ranges of parameter space to be viable.

\section{Discussion}
\label{sec:discussion}

\subsection{Limitations and future work}
\label{sec:discussion:limitations}

The aim of this work is to perform an initial exploration
of partial disruptive capture as a potential route
towards forming a proto-satellite disk around Mars.
The orbital evolution models here suggest that tens of percent
of a tidally disrupted asteroid can survive to beyond collision timescales.
Three key simplifications we make are:
(i) the further tidal disruptions of fragments
passing through the Roche limit are neglected;
(ii) only fragments larger than $3 \times 10^{15}$~kg are included;
and (iii) secondary fragments produced by subsequent collisions
are not accounted for.

Each of these leads us to underestimate
the number and spread of small bodies present in the system.
Provided that the majority of the debris remains of a large enough size
that its orbital evolution is not dominated by non-gravitational forces
(discussed further below),
these limitations mean that we are likely
underestimating the frequency of collisions,
and overestimating the timescale for the system to evolve
towards a more dynamically damped collisional disk.
As such, we consider these simplifications to be conservative rather than optimistic.
Nonetheless, future work will be crucial to test the effects
of improving on these simplifications
to make sophisticated predictions for the full evolution
of disruptive partial capture scenarios.

For the collisions between fragments, while we have analysed their general properties
and directly modelled some example subsets of collisions,
a goal for future work would be to simulate the detailed evolution
of the collision ejecta as the system evolves into a damped debris disk.
This would be followed by other models of satellite accretion,
as has been done following some giant-impact models
-- though often with similar approximations as made here for the initial disk formation
\citep[e.g.,][]{Canup+Salmon2018}.
The statistical properties of the collisions here may also change as,
for example, more fragments are produced on somewhat lower-eccentricity orbits
via subsequent tidal disruptions.
As the debris orbits are damped by disruptions and collisions
to typically lower $e$, $i$, and $a$,
the effects of the solar perturbations will also be reduced.
Other improvements and considerations,
such as material strength models and mass loss as an inclined disk
settles into the equatorial/Laplace plane,
will also be relevant for further constraining
the plausible parameter space and likelihood (see \S\ref{sec:discussion:comparison_caveats}).

Another potential loss mechanism for material evolving towards forming a disk
could be for collisional debris to be ground down to small enough grains
that non-gravitational forces become dominant.
Particles that drop to $\mu$m sizes can be removed by solar radiation pressure,
and Poynting--Robertson drag can cause the orbits of somewhat larger particles
to decay on a timescale of tens of kiloyears
up to a gigayear for $>$cm sizes \citep{Liang+Hyodo2023}.
As such, while this should be a consideration for future work
on the detailed outcomes of debris collisions
over the comparatively short timescale of disk formation,
it is probably more pertinent for predicting how long after satellite accretion
any remnant rings of tenuous debris could persist
-- which might help to constrain the timing of the moon-forming event \citep{Liang+Hyodo2023}.

The subsequent tidal disruption of fragments that pass close to Mars
will typically split them into smaller objects on new trajectories
in a similar way to the original parent's disruption
(\S\ref{sec:results:mass_disruption}).
However, the centre of mass is now on a bound and lower-eccentricity orbit,
so far more than half of the disrupted mass can remain bound \citep{Kegerreis+2022}.
Similarly, collisions between fragments will produce a range of
ejecta masses and trajectories.
These will depend sensitively on the masses, speed, and impact parameter,
but for most scenarios much of the material should circularise
and be redistributed across the system
\citep[\S\ref{sec:results:ang_mom_evol};][]{Teodoro+2023}.
In both cases, some of the new orbiting debris may be lost from the system.
However, just as in the orbital integrations we perform here,
a significant fraction should survive to experience further collisions.
That survival probability will increase as the orbits of the parent-fragments
become damped into a circularised disk,
as the number of debris objects grows,
and as the collision timescale decreases.
The collision velocity will also decrease with the objects' $e$ and $i$,
reducing the extent of disruption and the spread of post-collision orbits.
Therefore, we speculate that the large masses
we find can survive on collisional timescales
are a reasonable first estimate for the masses of the proto-satellite disks
that disruptive partial capture can produce.

A limitation of our SPH simulations also of note is that,
as in previous works \citep{Dones1991,Hyodo+2017c},
we neglected material strength and porosity.
Asteroids down to several orders of magnitude lower mass than we consider here
are found not to be dominated by strength
\citep{Holsapple+Michel2008,Harris+DAbramo2015}.
However, it is still plausible that a modest amount of strength
could reduce the fraction of mass that is tidally stripped
and thus adjust the range of periapses and speeds
for which this scenario is most effective \citep{Asphaug+Benz1996}.
So this will be important to test in the future,
but is unlikely to substantially alter the overall results in this regime.

The size distribution of fragments
might depend more sensitively on the disrupted parent asteroid's material properties,
affecting the fragments' evolution even if the total captured mass were similar.
For example, a population of fewer, larger fragments
could take longer to begin colliding with each other
than a distribution of many small objects.
However, unless this unexpectedly and dramatically affects
the subsequent survivability and evolution timescales of individual orbiting fragments,
we speculate that the following disruption and collisional events
would produce a broadly similar final outcome
to this first, exploratory investigation.

\subsection{Comparisons and caveats}
\label{sec:discussion:comparison_caveats}

Various observations could help to discriminate between formation scenarios
for Phobos and Deimos, when combined with sufficient predictions from simulations.
The detailed compositions of the moons would reveal
the nature of the parent body and/or the extent of mixing with Mars for a giant impact,
to distinguish between disruptive partial capture, with an asteroidal composition;
direct capture, with the two moons having differing compositions;
and an impact scenario, with a significant martian contribution
\citep{Lawrence+2019,Kuramoto+2022,Hirata+2024,Kuramoto2024}.
The volatile content and mineralogy could also constrain
the thermal history of Phobos and Deimos's building blocks.
Significant water and other volatiles could be lost
in a hot post-impact disk \citep{Hyodo+2018};
directly captured asteroids would retain their bulk content;
and the collisional but lower-energy disk evolution following disruptive partial capture
could, speculatively, result in a minor reduction of volatiles and mineralogical changes.
However, an incoming asteroid could also have undergone significant collisional processing
in its evolution prior to a direct capture -- or any -- scenario.
Finally, the moons' orbits continue to pose a challenge for direct capture,
unlike disruptive partial capture or a giant impact,
and their spectral features are thought to be
potentially explainable by all three scenarios (\S\ref{sec:intro}).
However, these opportunities face some additional challenges
to providing strong constraints or to conclusively distinguish between origin scenarios.
These challenges include a combination of modelling uncertainties
and the complexities of the real moons' Gyr histories \citep{Thomas+Veverka1980}.

In the two most recent simulation studies of giant impact scenarios,
only ``45'' and ``a few'' SPH particles represented
the proto-satellite disk's outer regions \citep{Hyodo+2017a,Canup+Salmon2018}
from which Phobos and Deimos would accrete.
This insufficient resolution precludes reliable estimates
of the outer disk's composition and thermodynamic state.
The moons are hypothesised to have accreted there
to explain both their orbits (on each side of synchronous)
and their spectra \citep{Ronnet+2016},
although the composition of the inner disk could also be relevant
for the Phobos-cycle scenario \citep{Hesselbrock+Minton2017}.

A further caveat for all compositional considerations is that
material from Mars is expected to be ejected by subsequent impacts
and accreted by Phobos and, in smaller amounts, by Deimos after they formed \citep{Hyodo+2019}.
This and additional asteroidal material accreted directly onto the moons
could add complexity to MMX and future observations, particularly surficial ones,
that attempt to discern the moons' original compositions.

The disks produced via disruptive partial capture
have a mass distribution less dominated by material close to the planet
than disks created by giant impacts (\S\ref{sec:results:ang_mom_evol}).
While a more comprehensive model that accounts for
the subsequent collisions and tidal disruptions
will be required to predict the detailed radial distribution of material,
this class of disk could alter or avoid the complications
induced by overly massive moons forming near the Roche limit
that can destabilise the accreting Phobos and Deimos
following a giant impact \citep{Canup+Salmon2018}.
On the other hand, one might speculate that some scenarios here
could also lead to the initial formation of additional moons outside Deimos
that would then need to be dynamically or collisionally removed.
However, this remains to be seen in future work.
It is also likely that a diversity of final-disk structures could arise
from the varied combinations of plausible initial disruption
and long-term evolution scenarios.

The evolution from the post-impact state of an SPH simulation
into a damped disk has not yet been studied in great detail either.
Models usually simulate only up to the formation of a long arm
of eccentric, potential proto-disk ejecta,
and then assume that the debris will evolve
directly onto their equivalent circular orbits with the same angular momentum,
as we do here \citep{Hyodo+2017a,Canup+Salmon2018}.
Therefore, in addition to the ongoing investigations into
the accretion of stable moons from a disk (\S\ref{sec:intro:hypotheses}),
the initial emergence of a disk from both disruptive partial capture
and giant impact scenarios will be valuable to pursue in future work.

\subsection{Scenario likelihood considerations}
\label{sec:discussion:likelihood}

We currently have only one Mars-like system to study,
so it is plausible that the formation of Phobos and Deimos
was not a high-probability event.
Nevertheless, in addition to how well the different origin scenarios
might explain the observations today,
their comparative likelihood is a useful consideration.
This is particularly relevant given the possibility
that MMX results may not immediately
distinguish unambiguously between scenarios (\S\ref{sec:discussion:comparison_caveats}).
However, while a complete and quantitative comparison would be valuable,
the large uncertainties behind many details of all the formation models
and the significant regions of parameter space that remain unexplored
restrict this discussion to mostly qualitative considerations.

From our current level of understanding,
the principal a priori difference in probability of
a successful disruptive partial capture and giant impact
may come from the required mass of the asteroid/impactor.
An impact in this regime produces a disk that has only a few percent
of the parent asteroid/impactor's mass
\citep{Hyodo+2017a,Canup+Salmon2018},
versus up to tens of percent from disruptive partial capture,
and a closer to linear dependence of the resulting mass on that of the parent
as we show here.
Furthermore, most of a post-impact disk's mass is within the Roche limit,
such that the total mass of a post-impact disk
must be far greater than that of the moons.

The requirements for a successful proto-satellite disk
remain uncertain for both scenarios,
but the mass needed for an impactor is proposed to be
in the range of $\sim$$10^{21}$--$10^{22}$~kg.
Under a pessimistic estimate that a similarly large $\gtrsim$$10^{19}$~kg disk mass
would still be required for disks produced by disruptive partial capture,
an order of magnitude smaller parent asteroid around $10^{20}$~kg would be sufficient.
However, the more efficient distribution out to the moons' accretion zones
of tidally disrupted debris here
-- where as much as several percent of the parent asteroid's mass
could circularise beyond the corotation radius (\S\ref{sec:results:ang_mom_evol}) --
means that a far lower total disk mass could be required.
As such, a parent asteroid mass of $\sim$$10^{18}$--$10^{19}$~kg
is likely a more appropriate range for the lower limit.
This could open up a far greater population of potential parents
in the early Solar System \citep{Ryan+2015,Lagain+2022}.

Unlike an impact, there is also not as straightforward a limit
on the parent's mass being larger,
as this would simply change which subsets
of disruption encounter scenarios are most viable,
requiring a smaller fraction of its mass to be initially stripped
and/or survive long-term.
However, we emphasise that making accurate estimates
will depend on improved models for both the evolution of the disk
and the subsequent satellite accretion,
in both disruptive partial capture and impact scenarios.

After the parent mass, the next consideration is
the extent of the viable region of parameter space for tidal encounters and impacts,
in terms of the speeds, impact parameters,
and orientations with respect to Mars and its orbit around the Sun.
We find that disruptive partial capture is effective around Mars
for speeds at infinity up to $\sim$1~km~s$^{-1}$.
This is towards the low end of the expected distributions of trajectories,
but a $v_\infty$ near zero and even slightly slower incoming speeds
to terrestrial planets still represents a significant population of objects
\citep{Quarles+Lissauer2015}.
However, the details of the populations and trajectories
of asteroids in the early Solar System are complex
and retain many uncertainties \citep[e.g.,][]{Brasser+2020}.
Even without any initial rotation, long-term partial capture can be successful
for periapses (impact parameters) from $\sim$1.1--1.8~$R_\smars$,
and up to $\sim$2.4~$R_\smars$ for favourable spin angular momenta.
In comparison, a giant impact might be viable
for a broader range of speeds at infinity, up to $\sim$14~km~s$^{-1}$,
but a far smaller range of impact parameters of $\sim$0.5--0.85~$R_\smars$
\citep{Canup+Salmon2018}.

The complex orbits enabled by chaotic Jacobi capture scenarios
could also further extend the range of initial speeds and impact parameters
for successful tidal disruption \citep{Boekholt+2023},
although the region of parameter space for an incoming trajectory
to lead to such orbits is relatively small.
In particular, the potential for multiple close periapsis passages in quick succession
could lead to an even greater proportion of the asteroid being captured,
allowing an even smaller mass for the parent asteroid.

With further reduced significance beyond the factors discussed above,
the longitudinal orientation and the inclination of the impactor/asteroid's orbit
are also relevant considerations.
For a giant impact, the former is probably not a limiting factor for its success,
perhaps excluding a retrograde impact.
For disruptive partial capture we find that, depending on the other variables,
the longitude could reduce the viable parameter space
-- but by far less than an order of magnitude.
Regarding the inclination and, relatedly, Mars's obliquity,
a large enough giant impact might affect the spin
and true polar wander of the planet \citep{Hyodo+2017b},
while the solar and oblate-Mars perturbations on the fragments
tend to spread their pre-damped inclinations regardless of the input value.
In both cases, a large tilt of the debris disk
away from Mars's equator could lead to significant mass loss
when disk material falls inwards as the inclinations to the Laplace plane are damped.

\section{Conclusions}
\label{sec:conclusions}

We have investigated the disruptive partial capture scenario
for the formation of Mars's moons using a combination of
smoothed particle hydrodynamics (SPH) simulations and orbital integrations.
Our results demonstrate that an unbound asteroid can be tidally disrupted
and tens of percent of its mass can survive in orbit around Mars
long enough to undergo significant collisions,
across a broad range of initial parameters.
These models represent the first stages of forming a proto-satellite disk,
and serve as a proof of concept for a
disruptive partial capture origin of Phobos and Deimos,
as a new alternative to the established options of direct capture and giant impact.

Furthermore, by assuming a simple collisional-damping evolution
for the surviving fragments with conserved angular momentum,
we find that several percent of the parent's mass could circularise
near and beyond synchronous orbit, where the moons could accrete.
This estimated value increases further when we directly model
the first collisions between fragments with additional SPH simulations,
with one example yielding over 7\% of the asteroid's mass settling in the outer disk.

Phobos and Deimos, if produced via disruptive partial capture as explored here,
would accrete from collisionally damped debris around Mars.
Their bulk compositions would thus reflect that of the parent asteroid,
potentially in line with their spectral properties \citep{Fraeman+2014}.
This also applies to direct capture scenarios,
but with the differing implication there that the base composition of Deimos
would not match that of Phobos \citep{Hunten1979,Higuchi+Ida2017}.
In contrast, after a giant impact, the proto-satellite disk would comprise
an as-yet-uncertain combination of impactor and martian material.

Unlike in direct capture scenarios \citep[e.g.,][]{Burns1972},
here the accreted moons would naturally form on
circular, low-inclination orbits.
The moons would also accrete from collisionally processed and ground-down material.
The comparatively low typical speeds we find for these collisions
reduces the amount of melting, vaporisation, and speculative volatile loss
compared with a hot and more highly shocked disk
produced by a giant impact \citep{Hyodo+2018}.

The imminent JAXA MMX mission aims to measure the compositional
and other properties of the moons to determine their origins,
via a combination of in-situ remote observations
and sample return from Phobos \citep{Kuramoto+2022,Hirata+2024,Kuramoto2024}.
This includes constraining, for example: the surface-level mineralogy
and the extent of hydration, which if high would support
direct capture or disruptive partial capture over a giant impact;
and the elemental abundance (and isotopic properties from later sample analysis),
which could distinguish between different chondritic and martian materials.
A substantial presence of the latter would support an impact origin,
and otherwise constrain the nature of the parent body
for an origin by direct or disruptive partial capture.

\vspace{\baselineskip}
\subsection*{Acknowledgments}
We thank K. J. Zahnle, M. \'Cuk, R. J. Massey, and L. F. A. Teodoro
for valuable input and discussion.
J.A.K. acknowledges support from a NASA Postdoctoral Program Fellowship
administered by Oak Ridge Associated Universities.
J.A.K. and J.J.L. acknowledge support from NASA Emerging Worlds grant 23-EW23-0035.
V.R.E. is supported by Science and Technology Facilities Council (STFC)
grant ST/X001075/1.
T.D.S. acknowledges support from STFC
grants ST/T506047/1 and ST/V506643/1.
The research in this paper made use of the \swift open-source simulation code
\citep{Schaller+2024}, version 0.9.0.
This work was supported by STFC 
grants ST/P000541/1 and ST/T000244/1,
and used the DiRAC@Durham facility managed by the
Institute for Computational Cosmology
on behalf of the STFC DiRAC HPC Facility (www.dirac.ac.uk).
This equipment was funded by BEIS via STFC capital grants
ST/K00042X/1, ST/P002293/1, ST/R002371/1, and ST/S002502/1,
Durham University and STFC operations grant ST/R000832/1.
DiRAC is part of the National e-Infrastructure.

\vspace{0.5\baselineskip}
\subsubsection*{Data availability}

The primary derived simulation data used for this study
are available in Tables~\ref{tab:results} and \ref{tab:reb_results}.
The raw simulation data are available from the corresponding author on reasonable request.

The \swift \citep{Schaller+2024} and \woma \citep{RuizBonilla+2021} codes
used in this study are publicly available
at \href{www.swiftsim.com}{www.swiftsim.com}
and \href{https://github.com/srbonilla/WoMa}{github.com/srbonilla/WoMa}.
The initial conditions and analysis scripts, the simulation input parameters,
and machine-readable versions of Tables~\ref{tab:results} and \ref{tab:reb_results}
are available at \href{https://github.com/jkeger/gihrpy}{github.com/jkeger/gihrpy}.

\vspace{0.5\baselineskip}
\subsubsection*{Competing interests}

The authors declare no competing interests.


\newpage
\appendix

\renewcommand\thefigure{\thesection\arabic{figure}}
\setcounter{figure}{0}

\section{Extended results}
\label{sec:extended_results}

\begin{figure}[t]
  \centering
  \includegraphics[
    width=\columnwidth, trim={8mm 7mm 7mm 7mm}, clip]{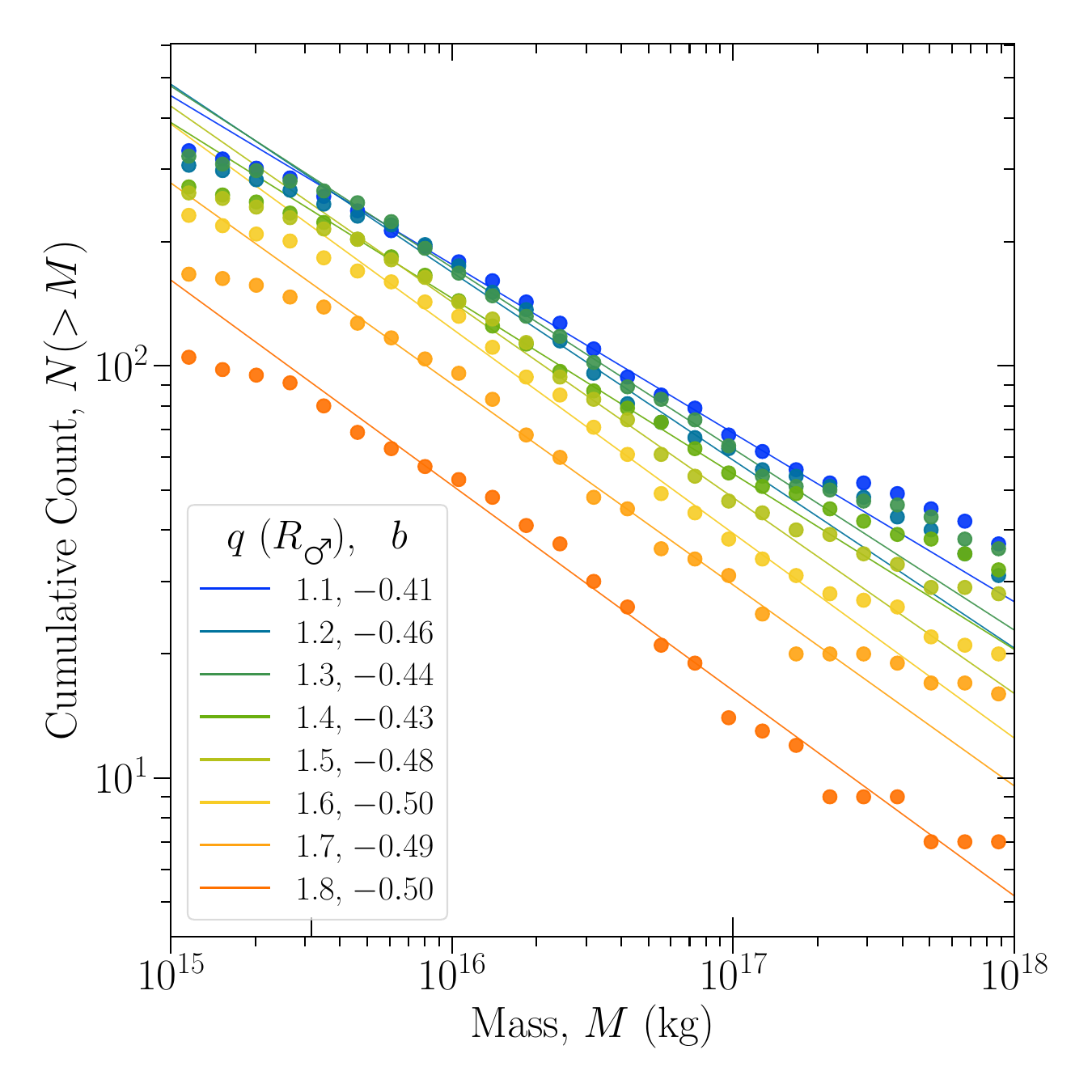}
  \\\vspace{-0.5em}
  \caption{
    The cumulative number distributions of fiducial-asteroid fragments
    for different periapsis distances,
    as also characterised by Fig.~\ref{fig:mass_funcs}.
    The lines show fitted power laws $N \propto M^b$
    for the mass range $10^{15.5}$--$10^{17}$~kg, as marked on the bottom axis.
    The fitted exponents are given in the legend,
    with uncertainties of $\sim$$0.01$.
    The fits exclude the smaller objects
    that are resolved less reliably,
    as well as the lower numbers of larger objects
    that do not as closely follow a power law in some cases.
    \label{fig:count_dist}}
  \vspace{-1em}
\end{figure}

\begin{figure}[t!]
  \centering
  \includegraphics[
    width=\columnwidth, trim={7mm 7mm 9mm 7mm}, clip]{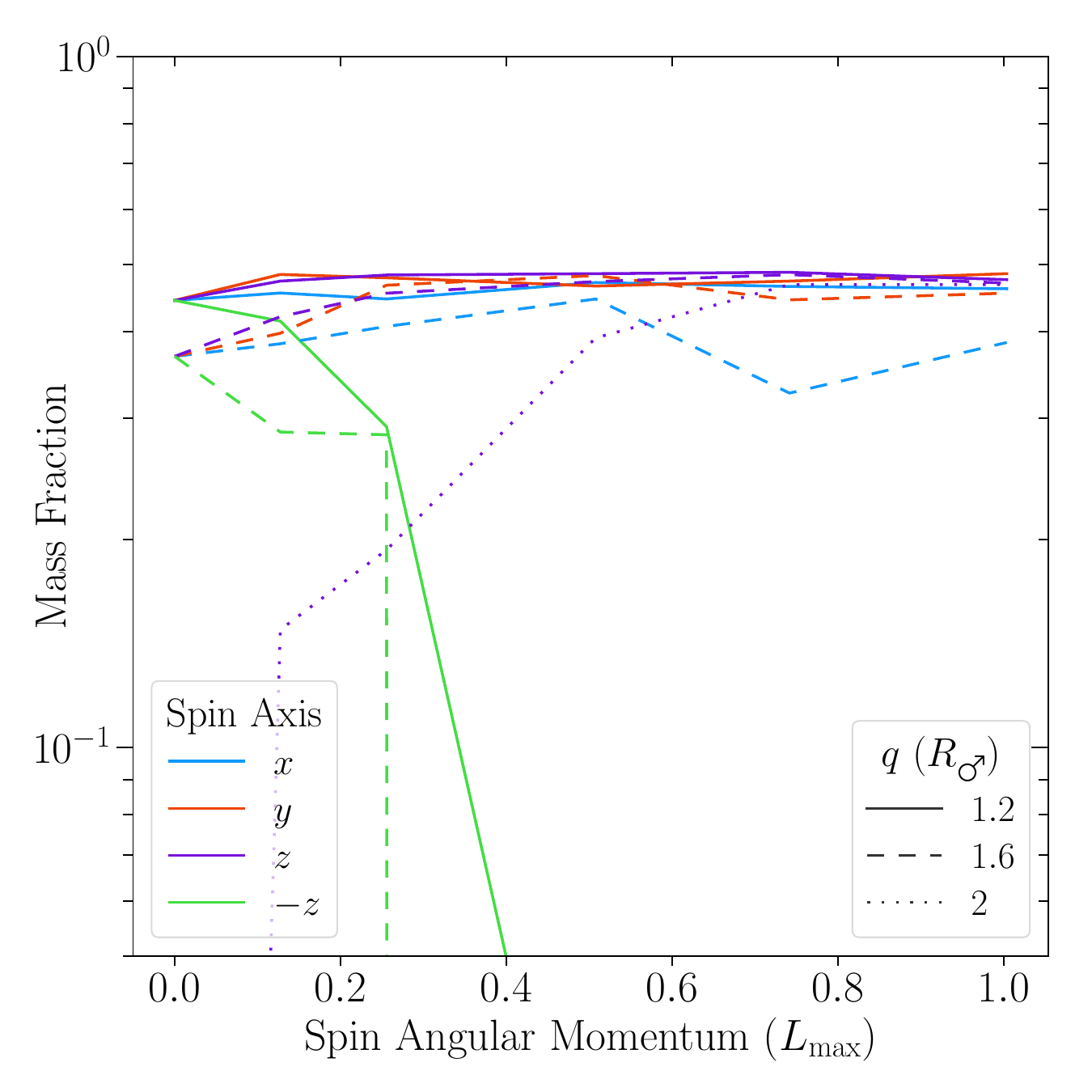}
  \\\vspace{-0.3em}
  \caption{
    The mass fraction of a $10^{20}$~kg asteroid with $v_\infty = 0$
    that is tidally stripped onto initially bound orbits within the Hill sphere,
    as a function of its spin angular momentum
    in units of the maximum stable spin.
    The line colours indicate the spin axis
    and the line styles indicate the periapsis,
    as detailed in the legends.
  \label{fig:m_capt_L_i}}
  \vspace{-1em}
\end{figure}

\begin{figure*}[t]
  \centering
  \includegraphics[
    width=\textwidth, trim={7mm 7mm 7mm 4mm}, clip]{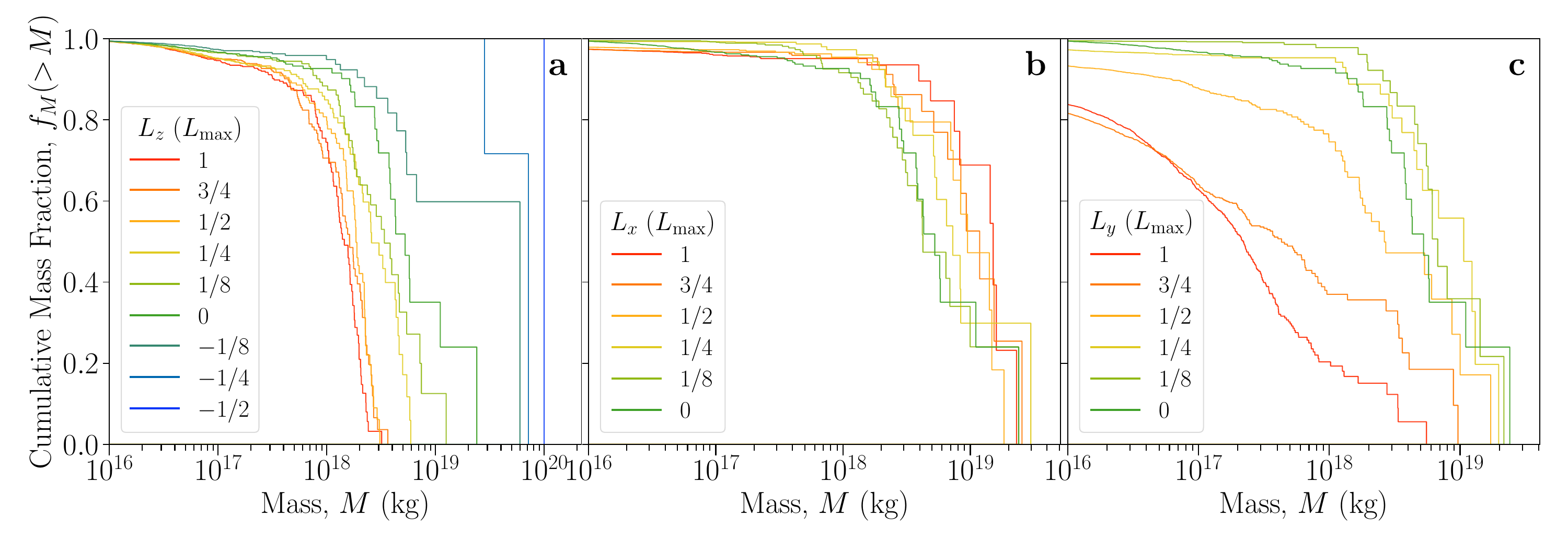}
  \\\vspace{-0.5em}
  \caption{
    The cumulative mass functions of fragments from spinning parent asteroids,
    as in Fig.~\ref{fig:mass_funcs},
    based on the reference tidal disruption scenario with
    $q = 1.6$~$R_\smars$, $v_\infty = 0$~km~s$^{-1}$.
    The lines show the results for $10^{20}$~kg asteroids
    with different spin angular momenta as detailed in the legends,
    as fractions of the maximum stable spin.
  \label{fig:mass_funcs_spin}}
  \vspace{-0.5em}
\end{figure*}

\begin{figure*}[t]
\centering
\begin{minipage}[t]{0.48\textwidth}
    \centering
    \includegraphics[
      width=\textwidth, trim={7mm 7mm 7mm 7mm}, clip]{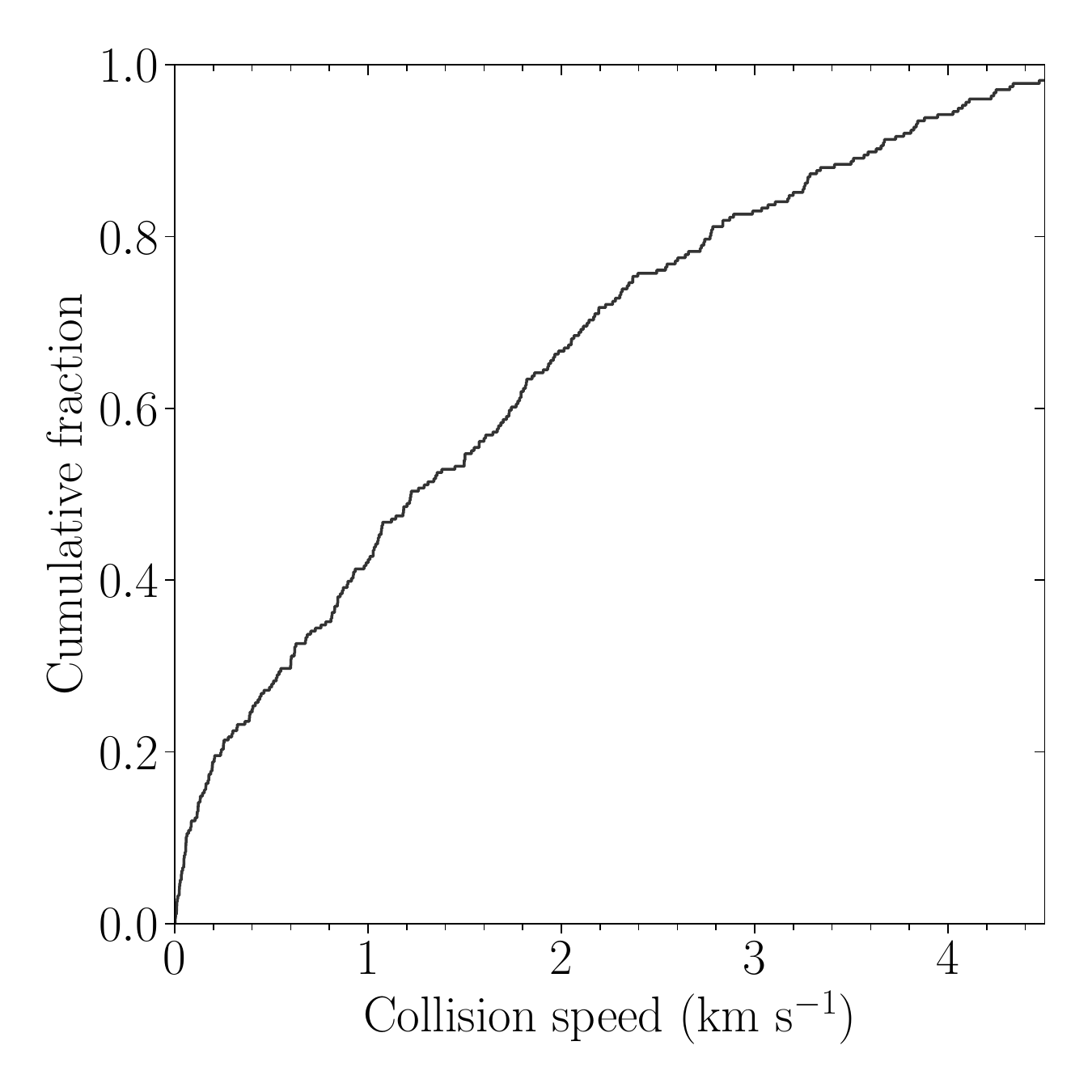}
    \\\vspace{-0.5em}
    \caption{
      The distribution of collision speeds between orbiting fragments around Mars
      from the representative example scenario of Fig.~\ref{fig:evol_e_i_r_p}
      with $\phi = 90^\circ$, and $i = 15^\circ$.
    \label{fig:v_rel_c_hist}}
\end{minipage}\hfill
\begin{minipage}[t]{0.48\textwidth}
    \centering
    \includegraphics[
      width=\textwidth, trim={7mm 7mm 7mm 7mm}, clip]{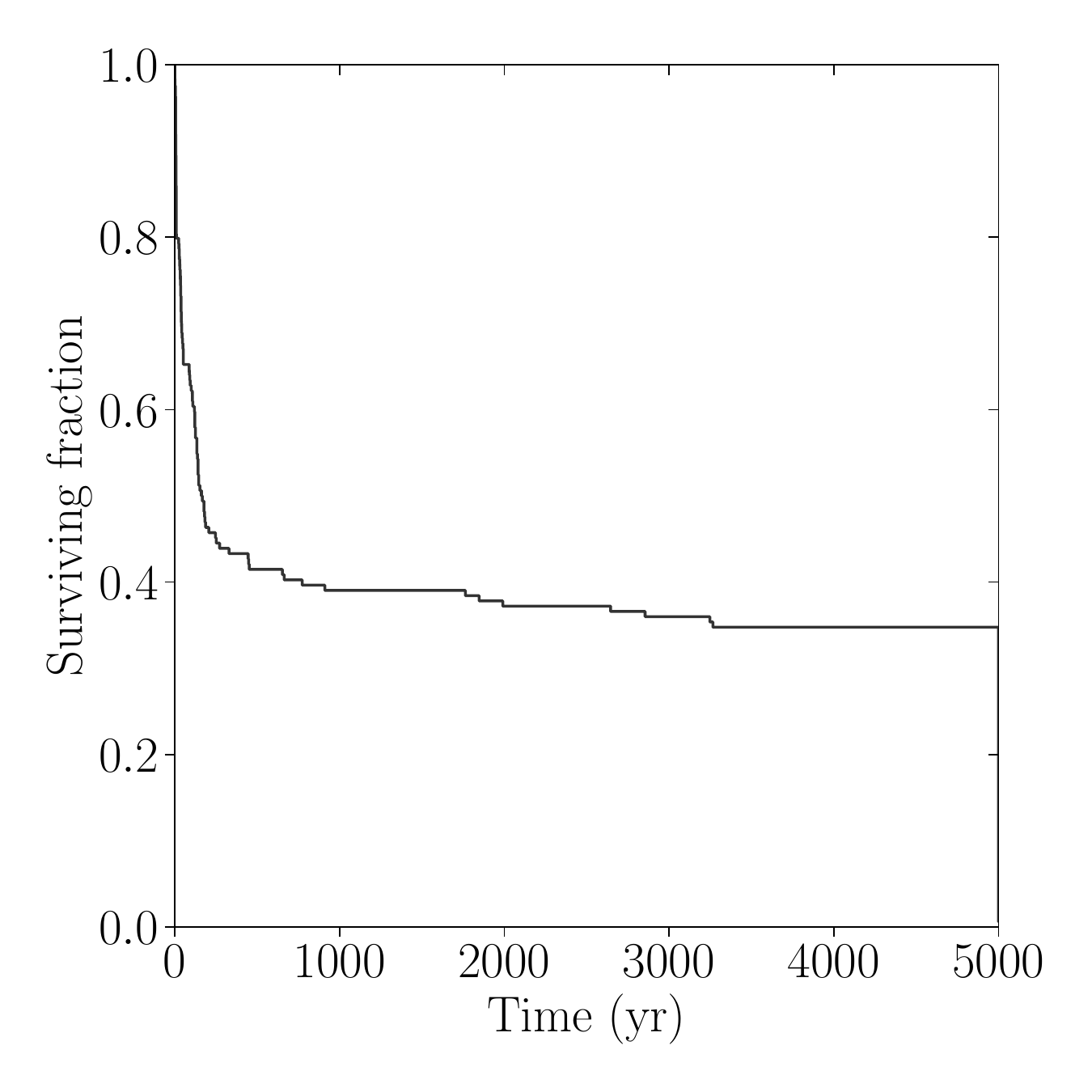}
    \\\vspace{-0.5em}
    \caption{
      The fraction of orbiting fragments that remain surviving
      in orbit around Mars before being removed by escaping or impacting the planet,
      in the Fig.~\ref{fig:evol_e_i_r_p} example scenario
      with $\phi = 90^\circ$, and $i = 15^\circ$.
    \label{fig:t_surv_hist}}
\end{minipage}
\vspace{-1em}
\end{figure*}

\begin{figure*}[t]
  \centering
  \includegraphics[
    width=\textwidth, trim={9mm 9mm 9mm 9mm}, clip]{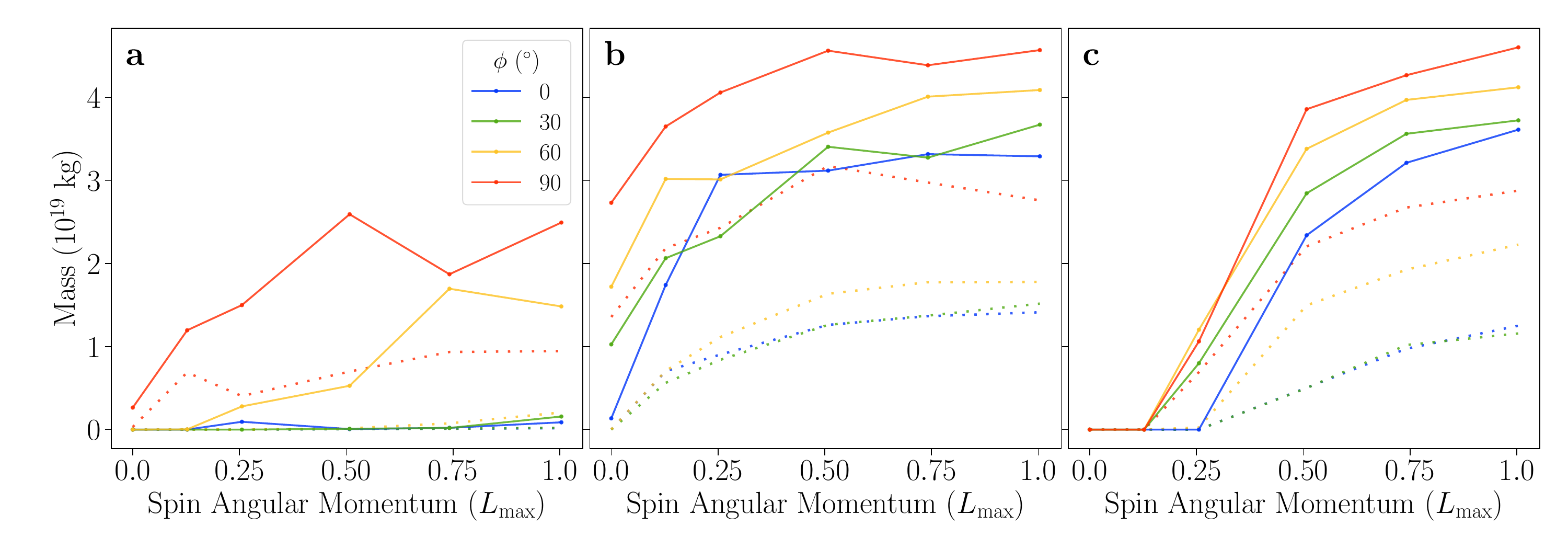}
  \\\vspace{-0.3em}
  \caption{
    The collisional and surviving mass of fragments,
    as in Fig.~\ref{fig:m_col_phi},
    as a function of the spin angular momentum
    of the $10^{20}$~kg asteroid,
    parallel to that of the orbit ($L_z$),
    for different longitude orientations as given in the legend,
    with the fiducial $i = 15^\circ$ and $o_\smars = 30^\circ$.
    Panels \textbf{a}, \textbf{b}, and \textbf{c}
    show input asteroids on parabolic orbits
    with periapses of $1.2$, $1.6$, and $2.0~R_\smars$, respectively.
  \label{fig:m_col_L_z_r_p}}
  \vspace{-1em}
\end{figure*}

The cumulative number distributions of self-gravitating fragments
produced by tidal disruption of the parent asteroid
are shown in Fig.~\ref{fig:count_dist} for different periapses,
with a power law $N \propto M^b$ fitted to each simulation,
as discussed in \S\ref{sec:results:disruption}.
There is a slight trend to shallower distributions at lower periapses,
but in all cases the contribution of larger objects here
is far greater than distributions from impact debris
that are dominated in mass by smaller objects \citep{Teodoro+2023}.

The mass fraction of material that is initially captured
from asteroids with initial spins
is shown in Fig.~\ref{fig:m_capt_L_i}.
As discussed in \S\ref{sec:results:spin_disruption},
while a spin AM in the $\pm z$ directions has a strong effect,
a spin axis in $x$ or $y$ has little influence,
at least in terms of the total initial capture
and for the asteroid sizes tested here.
Note that a positive and negative spin AM in $x$ or $y$
is the same by symmetry, unlike in $z$.
We also note that the apparent scatter in some of the trends of captured mass
often arise from the asteroid being disrupted into a pair or other small number
of massive fragments, rather than a chain of smaller fragments
with a smoother spread of orbital elements.

The distributions of fragment masses from the tidal disruption
of asteroids with increasing initial spin AM
are shown in Fig.~\ref{fig:mass_funcs_spin}.
As discussed in \S\ref{sec:results:spin_disruption},
the cumulative number distributions become shallower as $L_z$ increases,
with a greater number of large fragments.
The power-law exponents on the mass function increase
from the no-spin $\sim$$-0.5$ (Fig.~\ref{fig:count_dist})
to $\sim$$-0.36$ for $L_z \gtrsim \tfrac{1}{2}~L_{\rm max}$.

A spin AM in the $x$ direction has much less effect.
However, for high $L_x$, we do find a more uniform set of similar-size fragments
with close to the largest mass ($>$$10^{19}$~kg),
compared with the more diverse spread between $10^{18}$ and $10^{19}$~kg
in the no-spin case.
The number of smaller-size fragments does grow
with increasing $L_z \gtrsim \tfrac{1}{2}~L_{\rm max}$,
with far more insufficiently-resolved debris at this SPH resolution.
Unlike the $L_z$ case in this scenario,
no confluence of results appears as the spin approaches the maximum rate.

For the representative long-term evolution example shown in Fig.~\ref{fig:evol_e_i_r_p},
and as discussed in \S\ref{sec:results:orbit_evol},
the distribution of collision speeds between orbiting fragments
is shown in Fig.~\ref{fig:v_rel_c_hist},
with the fastest collisions reaching a few km~s$^{-1}$.
The distribution of how long the fragments survive in orbit
is shown in Fig.~\ref{fig:t_surv_hist}.
Most of the particles that do not survive to beyond the end of the simulation
are removed by colliding with Mars within the first few hundred years.

The mass of orbiting fragments that survive long-term
following the disruption of asteroids with increasing spin AM
in the $z$ direction are shown in Fig.~\ref{fig:m_col_L_z_r_p},
for different encounter periapses and longitude orientations.
As discussed in \S\ref{sec:results:evol_trends},
following the trends of greater initial disruption effects
of a $z$-direction spin AM for the parent asteroid,
more of the resulting fragments can also survive long-term in orbit
as $L_z$ increases.
The mass that survives beyond collisional timescales is significantly increased by
just $L_z \approx \tfrac{1}{4}$ or $\tfrac{1}{2}~L_{\rm max}$,
with additional but diminishing gain from more rapid rotation.
This includes encounter longitudes
that precluded any long-term survival with no spin, but now become viable.

\vspace{1em}
\section{Orbit equations}
\label{sec:orbit_equations}

\subsection{Orbital elements from trajectories}

For the analysis and interpretation of the fragment and particle data
produced by a tidal-disruption simulation or a long-term integration,
we estimate the instantaneous orbital elements
from the position, $\vec{r}$, and velocity, $\vec{v}$.
The semi-major axis, $a$, is obtained from the vis-viva equation:
\begin{equation}
  a = \left(\dfrac{2}{r} - \dfrac{v^2}{\mu}\right)^{-1} \;,
  \label{eqn:vis_viva}
\end{equation}
where $r \equiv |\vec{r}|$, $v \equiv |\vec{v}|$,
$\mu \equiv G (M_\smars + m)$, $m$ is the object's mass,
and $G$ is the gravitational constant.

We then compute the eccentricity, $e$, inclination, $i$,
and longitude of ascending node, $\Omega$,
from the specific angular momentum, $\vec{h} = \vec{r} \times \vec{v}\,$:
\begin{align}
  \vec{e} &= \dfrac{\vec{v} \times \vec{h}}{\mu} - \dfrac{\vec{r}}{r} \;, \\
  i &= \cos^{-1}\!\left( \dfrac{h_z}{h} \right) \;, \\
  \Omega &= \cos^{-1}\!\left( \dfrac{-h_x}{\sqrt{h_x^2 + h_y^2}} \right) \;,
\end{align}
where $\Omega$ is multiplied by $-1$ if $h_x < 0$.

\subsection{Scenario initial conditions}

To construct the initial conditions for an SPH tidal disruption simulation,
the primary input parameters are the periapsis, $q$,
and the speed at infinity, $v_\infty$,
as illustrated in Fig.~\ref{fig:disruption_setup}
(or, equivalently for a giant impact,
the impact parameter, $b$, and the speed at contact, $v_{\rm c}$,
with $v_\infty = \sqrt{v_{\rm c}^2 - 2 \mu / r_{\rm c}}\,$,
where $r_{\rm c} = R_\smars + R_{\rm A}$).
From these, we derive the starting position and velocity coordinates
of the asteroid (with Mars placed at the origin)
that yield the desired time or distance until periapsis (or contact),
and then shift to the centre-of-mass rest frame for the SPH simulation.
For convenience and aesthetic consistency,
we then rotate the coordinates such that the velocity at periapsis
(or contact) is in the $x$ direction%
\footnote{
  Wrappers for these calculations are included in the \href{https://github.com/srbonilla/WoMa}{\woma} package utilities.
}%
$^,$\footnote{
  A version of these equations for giant impacts
  appeared in Appx.~A of \citet{Kegerreis+2020},
  but with a typo in Eqn.~A7 (here Eqn.~\ref{eqn:q_quad})
  that would slightly alter the aesthetic rotation.
}.

\subsubsection{Starting position and velocity}
\label{sec:orbit_equations:start_pos_vel}

At periapsis, the speed is $v_q = \sqrt{v_\infty^2 + 2 \mu / q}\,$,
with which we compute the semi-major axis using Eqn.~\ref{eqn:vis_viva}
and the specific angular momentum, $h = r \,v = q \,v_q$.
At a chosen starting distance, $r$, the speed from vis-viva is
\begin{equation}
  v = \sqrt{\mu \left(\dfrac{2}{r} - \dfrac{1}{a}\right)}
  \label{eqn:vis_viva_v}
\end{equation}
for an elliptical or hyperbolic orbit, or
\begin{equation}
    v = \sqrt{2 \mu / r}
\end{equation}
for a parabolic orbit.
In the reference frame where the asteroid's initial velocity is $[-v, 0, 0]$,
its starting position is then
$x = \sqrt{r^2 - y^2}$, $y = h / v$, and $z = 0$.

\subsubsection{Frame rotation}

In order to rotate these coordinates such that
the velocity at periapsis defines the $x$ direction
(as in Fig.~\ref{fig:disruption_setup}; a purely aesthetic choice),
we calculate the true anomaly -- in this case its complement, $\theta$ --
from the eccentricity:
\begin{align}
  e &= 1 - q / a \\
  \theta &= \cos^{-1}\! \left(\dfrac{1 - \tfrac{a\left(1-e^2\right)}{r}}{e}\right) \;,
  \label{eqn:true_anomaly}
\end{align}
or, for a parabolic orbit,
\begin{align}
  \theta = \pi - \cos^{-1}\! \left(\dfrac{h^2}{\mu r}\right) \;.
\end{align}
The angle by which to rotate the starting coordinates about the $z$ axis is then
\begin{equation}
  \phi = -\dfrac{\pi}{2} + \theta - \sin^{-1}\! \left(\dfrac{y}{r}\right) \;.
  \label{eqn:rot_angle}
\end{equation}

In general, for an impact scenario configured as in \citet{Kegerreis+2020},
the periapsis is first found by setting $v = h / q$ in Eqn.~\ref{eqn:vis_viva_v}
at periapsis to eliminate the speed:
\begin{align}
  \dfrac{h^2}{q^2} &= \mu \left(\dfrac{2}{q} - \dfrac{1}{a}\right)
  \nonumber \\
  q^2 &- 2 a\, q + \tfrac{a h^2}{\mu} = 0
  \nonumber \\
  q &= a \pm \sqrt{a^2 - \tfrac{2 a h^2}{\mu}\,} \;. \label{eqn:q_quad}
\end{align}
We can then calculate the angle of the velocity at contact
away from the radial vector,
\begin{equation}
  \alpha' = \sin^{-1}\! \left(\dfrac{a^2 \left(1 - e^2\right)}{2 a r_{\rm c} - r_{\rm c}^2}\right) \;,
\end{equation}
or, for a parabolic orbit,
\begin{equation}
  \alpha' = \theta_{\rm c} / 2 \;,
\end{equation}
where $\theta_{\rm c}$ is Eqn.~\ref{eqn:true_anomaly} at contact.
Eqn.~\ref{eqn:rot_angle} for an impact then becomes
\begin{equation}
  \phi = \alpha' - \theta_{\rm c} + \theta - \sin^{-1}\left(\dfrac{y}{r}\right) \;.
  \label{eqn:rot_angle_impact}
\end{equation}

\subsubsection{Starting time}

The time taken from the initial position to periapsis, $t_q$,
for elliptical, hyperbolic, and parabolic orbits, respectively,
can be found by using the eccentric anomaly, $E$,
\begin{align}
  E_{\rm ell} &= \cos^{-1}\! \left(\dfrac{e + \cos \theta}{1 + e \cos \theta}\right) \\
  E_{\rm hyp} &= \cosh^{-1}\! \left(\dfrac{e + \cos \theta}{1 + e \cos \theta}\right) \\
  E_{\rm par} &= \tan\! \left(\theta / 2\right)
\end{align}
and mean anomaly, $M$,
\begin{align}
  M_{\rm ell} &= E - e \sin E \label{eqn:M_E_ell} \\
  M_{\rm hyp} &= -E + e \sinh E \\
  M_{\rm par} &= E + \tfrac{1}{3} E^3 \label{eqn:M_E_par}
\end{align}
to find the time since periapsis:
\begin{align}
  t_{q, \rm ell, hyp} &= \sqrt{|a|^3 / \mu\,} \,M \\
  t_{q, \rm par} &= \sqrt{2 \, q^3 / \mu\,} \,M \;.
\end{align}
Then, for time until contact for an impact,
$t_{\rm c} = t_q(\theta) - t_q(\theta_{\rm c})$.

For completeness, in the case of a radial orbit,
the equivalent time until the objects as point masses would overlap,
$t_q'$, is
\begin{align}
  t_{q, \rm par}' &= \sqrt{\dfrac{2 r^3}{9 \mu}\,} \\
  t_{q, \rm ell}' &= \dfrac{\sin^{-1} \left(\sqrt{w r\,}\right) - \sqrt{w r (1 - w r)\,}}
    {\sqrt{2 \mu w^3\,}} \\
  t_{q, \rm hyp}' &= \left[\sqrt{(|w| r)^2 + |w| r\,}\right. \nonumber\\
    &\phantom{=}\;\;
    \dfrac{\;\left. - \ln\left(\sqrt{|w| r\,} + \sqrt{1 + |w| r\,}\right)\right]}
    {\sqrt{2 \mu |w|^3\,}} \;,
\end{align}
where $w$ is the standard constant
\begin{align}
  w &= \dfrac{1}{r} - \dfrac{v^2}{2\mu} \;.
\end{align}
Eqns.~\ref{eqn:M_E_ell}--\ref{eqn:M_E_par} cannot be inverted
to compute $E$ from $M$ analytically.
Therefore, the equations above are solved iteratively
to find the initial separation that yields
the desired time until periapsis or contact.
We then use this distance as described in \S\ref{sec:orbit_equations:start_pos_vel}
to set up the simulation.

\vspace{1em}
\section{Results tables}
\label{sec:results_tables}

The primary results from the SPH simulations
are listed in Table~\ref{tab:results},
arranged in groups by subset, periapsis, and speed.
The primary results from the orbit integrations
are listed in Table~\ref{tab:reb_results},
arranged in groups by subset, input disruption scenario,
and encounter orientation.
The raw simulation data are available on reasonable request.


\newpage

{
\footnotesize
\bibliography{./gihr.bib}

\begin{thebibliography}{}
\makeatletter
\relax
\def\mn@urlcharsother{\let\do\@makeother \do\$\do\&\do\#\do\^\do\_\do\%\do\~}
\def\mn@doi{\begingroup\mn@urlcharsother \@ifnextchar [ {\mn@doi@}
  {\mn@doi@[]}}
\def\mn@doi@[#1]#2{\def\@tempa{#1}\ifx\@tempa\@empty \href
  {http://dx.doi.org/#2} {doi:#2}\else \href {http://dx.doi.org/#2} {#1}\fi
  \endgroup}
\def\mn@eprint#1#2{\mn@eprint@#1:#2::\@nil}
\def\mn@eprint@arXiv#1{\href {http://arxiv.org/abs/#1} {{\tt arXiv:#1}}}
\def\mn@eprint@dblp#1{\href {http://dblp.uni-trier.de/rec/bibtex/#1.xml}
  {dblp:#1}}
\def\mn@eprint@#1:#2:#3:#4\@nil{\def\@tempa {#1}\def\@tempb {#2}\def\@tempc
  {#3}\ifx \@tempc \@empty \let \@tempc \@tempb \let \@tempb \@tempa \fi \ifx
  \@tempb \@empty \def\@tempb {arXiv}\fi \@ifundefined
  {mn@eprint@\@tempb}{\@tempb:\@tempc}{\expandafter \expandafter \csname
  mn@eprint@\@tempb\endcsname \expandafter{\@tempc}}}

\bibitem[\protect\citeauthoryear{{Ahrens} \& {O'Keefe}}{{Ahrens} \&
  {O'Keefe}}{1972}]{Ahrens+OKeefe1972}
{Ahrens} T.~J.,  {O'Keefe} J.~D.,  1972, \mn@doi [Moon] {10.1007/BF00562927},
  \href {https://ui.adsabs.harvard.edu/abs/1972Moon....4..214A} {4, 214}

\bibitem[\protect\citeauthoryear{{Asphaug} \& {Benz}}{{Asphaug} \&
  {Benz}}{1996}]{Asphaug+Benz1996}
{Asphaug} E.,  {Benz} W.,  1996, \mn@doi [\icarus] {10.1006/icar.1996.0083},
  \href {https://ui.adsabs.harvard.edu/abs/1996Icar..121..225A} {121, 225}

\bibitem[\protect\citeauthoryear{{Bagheri}, {Khan}, {Efroimsky}, {Kruglyakov}
  \& {Giardini}}{{Bagheri} et~al.}{2021}]{Bagheri+2021}
{Bagheri} A.,  {Khan} A.,  {Efroimsky} M.,  {Kruglyakov} M.,   {Giardini} D.,
  2021, \mn@doi [Nature Astronomy] {10.1038/s41550-021-01306-2}, \href
  {https://ui.adsabs.harvard.edu/abs/2021NatAs...5..539B} {5, 539}

\bibitem[\protect\citeauthoryear{{Balsara}}{{Balsara}}{1995}]{Balsara1995}
{Balsara} D.~S.,  1995, \mn@doi [J. Comput. Phys.]
  {10.1016/S0021-9991(95)90221-X}, \href
  {https://ui.adsabs.harvard.edu/\#abs/1995JCoPh.121..357B} {121, 357}

\bibitem[\protect\citeauthoryear{{Benner} \& {McKinnon}}{{Benner} \&
  {McKinnon}}{1995}]{Benner+McKinnon1995}
{Benner} L. A.~M.,  {McKinnon} W.~B.,  1995, \mn@doi [\icarus]
  {10.1006/icar.1995.1039}, \href
  {https://ui.adsabs.harvard.edu/abs/1995Icar..114....1B} {114, 1}

\bibitem[\protect\citeauthoryear{{Boekholt}, {Rowan}  \& {Kocsis}}{{Boekholt}
  et~al.}{2023}]{Boekholt+2023}
{Boekholt} T. C.~N.,  {Rowan} C.,   {Kocsis} B.,  2023, \mn@doi [\mnras]
  {10.1093/mnras/stac3495}, \href
  {https://ui.adsabs.harvard.edu/abs/2023MNRAS.518.5653B} {518, 5653}

\bibitem[\protect\citeauthoryear{{Brasser}, {Werner}  \& {Mojzsis}}{{Brasser}
  et~al.}{2020}]{Brasser+2020}
{Brasser} R.,  {Werner} S.~C.,   {Mojzsis} S.~J.,  2020, \mn@doi [\icarus]
  {10.1016/j.icarus.2019.113514}, \href
  {https://ui.adsabs.harvard.edu/abs/2020Icar..33813514B} {338, 113514}

\bibitem[\protect\citeauthoryear{{Burns}}{{Burns}}{1972}]{Burns1972}
{Burns} J.~A.,  1972, \mn@doi [Rev. Geophys. and Space Phys.]
  {10.1029/RG010i002p00463}, \href
  {https://ui.adsabs.harvard.edu/abs/1972RvGSP..10..463B} {10, 463}

\bibitem[\protect\citeauthoryear{{Canup} \& {Salmon}}{{Canup} \&
  {Salmon}}{2018}]{Canup+Salmon2018}
{Canup} R.,  {Salmon} J.,  2018, \mn@doi [Science Advances]
  {10.1126/sciadv.aar6887}, \href
  {https://ui.adsabs.harvard.edu/#abs/2018SciA....4R6887C} {4, eaar6887}

\bibitem[\protect\citeauthoryear{{Citron}, {Genda}  \& {Ida}}{{Citron}
  et~al.}{2015}]{Citron+2015}
{Citron} R.~I.,  {Genda} H.,   {Ida} S.,  2015, \mn@doi [\icarus]
  {10.1016/j.icarus.2015.02.011}, \href
  {https://ui.adsabs.harvard.edu/#abs/2015Icar..252..334C} {252, 334}

\bibitem[\protect\citeauthoryear{{Craddock}}{{Craddock}}{2011}]{Craddock2011}
{Craddock} R.~A.,  2011, \mn@doi [\icarus] {10.1016/j.icarus.2010.10.023},
  \href {https://ui.adsabs.harvard.edu/#abs/2011Icar..211.1150C} {211, 1150}

\bibitem[\protect\citeauthoryear{{{\'C}uk}, {Minton}, {Pouplin}  \&
  {Wishard}}{{{\'C}uk} et~al.}{2020}]{Cuk+2020b}
{{\'C}uk} M.,  {Minton} D.~A.,  {Pouplin} J. L.~L.,   {Wishard} C.,  2020,
  \mn@doi [\apjl] {10.3847/2041-8213/ab974f}, \href
  {https://ui.adsabs.harvard.edu/abs/2020ApJ...896L..28C} {896, L28}

\bibitem[\protect\citeauthoryear{{Davis}, {Efstathiou}, {Frenk}  \&
  {White}}{{Davis} et~al.}{1985}]{Davis+1985}
{Davis} M.,  {Efstathiou} G.,  {Frenk} C.~S.,   {White} S.~D.~M.,  1985,
  \mn@doi [\apj] {10.1086/163168}, \href
  {http://ukads.nottingham.ac.uk/abs/1985ApJ...292..371D} {292, 371}

\bibitem[\protect\citeauthoryear{{Dones}}{{Dones}}{1991}]{Dones1991}
{Dones} L.,  1991, \mn@doi [\icarus] {10.1016/0019-1035(91)90045-U}, \href
  {https://ui.adsabs.harvard.edu/abs/1991Icar...92..194D} {92, 194}

\bibitem[\protect\citeauthoryear{{Ernst}, {Daly}, {Gaskell}, {Barnouin},
  {Nair}, {Hyatt}, {Al Asad}  \& {Hoch}}{{Ernst} et~al.}{2023}]{Ernst+2023}
{Ernst} C.~M.,  {Daly} R.~T.,  {Gaskell} R.~W.,  {Barnouin} O.~S.,  {Nair} H.,
  {Hyatt} B.~A.,  {Al Asad} M.~M.,   {Hoch} K. K.~W.,  2023, \mn@doi [Earth,
  Planets and Space] {10.1186/s40623-023-01814-7}, \href
  {https://ui.adsabs.harvard.edu/abs/2023EP&S...75..103E} {75, 103}

\bibitem[\protect\citeauthoryear{{Fornasier}, {Wargnier}, {Hasselmann},
  {Tirsch}, {Matz}, {Doressoundiram}, {Gautier}  \& {Barucci}}{{Fornasier}
  et~al.}{2024}]{Fornasier+2024}
{Fornasier} S.,  {Wargnier} A.,  {Hasselmann} P.~H.,  {Tirsch} D.,  {Matz}
  K.~D.,  {Doressoundiram} A.,  {Gautier} T.,   {Barucci} M.~A.,  2024, \mn@doi
  [\aap] {10.1051/0004-6361/202449220}, \href
  {https://ui.adsabs.harvard.edu/abs/2024A&A...686A.203F} {686, A203}

\bibitem[\protect\citeauthoryear{{Fraeman}, {Murchie}, {Arvidson}, {Clark},
  {Morris}, {Rivkin}  \& {Vilas}}{{Fraeman} et~al.}{2014}]{Fraeman+2014}
{Fraeman} A.~A.,  {Murchie} S.~L.,  {Arvidson} R.~E.,  {Clark} R.~N.,  {Morris}
  R.~V.,  {Rivkin} A.~S.,   {Vilas} F.,  2014, \mn@doi [\icarus]
  {10.1016/j.icarus.2013.11.021}, \href
  {https://ui.adsabs.harvard.edu/#abs/2014Icar..229..196F} {229, 196}

\bibitem[\protect\citeauthoryear{{Genda}, {Fujita}, {Kobayashi}, {Tanaka}  \&
  {Abe}}{{Genda} et~al.}{2015}]{Genda+2015}
{Genda} H.,  {Fujita} T.,  {Kobayashi} H.,  {Tanaka} H.,   {Abe} Y.,  2015,
  \mn@doi [\icarus] {10.1016/j.icarus.2015.08.029}, \href
  {https://ui.adsabs.harvard.edu/#abs/2015Icar..262...58G} {262, 58}

\bibitem[\protect\citeauthoryear{{Gingold} \& {Monaghan}}{{Gingold} \&
  {Monaghan}}{1977}]{Gingold+Monaghan1977}
{Gingold} R.~A.,  {Monaghan} J.~J.,  1977, \mn@doi [\mnras]
  {10.1093/mnras/181.3.375}, \href
  {http://adsabs.harvard.edu/abs/1977MNRAS.181..375G} {181, 375}

\bibitem[\protect\citeauthoryear{{Harris} \& {D'Abramo}}{{Harris} \&
  {D'Abramo}}{2015}]{Harris+DAbramo2015}
{Harris} A.~W.,  {D'Abramo} G.,  2015, \mn@doi [\icarus]
  {10.1016/j.icarus.2015.05.004}, \href
  {https://ui.adsabs.harvard.edu/abs/2015Icar..257..302H} {257, 302}

\bibitem[\protect\citeauthoryear{{Hesselbrock} \& {Minton}}{{Hesselbrock} \&
  {Minton}}{2017}]{Hesselbrock+Minton2017}
{Hesselbrock} A.~J.,  {Minton} D.~A.,  2017, \mn@doi [Nature Geoscience]
  {10.1038/ngeo2916}, \href
  {https://ui.adsabs.harvard.edu/#abs/2017NatGe..10..266H} {10, 266}

\bibitem[\protect\citeauthoryear{{Higuchi} \& {Ida}}{{Higuchi} \&
  {Ida}}{2017}]{Higuchi+Ida2017}
{Higuchi} A.,  {Ida} S.,  2017, \mn@doi [\aj] {10.3847/1538-3881/aa5daa}, \href
  {https://ui.adsabs.harvard.edu/abs/2017AJ....153..155H} {153, 155}

\bibitem[\protect\citeauthoryear{{Hirata} et~al.,}{{Hirata}
  et~al.}{2024}]{Hirata+2024}
{Hirata} K.,  et~al., 2024, \mn@doi [\icarus] {10.1016/j.icarus.2023.115891},
  \href {https://ui.adsabs.harvard.edu/abs/2024Icar..41015891H} {410, 115891}

\bibitem[\protect\citeauthoryear{{Holsapple} \& {Michel}}{{Holsapple} \&
  {Michel}}{2008}]{Holsapple+Michel2008}
{Holsapple} K.~A.,  {Michel} P.,  2008, \mn@doi [\icarus]
  {10.1016/j.icarus.2007.09.011}, \href
  {https://ui.adsabs.harvard.edu/abs/2008Icar..193..283H} {193, 283}

\bibitem[\protect\citeauthoryear{{Hosono}, {Iwasawa}, {Tanikawa}, {Nitadori},
  {Muranushi}  \& {Makino}}{{Hosono} et~al.}{2017}]{Hosono+2017}
{Hosono} N.,  {Iwasawa} M.,  {Tanikawa} A.,  {Nitadori} K.,  {Muranushi} T.,
  {Makino} J.,  2017, \mn@doi [Publ. Astron. Soc. Jpn.] {10.1093/pasj/psw131},
  \href {https://ui.adsabs.harvard.edu/#abs/2017PASJ...69...26H} {69, 26}

\bibitem[\protect\citeauthoryear{{Hunten}}{{Hunten}}{1979}]{Hunten1979}
{Hunten} D.~M.,  1979, \mn@doi [\icarus] {10.1016/0019-1035(79)90119-2}, \href
  {https://ui.adsabs.harvard.edu/abs/1979Icar...37..113H} {37, 113}

\bibitem[\protect\citeauthoryear{{Hyodo}, {Charnoz}, {Ohtsuki}  \&
  {Genda}}{{Hyodo} et~al.}{2017a}]{Hyodo+2017c}
{Hyodo} R.,  {Charnoz} S.,  {Ohtsuki} K.,   {Genda} H.,  2017a, \mn@doi
  [\icarus] {10.1016/j.icarus.2016.09.012}, \href
  {https://ui.adsabs.harvard.edu/abs/2017Icar..282..195H} {282, 195}

\bibitem[\protect\citeauthoryear{{Hyodo}, {Genda}, {Charnoz}  \&
  {Rosenblatt}}{{Hyodo} et~al.}{2017b}]{Hyodo+2017a}
{Hyodo} R.,  {Genda} H.,  {Charnoz} S.,   {Rosenblatt} P.,  2017b, \mn@doi
  [\apj] {10.3847/1538-4357/aa81c4}, \href
  {http://adsabs.harvard.edu/abs/2017ApJ...845..125H} {845, 125}

\bibitem[\protect\citeauthoryear{{Hyodo}, {Rosenblatt}, {Genda}  \&
  {Charnoz}}{{Hyodo} et~al.}{2017c}]{Hyodo+2017b}
{Hyodo} R.,  {Rosenblatt} P.,  {Genda} H.,   {Charnoz} S.,  2017c, \mn@doi
  [\apj] {10.3847/1538-4357/aa9984}, \href
  {https://ui.adsabs.harvard.edu/#abs/2017ApJ...851..122H} {851, 122}

\bibitem[\protect\citeauthoryear{{Hyodo}, {Genda}, {Charnoz}, {Pignatale}  \&
  {Rosenblatt}}{{Hyodo} et~al.}{2018}]{Hyodo+2018}
{Hyodo} R.,  {Genda} H.,  {Charnoz} S.,  {Pignatale} F. C.~F.,   {Rosenblatt}
  P.,  2018, \mn@doi [\apj] {10.3847/1538-4357/aac024}, \href
  {https://ui.adsabs.harvard.edu/#abs/2018ApJ...860..150H} {860, 150}

\bibitem[\protect\citeauthoryear{{Hyodo}, {Kurosawa}, {Genda}, {Usui}  \&
  {Fujita}}{{Hyodo} et~al.}{2019}]{Hyodo+2019}
{Hyodo} R.,  {Kurosawa} K.,  {Genda} H.,  {Usui} T.,   {Fujita} K.,  2019,
  \mn@doi [Scientific Reports] {10.1038/s41598-019-56139-x}, \href
  {https://ui.adsabs.harvard.edu/abs/2019NatSR...919833H} {9, 19833}

\bibitem[\protect\citeauthoryear{{Hyodo}, {Genda}, {Sekiguchi}, {Madeira}  \&
  {Charnoz}}{{Hyodo} et~al.}{2022}]{Hyodo+2022}
{Hyodo} R.,  {Genda} H.,  {Sekiguchi} R.,  {Madeira} G.,   {Charnoz} S.,  2022,
  \mn@doi [\psj] {10.3847/PSJ/ac88d2}, \href
  {https://ui.adsabs.harvard.edu/abs/2022PSJ.....3..204H} {3, 204}

\bibitem[\protect\citeauthoryear{{Ito} \& {Ohtsuka}}{{Ito} \&
  {Ohtsuka}}{2019}]{Ito+Ohtsuka2019}
{Ito} T.,  {Ohtsuka} K.,  2019, \mn@doi [Monographs on Environment, Earth and
  Planets] {10.5047/meep.2019.00701.0001}, \href
  {https://ui.adsabs.harvard.edu/abs/2019MEEP....7....1I} {7, 1}

\bibitem[\protect\citeauthoryear{{Kaula}}{{Kaula}}{1966}]{Kaula1966}
{Kaula} W.~M.,  1966, {Theory of satellite geodesy. Applications of satellites
  to geodesy}.
Blaisdell Publishing Company

\bibitem[\protect\citeauthoryear{{Kegerreis}, {Eke}, {Gonnet}, {Korycansky},
  {Massey}, {Schaller}  \& {Teodoro}}{{Kegerreis}
  et~al.}{2019}]{Kegerreis+2019}
{Kegerreis} J.~A.,  {Eke} V.~R.,  {Gonnet} P.,  {Korycansky} D.~G.,  {Massey}
  R.~J.,  {Schaller} M.,   {Teodoro} L.~F.~A.,  2019, \mn@doi [\mnras]
  {10.1093/mnras/stz1606}, \href
  {https://ui.adsabs.harvard.edu/abs/2019MNRAS.tmp.1536K} {487, 1536}

\bibitem[\protect\citeauthoryear{{Kegerreis}, {Eke}, {Massey}  \&
  {Teodoro}}{{Kegerreis} et~al.}{2020}]{Kegerreis+2020}
{Kegerreis} J.~A.,  {Eke} V.~R.,  {Massey} R.~J.,   {Teodoro} L.~F.~A.,  2020,
  \mn@doi [\apj] {10.3847/1538-4357/ab9810}, \href
  {https://ui.adsabs.harvard.edu/abs/2020ApJ...897..161K} {897, 161}

\bibitem[\protect\citeauthoryear{{Kegerreis}, {Ruiz-Bonilla}, {Eke}, {Massey},
  {Sandnes}  \& {Teodoro}}{{Kegerreis} et~al.}{2022}]{Kegerreis+2022}
{Kegerreis} J.~A.,  {Ruiz-Bonilla} S.,  {Eke} V.~R.,  {Massey} R.~J.,
  {Sandnes} T.~D.,   {Teodoro} L. F.~A.,  2022, \mn@doi [\apjl]
  {10.3847/2041-8213/abb5fb}, 937, L40

\bibitem[\protect\citeauthoryear{{Kozai}}{{Kozai}}{1962}]{Kozai1962}
{Kozai} Y.,  1962, \mn@doi [\aj] {10.1086/108790}, \href
  {https://ui.adsabs.harvard.edu/abs/1962AJ.....67..591K} {67, 591}

\bibitem[\protect\citeauthoryear{{Kuramoto}}{{Kuramoto}}{2024}]{Kuramoto2024}
{Kuramoto} K.,  2024, \mn@doi [Annual Review of Earth and Planetary Sciences]
  {10.1146/annurev-earth-040522-110615}, \href
  {https://ui.adsabs.harvard.edu/abs/2024AREPS..52..495K} {52, 495}

\bibitem[\protect\citeauthoryear{{Kuramoto} et~al.,}{{Kuramoto}
  et~al.}{2022}]{Kuramoto+2022}
{Kuramoto} K.,  et~al., 2022, \mn@doi [Earth, Planets and Space]
  {10.1186/s40623-021-01545-7}, \href
  {https://ui.adsabs.harvard.edu/abs/2022EP&S...74...12K} {74, 12}

\bibitem[\protect\citeauthoryear{{Lagain} et~al.,}{{Lagain}
  et~al.}{2022}]{Lagain+2022}
{Lagain} A.,  et~al., 2022, \mn@doi [Earth Planet. Sci. Lett.]
  {10.1016/j.epsl.2021.117362}, \href
  {https://ui.adsabs.harvard.edu/abs/2022E&PSL.57917362L} {579, 117362}

\bibitem[\protect\citeauthoryear{{Landis}}{{Landis}}{2009}]{Landis2009}
{Landis} G.~A.,  2009, \mn@doi [arXiv e-prints] {10.48550/arXiv.0903.3434},
  \href {https://ui.adsabs.harvard.edu/abs/2009arXiv0903.3434L} {p.
  arXiv:0903.3434}

\bibitem[\protect\citeauthoryear{{Laskar}, {Correia}, {Gastineau}, {Joutel},
  {Levrard}  \& {Robutel}}{{Laskar} et~al.}{2004}]{Laskar+2004}
{Laskar} J.,  {Correia} A.~C.~M.,  {Gastineau} M.,  {Joutel} F.,  {Levrard} B.,
    {Robutel} P.,  2004, \mn@doi [\icarus] {10.1016/j.icarus.2004.04.005},
  \href {https://ui.adsabs.harvard.edu/abs/2004Icar..170..343L} {170, 343}

\bibitem[\protect\citeauthoryear{{Lawrence} et~al.,}{{Lawrence}
  et~al.}{2019}]{Lawrence+2019}
{Lawrence} D.~J.,  et~al., 2019, \mn@doi [Earth and Space Science]
  {10.1029/2019EA000811}, \href
  {https://ui.adsabs.harvard.edu/abs/2019E&SS....6.2605L} {6, 2605}

\bibitem[\protect\citeauthoryear{{Leinhardt} \& {Stewart}}{{Leinhardt} \&
  {Stewart}}{2012}]{Leinhardt+Stewart2012}
{Leinhardt} Z.~M.,  {Stewart} S.~T.,  2012, \mn@doi [\apj]
  {10.1088/0004-637X/745/1/79}, \href
  {https://ui.adsabs.harvard.edu/#abs/2012ApJ...745...79L} {745, 79}

\bibitem[\protect\citeauthoryear{{Liang} \& {Hyodo}}{{Liang} \&
  {Hyodo}}{2023}]{Liang+Hyodo2023}
{Liang} Y.,  {Hyodo} R.,  2023, \mn@doi [\icarus]
  {10.1016/j.icarus.2022.115335}, \href
  {https://ui.adsabs.harvard.edu/abs/2023Icar..39115335L} {391, 115335}

\bibitem[\protect\citeauthoryear{{Lidov}}{{Lidov}}{1962}]{Lidov1962}
{Lidov} M.~L.,  1962, \mn@doi [\planss] {10.1016/0032-0633(62)90129-0}, \href
  {https://ui.adsabs.harvard.edu/abs/1962P&SS....9..719L} {9, 719}

\bibitem[\protect\citeauthoryear{{Lucy}}{{Lucy}}{1977}]{Lucy1977}
{Lucy} L.~B.,  1977, \mn@doi [\aj] {10.1086/112164}, \href
  {http://adsabs.harvard.edu/abs/1977AJ.....82.1013L} {82, 1013}

\bibitem[\protect\citeauthoryear{{Madeira}, {Charnoz}, {Zhang}, {Hyodo},
  {Michel}, {Genda}  \& {Giuliatti Winter}}{{Madeira}
  et~al.}{2023}]{Madeira+2023}
{Madeira} G.,  {Charnoz} S.,  {Zhang} Y.,  {Hyodo} R.,  {Michel} P.,  {Genda}
  H.,   {Giuliatti Winter} S.,  2023, \mn@doi [\aj] {10.3847/1538-3881/acbf53},
  \href {https://ui.adsabs.harvard.edu/abs/2023AJ....165..161M} {165, 161}

\bibitem[\protect\citeauthoryear{{Movshovitz}, {Asphaug}  \&
  {Korycansky}}{{Movshovitz} et~al.}{2012}]{Movshovitz+2012}
{Movshovitz} N.,  {Asphaug} E.,   {Korycansky} D.,  2012, \mn@doi [\apj]
  {10.1088/0004-637X/759/2/93}, \href
  {https://ui.adsabs.harvard.edu/abs/2012ApJ...759...93M} {759, 93}

\bibitem[\protect\citeauthoryear{{Murray} \& {Dermott}}{{Murray} \&
  {Dermott}}{1999}]{Murray+Dermott1999}
{Murray} C.~D.,  {Dermott} S.~F.,  1999, {Solar System Dynamics}.
Cambridge University Press, \mn@doi{10.1017/CBO9781139174817}

\bibitem[\protect\citeauthoryear{{Pajola} et~al.,}{{Pajola}
  et~al.}{2012}]{Pajola+2012}
{Pajola} M.,  et~al., 2012, \mn@doi [\mnras]
  {10.1111/j.1365-2966.2012.22026.x}, \href
  {https://ui.adsabs.harvard.edu/abs/2012MNRAS.427.3230P} {427, 3230}

\bibitem[\protect\citeauthoryear{{Pajola} et~al.,}{{Pajola}
  et~al.}{2013}]{Pajola+2013}
{Pajola} M.,  et~al., 2013, \mn@doi [\apj] {10.1088/0004-637X/777/2/127}, \href
  {https://ui.adsabs.harvard.edu/#abs/2013ApJ...777..127P} {777, 127}

\bibitem[\protect\citeauthoryear{{Pouplin}, {Wishard}, {Minton}, {Elliott}  \&
  {Singh}}{{Pouplin} et~al.}{2021}]{Pouplin+2021}
{Pouplin} J.,  {Wishard} C.~A.,  {Minton} D.~A.,  {Elliott} J.~R.,   {Singh}
  D.,  2021, in AAS/Division for Planetary Sciences Meeting Abstracts. p.
  506.09

\bibitem[\protect\citeauthoryear{{Quarles} \& {Lissauer}}{{Quarles} \&
  {Lissauer}}{2015}]{Quarles+Lissauer2015}
{Quarles} B.~L.,  {Lissauer} J.~J.,  2015, \mn@doi [\icarus]
  {10.1016/j.icarus.2014.10.044}, \href
  {https://ui.adsabs.harvard.edu/abs/2015Icar..248..318Q} {248, 318}

\bibitem[\protect\citeauthoryear{{Rein} \& {Liu}}{{Rein} \&
  {Liu}}{2012}]{Rein+Liu2012}
{Rein} H.,  {Liu} S.~F.,  2012, \mn@doi [\aap] {10.1051/0004-6361/201118085},
  \href {https://ui.adsabs.harvard.edu/abs/2012A&A...537A.128R} {537, A128}

\bibitem[\protect\citeauthoryear{{Rein} \& {Spiegel}}{{Rein} \&
  {Spiegel}}{2015}]{Rein+Spiegel2015}
{Rein} H.,  {Spiegel} D.~S.,  2015, \mn@doi [\mnras] {10.1093/mnras/stu2164},
  \href {https://ui.adsabs.harvard.edu/abs/2015MNRAS.446.1424R} {446, 1424}

\bibitem[\protect\citeauthoryear{{Ronnet}, {Vernazza}, {Mousis}, {Brugger},
  {Beck}, {Devouard}, {Witasse}  \& {Cipriani}}{{Ronnet}
  et~al.}{2016}]{Ronnet+2016}
{Ronnet} T.,  {Vernazza} P.,  {Mousis} O.,  {Brugger} B.,  {Beck} P.,
  {Devouard} B.,  {Witasse} O.,   {Cipriani} F.,  2016, \mn@doi [\apj]
  {10.3847/0004-637X/828/2/109}, \href
  {https://ui.adsabs.harvard.edu/#abs/2016ApJ...828..109R} {828, 109}

\bibitem[\protect\citeauthoryear{{Rosenblatt} \& {Charnoz}}{{Rosenblatt} \&
  {Charnoz}}{2012}]{Rosenblatt+Charnoz2012}
{Rosenblatt} P.,  {Charnoz} S.,  2012, \mn@doi [\icarus]
  {10.1016/j.icarus.2012.09.009}, \href
  {https://ui.adsabs.harvard.edu/abs/2012Icar..221..806R} {221, 806}

\bibitem[\protect\citeauthoryear{{Rosenblatt}, {Charnoz}, {Dunseath},
  {Terao-Dunseath}, {Trinh}, {Hyodo}, {Genda}  \& {Toupin}}{{Rosenblatt}
  et~al.}{2016}]{Rosenblatt+2016}
{Rosenblatt} P.,  {Charnoz} S.,  {Dunseath} K.~M.,  {Terao-Dunseath} M.,
  {Trinh} A.,  {Hyodo} R.,  {Genda} H.,   {Toupin} S.,  2016, \mn@doi [Nature
  Geoscience] {10.1038/ngeo2742}, \href
  {https://ui.adsabs.harvard.edu/#abs/2016NatGe...9..581R} {9, 581}

\bibitem[\protect\citeauthoryear{{Ruiz-Bonilla}, {Eke}, {Kegerreis}, {Massey}
  \& {Teodoro}}{{Ruiz-Bonilla} et~al.}{2021}]{RuizBonilla+2021}
{Ruiz-Bonilla} S.,  {Eke} V.~R.,  {Kegerreis} J.~A.,  {Massey} R.~J.,
  {Teodoro} L.~F.~A.,  2021, \mn@doi [\mnras] {10.1093/mnras/staa3385}, \href
  {https://ui.adsabs.harvard.edu/abs/2021MNRAS.500.2861R} {500, 2861}

\bibitem[\protect\citeauthoryear{{Ryan}, {Mizuno}, {Shenoy}, {Woodward},
  {Carey}, {Noriega-Crespo}, {Kraemer}  \& {Price}}{{Ryan}
  et~al.}{2015}]{Ryan+2015}
{Ryan} E.~L.,  {Mizuno} D.~R.,  {Shenoy} S.~S.,  {Woodward} C.~E.,  {Carey}
  S.~J.,  {Noriega-Crespo} A.,  {Kraemer} K.~E.,   {Price} S.~D.,  2015,
  \mn@doi [\aap] {10.1051/0004-6361/201321375}, \href
  {https://ui.adsabs.harvard.edu/abs/2015A&A...578A..42R} {578, A42}

\bibitem[\protect\citeauthoryear{{Schaller} et~al.,}{{Schaller}
  et~al.}{2024}]{Schaller+2024}
{Schaller} M.,  et~al., 2024, \mn@doi [\mnras] {10.1093/mnras/stae922}, \href
  {https://ui.adsabs.harvard.edu/abs/2024MNRAS.530.2378S} {530, 2378}

\bibitem[\protect\citeauthoryear{{Stewart} et~al.,}{{Stewart}
  et~al.}{2020}]{Stewart+2020}
{Stewart} S.,  et~al., 2020, in American Institute of Physics Conference
  Series. p. 080003 (\mn@eprint {arXiv} {1910.04687}),
  \mn@doi{10.1063/12.0000946}

\bibitem[\protect\citeauthoryear{{Szab{\'o}} et~al.,}{{Szab{\'o}}
  et~al.}{2022}]{Szabo+2022}
{Szab{\'o}} G.~M.,  et~al., 2022, \mn@doi [\aap] {10.1051/0004-6361/202142223},
  \href {https://ui.adsabs.harvard.edu/abs/2022A&A...661A..48S} {661, A48}

\bibitem[\protect\citeauthoryear{{Teodoro}, {Kegerreis}, {Estrada}, {{\'C}uk},
  {Eke}, {Cuzzi}, {Massey}  \& {Sandnes}}{{Teodoro}
  et~al.}{2023}]{Teodoro+2023}
{Teodoro} L.~F.~A.,  {Kegerreis} J.~A.,  {Estrada} P.~R.,  {{\'C}uk} M.,  {Eke}
  V.~R.,  {Cuzzi} J.~N.,  {Massey} R.~J.,   {Sandnes} T.~D.,  2023, \mn@doi
  [\apj] {10.3847/1538-4357/acf4ed}, \href
  {https://ui.adsabs.harvard.edu/abs/2023ApJ...955..137T} {955, 137}

\bibitem[\protect\citeauthoryear{{Thomas} \& {Veverka}}{{Thomas} \&
  {Veverka}}{1980}]{Thomas+Veverka1980}
{Thomas} P.,  {Veverka} J.,  1980, \mn@doi [\icarus]
  {10.1016/0019-1035(80)90221-3}, \href
  {https://ui.adsabs.harvard.edu/abs/1980Icar...41..365T} {41, 365}

\bibitem[\protect\citeauthoryear{{Walsh} \& {Levison}}{{Walsh} \&
  {Levison}}{2015}]{Walsh+Levison2015}
{Walsh} K.~J.,  {Levison} H.~F.,  2015, \mn@doi [\aj]
  {10.1088/0004-6256/150/1/11}, \href
  {https://ui.adsabs.harvard.edu/abs/2015AJ....150...11W} {150, 11}

\bibitem[\protect\citeauthoryear{{Zel'dovich} \& {Raizer}}{{Zel'dovich} \&
  {Raizer}}{1967}]{Zeldovich+Raizer1967}
{Zel'dovich} Y.~B.,  {Raizer} Y.~P.,  1967, {Physics of shock waves and
  high-temperature hydrodynamic phenomena}.
Academic Press, New York

\bibitem[\protect\citeauthoryear{{{\v{D}}urech} \&
  {Hanu{\v{s}}}}{{{\v{D}}urech} \& {Hanu{\v{s}}}}{2023}]{Durech+Hanus2023}
{{\v{D}}urech} J.,  {Hanu{\v{s}}} J.,  2023, \mn@doi [\aap]
  {10.1051/0004-6361/202345889}, \href
  {https://ui.adsabs.harvard.edu/abs/2023A&A...675A..24D} {675, A24}

\bibitem[\protect\citeauthoryear{{von Zeipel}}{{von
  Zeipel}}{1910}]{vonZeipel1910}
{von Zeipel} H.,  1910, \mn@doi [Astronomische Nachrichten]
  {10.1002/asna.19091832202}, \href
  {https://ui.adsabs.harvard.edu/abs/1910AN....183..345V} {183, 345}

\makeatother
\end{thebibliography}
}


\onecolumn
\begin{center}
  {
  \small
\begin{longtable}{cccllrc | cccccc}
    \caption{
      Outcomes of the tidal disruption SPH simulations.
      Columns list the simulation ID, for reference in Table~\ref{tab:reb_results};
      the periapsis, $q$;
      the speed at infinity, $v_\infty$;
      the mass, $M_{\rm A}$, where `d' indicates the differentiated asteroid;
      the number of particles, $N$;
      the spin angular momentum direction and magnitude, $\hat{L}$ and $L$;
      the number of resolved fragments larger than $3 \times 10^{15}$~kg, $n^{\rm frag}$;
      the mass fraction of material stripped onto unbound orbits, $f_{\rm unb}$;
      the mass fraction stripped onto bound orbits, $f_{\rm bnd}$;
      the mass fraction initially captured
      onto bound orbits with $Q < R_{\rm Hill}$, $f_{\rm capt}$;
      and the mass fraction and number of initially captured fragments
      larger than $3 \times 10^{15}$~kg,
      which are used as inputs for the long-term integrations,
      $f_{\rm capt}^{\rm frag}$ and $n_{\rm capt}^{\rm frag}$.
      Line breaks indicate subsets of scenarios
      in terms of which input parameters are fixed and varied.
    } \label{tab:results} \\

    \hline
    \hline\\[-0.4em]
    ID & $q$ & $v_\infty$ & $M_{\rm A}$ & $\;N$ & $\hat{L}$ & $L$
    & $n^{\rm frag}$ & $f_{\rm unb}$ & $f_{\rm bnd}$ & $f_{\rm capt}$ & $f_{\rm capt}^{\rm frag}$ & $n_{\rm capt}^{\rm frag}$
    \\
     & ($R_\smars$) & (km s$^{-1}$) & (kg) &  &  & ($L^{\rm max}$) &  & (\%) & (\%) & (\%) & (\%) &
    \\
    \hline\\[-1.05em]  
    \endfirsthead

    \multicolumn{4}{l}{\textbf{\tablename\ \thetable{}} -- continued} \\
    \hline\\[-1.05em]
    ID & $q$ & $v_\infty$ & $M_{\rm A}$ & $\;N$ & $\hat{L}$ & $L$
    & $n^{\rm frag}$ & $f_{\rm unb}$ & $f_{\rm bnd}$ & $f_{\rm capt}$ & $f_{\rm capt}^{\rm frag}$ & $n_{\rm capt}^{\rm frag}$
    \\
     & ($R_\smars$) & (km s$^{-1}$) & (kg) &  &  & ($L^{\rm max}$) &  & (\%) & (\%) & (\%) & (\%) &
    \\
    \hline\\[-1.05em]  
    \endhead

    \hline
    \endfoot

    \hline
    \hline
    \endlastfoot

    1 & $1.1$ & $0.0$ & $10^{20}$ & $10^{6.5}$ &  &  & 259 & $47.4$ & $52.6$ & $47.9$ & $47.7$ & 146  \\
      2 & $1.2$ & $0.0$ & $10^{20}$ & $10^{6.5}$ &  &  & 247 & $48.4$ & $51.6$ & $44.4$ & $44.3$ & 139  \\
      3 & $1.3$ & $0.0$ & $10^{20}$ & $10^{6.5}$ &  &  & 268 & $48.4$ & $51.6$ & $42.4$ & $42.2$ & 142  \\
      4 & $1.4$ & $0.0$ & $10^{20}$ & $10^{6.5}$ &  &  & 223 & $48.7$ & $51.3$ & $44.2$ & $44.1$ & 133  \\
      5 & $1.5$ & $0.0$ & $10^{20}$ & $10^{6.5}$ &  &  & 215 & $56.6$ & $43.4$ & $42.8$ & $42.7$ & 129  \\
      6 & $1.6$ & $0.0$ & $10^{20}$ & $10^{6.5}$ &  &  & 183 & $53.4$ & $46.6$ & $36.8$ & $36.7$ & 102  \\
      7 & $1.7$ & $0.0$ & $10^{20}$ & $10^{6.5}$ &  &  & 139 & $61.2$ & $38.8$ & $33.6$ & $33.5$ & 95  \\
      8 & $1.8$ & $0.0$ & $10^{20}$ & $10^{6.5}$ &  &  & 80 & $76.2$ & $23.8$ & $23.5$ & $23.4$ & 49  \\
\hline
  \end{longtable}
  }
\end{center}

\begin{center}
  {
  \small
  \begin{longtable}{ccccc | cccccrr}
    \caption{
      Outcomes of the long-term orbital integrations.
      Columns list the input disruption simulation's ID
      as listed in Table~\ref{tab:results};
      the encounter orientation longitude, $\phi$;
      the encounter inclination, $i$;
      Mars's obliquity, $o$;
      Mars's starting true anomaly, $\nu$,
      for scenarios with $e_\smars = 0.1$ (otherwise $e_\smars = 0$);
      the mass removed by colliding with Mars at a low periapsis, $M_{\rm rem}$;
      the mass lost by escaping far beyond Mars's Hill sphere, $M_{\rm esc}$;
      the mass that experienced a low-energy, non-disruptive collision, $M_{\rm bump}$;
      the mass still in orbit after 5~kyr, $M_{\rm orb}$;
      and the mass that survived in orbit or suffered a significant collision, $M_{\rm coll}$.
      Line breaks indicate subsets of scenarios
      in terms of which input parameters are fixed and varied.
    } \label{tab:reb_results} \\

    \hline
    \hline
    ID$_{\rm in}$ & $\phi$ & $i$ & $o$ & $\nu$ & $M_{\rm rem}$ & $M_{\rm esc}$ & $M_{\rm bump}$ & $M_{\rm orb}$ & $M_{\rm coll}$
    \\
     & ($^\circ$) & ($^\circ$) & ($^\circ$) & ($^\circ$) & ($10^{19}$ kg) & ($10^{19}$ kg) & ($10^{19}$ kg) & ($10^{19}$ kg) & ($10^{19}$ kg)
    \\
    \hline
    \endfirsthead

    \multicolumn{4}{l}{\textbf{\tablename\ \thetable{}} -- continued} \\
    \hline
    ID$_{\rm in}$ & $\phi$ & $i$ & $o$ & $\nu$ & $M_{\rm rem}$ & $M_{\rm esc}$ & $M_{\rm bump}$ & $M_{\rm orb}$ & $M_{\rm coll}$
    \\
     & ($^\circ$) & ($^\circ$) & ($^\circ$) & ($^\circ$) & ($10^{19}$ kg) & ($10^{19}$ kg) & ($10^{19}$ kg) & ($10^{19}$ kg) & ($10^{19}$ kg)
    \\
    \hline
    \endhead

    \hline
    \endfoot

    \hline
    \hline
    \endlastfoot

    6 & $0$ & $15$ & $30$ &  & $4.06$ & $0.00$ & $0.71$ & $0.00$ & $0.14$  \\
      6 & $15$ & $15$ & $30$ &  & $4.06$ & $0.00$ & $0.63$ & $0.00$ & $0.39$  \\
      6 & $30$ & $15$ & $30$ &  & $3.67$ & $0.39$ & $0.17$ & $0.00$ & $1.03$  \\
      6 & $45$ & $15$ & $30$ &  & $3.67$ & $0.39$ & $0.15$ & $0.00$ & $1.69$  \\
      6 & $60$ & $15$ & $30$ &  & $3.67$ & $0.39$ & $0.17$ & $0.00$ & $1.72$  \\
      6 & $75$ & $15$ & $30$ &  & $2.88$ & $0.39$ & $0.00$ & $0.79$ & $2.55$  \\
      6 & $90$ & $15$ & $30$ &  & $2.31$ & $0.39$ & $0.32$ & $1.36$ & $2.73$  \\
      6 & $105$ & $15$ & $30$ &  & $2.64$ & $0.39$ & $0.02$ & $1.03$ & $2.70$  \\
\hline
  \end{longtable}
  }
\end{center}

\let\clearpage\relax

\end{document}